\documentclass[12pt]{article}
\pdfoutput=1

\usepackage{amsmath,amssymb,amsfonts,amsfonts,bm,supertabular}
\usepackage{color}
\usepackage[hidelinks]{hyperref}
\usepackage{graphicx}
\usepackage{subcaption}

\newcommand{\N}{{\mathcal{N}}}
\newcommand{\beq}{\begin{equation}}
\newcommand{\eeq}{\end{equation}}
\newcommand{\bea}{\begin{eqnarray}}
\newcommand{\eea}{\end{eqnarray}}

\allowdisplaybreaks[4]

\date{}
\begin{document}
\begin{flushright}
DMUS-MP-21/11
\end{flushright}

\vspace{0.1cm}

\begin{center}
  {\large
Global Symmetries and Partial Confinement}
\end{center}
\vspace{0.1cm}
\vspace{0.1cm}
\begin{center}

Masanori Hanada,$^a$ Jack Holden,$^b$ Matthew Knaggs$^{c,d}$, Andy O'Bannon$^b$ 
\vspace{0.3cm}

$^a$
Department of Mathematics, University of Surrey, Guildford, Surrey, GU2 7XH, UK\\
$^b$
STAG Research Centre, University of Southampton, Southampton, SO17 1BJ, UK\\
$^c$
Department of Mathematics, King's College London, Strand, London WC2R 2LS, UK\\
$^d$
Department of Mathematical Sciences, University of Liverpool, Liverpool, L69 7ZL, UK\\

\end{center}

\vspace{1.5cm}

\begin{center}
  {\bf Abstract}
\end{center}

In gauge theories, spontaneous breaking of the centre symmetry provides a precise definition of deconfinement. In large-$N$ gauge theories, evidence has emerged recently that between confined and deconfined phases a partially-deconfined phase can appear, in which only a subset of colours deconfine. In the partially-deconfined phase, the centre symmetry is spontaneously broken, raising the question of whether an order parameter exists that can distinguish completely- and partially-deconfined phases. We present two examples in gauge theories of global symmetries that are spontaneously broken in the confined phase and preserved in the deconfined phase, and we show that this symmetry is spontaneously broken in the partially-deconfined phase. As a result, in these theories the transition from complete to partial deconfinement is accompanied by the spontaneous breaking of a global symmetry. The two examples are CP symmetry in $\N=1$ super-Yang-Mills with a massive gluino and theta-angle $\theta=\pi$, and chiral symmetry in a strongly-coupled lattice gauge theory. For $\N=1$ SYM we also present numerical evidence that the same phenomenon occurs at finite $N \geq 30$. We thus conjecture that global symmetries may provide order parameters to distinguish completely and partially deconfined phases generically, including at finite $N$.

\newpage
\setcounter{tocdepth}{2}
\tableofcontents
\section{Introduction}
\label{sec:intro}
\hspace{0.51cm}

In gauge theories, understanding the transition between confinement and deconfinement is a long-standing problem of crucial importance for heavy ion collisions, neutron stars, the early universe, and more. Via holography, any insight into the confinement/deconfinement transition can also reveal properties of black holes in the dual gravitational systems.

The transition can be characterised in many ways. For example, in the 't Hooft limit of infinite number of colours $N$, the confined phase typically has entropy of order $N^0$, arising from glueballs and other gauge-singlet degrees of freedom, while the deconfined phase typically has entropy of order $N^2$, arising from deconfined gluons and other adjoint degrees of freedom. Such behavior is generic, appearing for example in the strongly-coupled large-$N$ gauge theories of holography~\cite{Witten:1998zw}, as well as weakly-coupled gauge theories on spatial spheres, described by matrix models~\cite{Sundborg:1999ue,Aharony:2003sx}, among others.

However, a precise and widely accepted definition of the transition is in theories with a centre symmetry: the centre symmetry is preserved in the confined phase and spontaneously broken in the deconfined phase. The order parameter is the Polyakov loop, defined as follows. In quantum field theory, the thermal partition function can be represented as a path integral of the theory in Euclidean signature with the time direction compactified into a circle of circumference $1/T$. The Polyakov loop is then a Wilson loop around this circle, in the fundamental representation of the gauge group. The Polyakov loop is charged under the centre symmetry, and its expectation value, $P$, is zero in the confined phase and non-zero in the deconfined phase, indicating the spontaneous breaking of the centre symmetry.

In recent years, evidence has emerged that, quite generally in large-$N$ gauge theories, a partially-confined, or equivalently partially-deconfined, phase can appear between the completely-confined and completely-deconfined phases~\cite{Hanada:2016pwv,Berenstein:2018lrm,Hanada:2018zxn,Hanada:2019czd,Hanada:2019kue,Hanada:2019rzv,Hanada:2020uvt,Watanabe:2020ufk}. More specifically, in $SU(N)$ gauge theories at large $N$ a phase can appear in which only an $SU(M)$ sub-sector deconfines, where as the energy $E$ increases, $M$ increases from $M=0$ to $M=N$, at which points the phase connects to the completely-confined and completely-deconfined phases, respectively. Rather than commit to the name ``partially-confined phase'' or ``partially-deconfined phase'', we will henceforth refer to this phase simply as ``the partial phase''.
Moreover, instead of ``completely-deconfined" and ``completely-confined" we will refer to these phases simply as ``deconfined" and ``confined".

In the partial phase, confined and deconfined degrees of freedom coexist, similar to the coexistence of liquid water and ice at water's freezing temperature. However, unlike water, in the partial phase the coexistence occurs in the ``internal'' colour space, where interactions are non-local, and hence can lead to non-trivial $T$ dependence.\footnote{In QCD, the change of the ratio between the numbers of flavors and confined colour degrees of freedom can lead to a nontrivial $T$ dependence, even in the free limit~\cite{Hanada:2019kue}.} In fact, the underlying physics is more similar to Bose-Einstein condensate (BEC), where the confined and deconfined phases are like the superfluid and normal fluid phases, respectively, and the partial phase is like a two-fluid phase of $N$ bosons split between $N-M$ in the superfluid and $M$ in the normal fluid. The similarity to a BEC can be made precise, because a BEC can be treated in terms of gauge theory: the indistinguishability of the $N$ bosons means their permutation symmetry, $S_N$, is gauged~\cite{Hanada:2020uvt}.

\begin{figure}
	\begin{center}
		\includegraphics[width=0.8\textwidth]{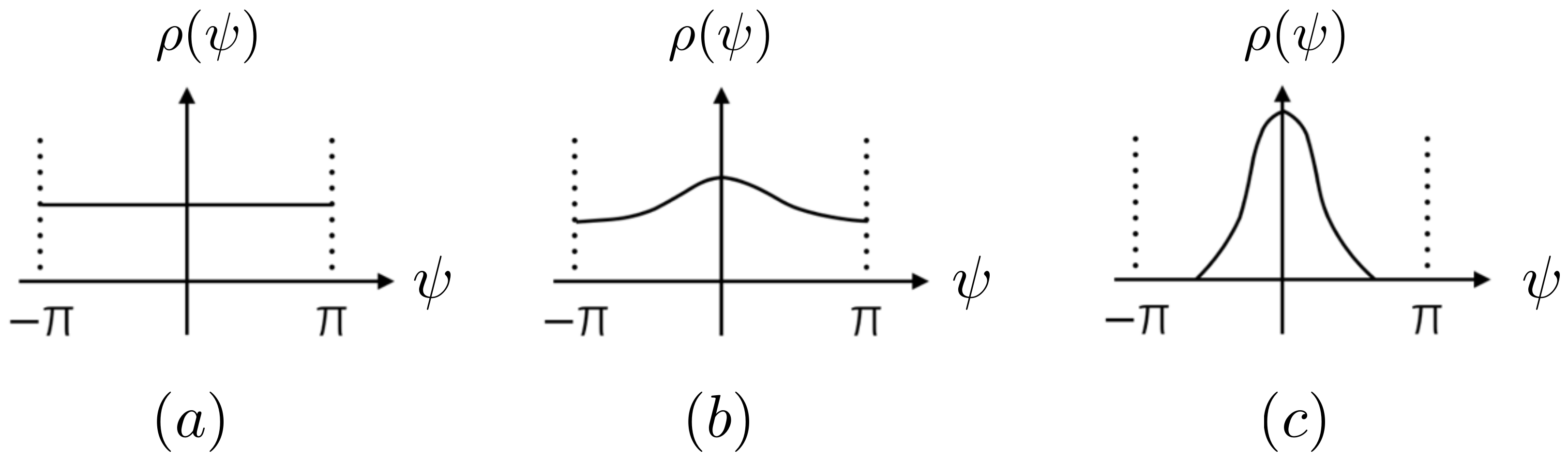} 
		\caption{Schematic depictions of the eigenvalue distributions, $\rho(\psi)$, of the path-ordered exponential of the gauge holonomy around the thermal circle in a Euclidean gauge theory, in the 't Hooft large-$N$ limit. We assume an $SU(N)$ gauge group, so the eigenvalues are complex phases distributed on $\psi\in[-\pi,\pi]$. For such theories with a centre symmetry, $\rho(\psi)$ provides a gauge-invariant way to distinguish three phases, as follows. (a) The confined phase, in which $\rho(\psi)$ is uniform, so the centre symmetry is unbroken, and hence the Polyakov loop expectation value vanishes, $P=0$. (b) The partial phase, in which $\rho(\psi)$ is non-uniform, so the centre symmetry is spontaneously broken and $P\neq0$. However, $\rho(\psi)$ is ungapped. (c) The deconfined phase, in which $\rho(\psi)$ is non-uniform, so again the centre symmetry is broken and $P\neq0$, but now $\rho(\psi)$ is gapped. (a) and (b) are separated by a Hagedorn transition, while (b) and (c) are separated by a Gross-Witten-Wadia (GWW) transition.}
		\label{fig:evalcartoon}
	\end{center}
\end{figure}

Crucially, the partial phase can be characterised in a gauge-invariant manner. Perhaps the most precise characterisation uses the eigenvalues of the path-ordered exponential of the gauge holonomy around the thermal circle, a.k.a. the Polyakov loop without the colour trace. In the large-$N$ limit these eigenvalues can be treated as a continuum, and their distribution, $\rho(\psi)$, provides precise definitions of the confined, partial, and deconfined phases. Cartoons of $\rho(\psi)$ in these three phases appear in Fig.~\ref{fig:evalcartoon} for gauge group $SU(N)$, where the eigenvalues are all complex phases distributed on $\psi\in[-\pi,\pi]$.

In the confined phase, shown in Fig.~\ref{fig:evalcartoon} (a), the distribution is uniform: $\rho(\psi)=\frac{1}{2\pi}$ when properly normalised. The centre symmetry acts as a permutation of the eigenvalues, which in Fig.~\ref{fig:evalcartoon} is a discrete translational symmetry acting horizontally, so a constant $\rho(\psi)$ preserves the centre symmetry. Correspondingly, the Polyakov loop expectation value vanishes, $P=0$, because upon taking the trace, all the phases cancel one another. In the partial phase, shown in Fig.~\ref{fig:evalcartoon} (b), $\rho(\psi)$ becomes non-uniform, indicating spontaneous breaking of the centre symmetry, and correspondingly, $P\neq0$. The distribution remains ungapped, however, meaning eigenvalues appear for all possible values on the interval $\psi \in [-\pi,\pi]$. Crucially, $\textrm{min}\left[\rho(\psi)\right] = \frac{1}{2\pi}\left(1-\frac{M}{N}\right)$ provides a gauge-invariant definition of the size $M$ of the deconfined subsector~\cite{Hanada:2020uvt}. In the deconfined phase, shown in Fig.~\ref{fig:evalcartoon} (c), the distribution is non-uniform, so again centre symmetry is spontaneously broken and $P\neq0$, but now the distribution is gapped, as some eigenvalues near $-\pi$ and $\pi$ disappear.

\begin{figure}[ht!]
\begin{center}
\scalebox{0.6}{
\includegraphics{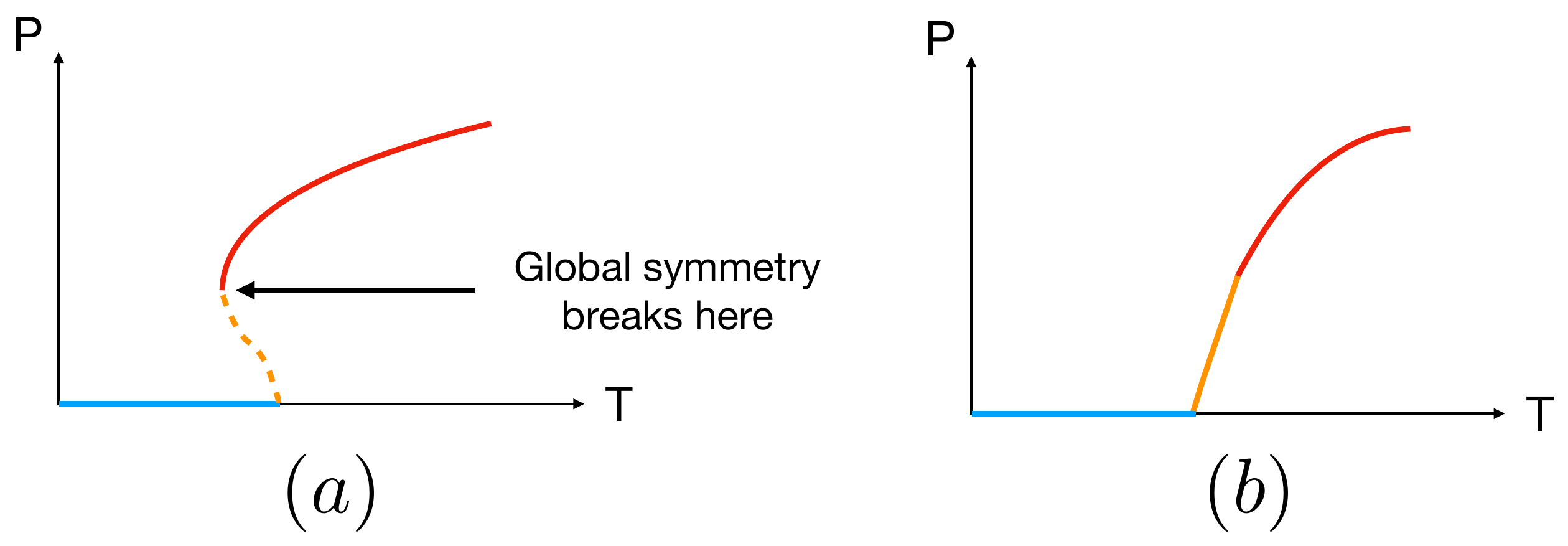}}
\end{center}
\caption{
Schematic depiction of the Polyakov loop expectation value, $P$, as a function of temperature $T$ for a confinement/deconfinement transition that is (a) first order or (b) second order or higher, i.e. continuous. The blue, orange and red curves represent the confined, partial, and deconfined phases, respectively. The theories we study in this paper have first-order transitions, as in (a). We find that the global symmetry that is spontaneously broken in the confined phase, namely CP or chiral symmetry, is also broken in the partial phase. As a result, the transition from the deconfined to partial phases is characterised by the spontaneous breaking of this global symmetry, as indicated in (a). 
}\label{fig:partial-confinement-cartoon}
\end{figure}

The order of the transition from confined to deconfined phase depends on the detailed dynamics of each particular gauge theory. The two logical possibilities are either first order, or second order or higher, i.e. continuous. By extension, where the partial phase appears in the thermodynamic phase diagram depends on detailed dynamics. For each of the two logical possibilities, Fig.~\ref{fig:partial-confinement-cartoon} shows cartoons of the Polyakov loop expectation value as a function of $T$. In the first order case, depicted in Fig.~\ref{fig:partial-confinement-cartoon} (a), the partial phase is the thermodynamically unstable phase between the confined and deconfined phases. In this case, in the free energy the partial phase is the local maximum between two local minima, and the transition occurs when the local minima exchange roles as the global minimum. In the continuous case, depicted in Fig.~\ref{fig:partial-confinement-cartoon} (b), the partial phase is a stable phase that connects the confined and deconfined phases. The same theory can exhibit both cases depending on the details of the parameters, such as quark mass. 
In either case, the partial phase connects to the confined phase via a Hagedorn transition, characterised by an exponential growth of the density of states when approaching the transition from the confined phase, and connects to the deconfined phase via a Gross-Witten-Wadia (GWW) transition, a transition defined by the appearance of the gap in $\rho(\psi)$. The order of the GWW transition depends on the details of the dynamics.\footnote{The original example of the GWW transition~\cite{Gross:1980he,Wadia:2012fr} was in two-dimensional lattice Yang-Mills theory, and the transition was of third order.
} 

Because the centre symmetry is preserved in the confined phase and broken in the partial and deconfined phases, $P$ can act as an order parameter to distinguish the former from the latter. The gap in $\rho(\psi)$ can distinguish the partial and deconfined phases; however this gap is not associated with any symmetry, and hence is not an order parameter. The question therefore arises of whether any order parameter can distinguish the partial and deconfined phases.\footnote{
An intuitive way to understand this transition in terms of symmetry is to relate it to the breaking and restoration of gauge symmetry~\cite{Hanada:2019czd,Hanada:2020uvt,Hanada:2021ipb}. However, note that gauge symmetry is a sort of redundancy, unlike a global symmetry. Here we want to find a more conventional (and less-controversial) characterisation, based on global symmetry.} To date, the only proposals for such order parameters are in holography, and rely in part on arguments about black holes (namely Gregory-Laflamme instabilities of small black holes in AdS)~\cite{Asplund:2008xd,Berenstein:2018lrm}, but a completely generic proposal remains absent.

In this paper, we propose an order parameter that can distinguish the partial and deconfined phases. In short, we argue that if a global symmetry is spontaneously broken in the confined phase and preserved in the deconfined phase, then that global symmetry should be broken in the partial phase. As a result, the transition from the deconfined to the partial phase should be accompanied by spontaneous breaking of the global symmetry, and hence these two phases can be distinguished by an order parameter. We present two examples in which this occurs, and then argue that such behaviour is generic.

Our first example is $\N=1$ supersymmetric (SUSY) Yang-Mills (SYM) theory with gauge group $SU(N)$ on $S^1 \times \mathbb{R}^3$ with periodic boundary conditions for both the gluons and gluino, and with a non-zero gluino mass. We also introduce a non-zero $\theta$-angle, $\theta=\pi$, in which case a mixed CP-centre symmetry anomaly forces CP symmetry to be spontaneously broken in the confined phase~\cite{Gaiotto:2017yup}. When $\theta=0$, CP symmetry is preserved in both the confined and deconfined phases~\cite{Gaiotto:2017yup}. We work at weak coupling, in the so-called Abelian large-$N$ limit, which is distinct from the 't Hooft large limit, as we discuss in section~\ref{sec:4dSYM}. We also present some finite-$N$ numerical results, though we consider relatively large $N \geq 30$.

Our second example is a strongly-coupled $SU(N)$ YM theory on $S^1$ times a three-dimensional spatial lattice. We work in the 't Hooft large-$N$ limit and employ Eguchi-Kawai reduction to a single site in space. We then add quarks in a probe limit, so that to leading approximation the centre symmetry is not explicitly broken, while chiral symmetry is spontaneously broken in the confined phase but restored in the deconfined phase.

In each case we perform numerical calculations, which for the lattice YM theory are in the 't Hooft large-$N$ limit, while for $\N=1$ SYM, some are genuine finite-$N$ calculations. In each case, we reproduce known results for the confinement/deconfinement transition, which is first order in both theories, so that $P$ behaves as shown in Fig.~\ref{fig:partial-confinement-cartoon} (a). Our new results are numerical evaluations for the unstable saddles connecting the confined and deconfined phases, where we identify the partial phase using the Polyakov eigenvalue distribution. We also numerically compute the order parameters for CP and chiral symmetries, respectively, and show that in each case this global symmetry is spontaneously broken in the partial phase. As a result, in the transition from deconfined to partial phase, the global symmetry breaks spontaneously, as we have indicated in Fig.~\ref{fig:partial-confinement-cartoon} (a).

In short, our two examples share the following behaviour. In the confined phase, the centre symmetry is preserved but another global symmetry, namely CP or chiral symmetry, is spontaneously broken. In the deconfined phase, the centre symmetry is spontaneously broken, but the other global symmetry is preserved. In the partial phase, both symmetries are spontaneously broken. In these examples, we therefore prove that the partial phase is distinguished by global symmetries from both the confined and deconfined phases, and in particular, we identify an order parameter to distinguish the deconfined and partial phases, as mentioned above.

Furthermore, we conjecture that such behaviour may be generic, and can be used to identify, or even define, partial phases, in theories with both a centre symmetry and a global symmetry that is spontaneously broken in the confined phase but preserved in the deconfined phase. Specifically, the partial phase could be defined as the phase in which both of these symmetries are spontaneously broken. Such a definition, using symmetries alone, would obviously have advantages. In particular, such a definition could apply for any $N$, not just the large-$N$ limit. Indeed, our finite-$N$ numerical results for $\N=1$ SYM provide some compelling evidence for this.

However, we can provide more general plausibility arguments for this conjecture. For instance, we expect that whatever strong-coupling physics breaks chiral symmetry in a confined phase should occur also in a partial phase's confined subsector. As a result, if chiral symmetry is preserved in the deconfined phase, then chiral symmetry can provide an order parameter distinguishing deconfined and partial phases. Anomalies may provide a more rigorous argument, being independent of coupling strengths and invariant under renormalisation group flows, including flows induced by $T$. For an anomaly-matching argument that chiral symmetry should be partially broken in the partial phase, see ref.~\cite{Hanada:2019kue}. Another example is $\N=1$ SYM with $\theta=\pi$, where an anomaly forces CP symmetry to be spontaneously broken at $T=0$~\cite{Gaiotto:2017yup}. As a result, as we raise $T$, CP symmetry must remain spontaneously broken in the confined phase. Presumably, in the partial phase the same anomaly arguments extend to the confined subsector, and hence CP symmetry must be spontaneously broken, as indeed we observe. Such arguments should generalise to any theory where an anomaly forces spontaneous symmetry breaking in the confining vacuum: we expect the symmetry to remain spontaneously broken in any $T>0$ phase that has a confined subsector, and if the symmetry is preserved in the deconfined phase, then this symmetry can provide an order parameter distinguishing deconfined and partial phases.

This paper is organised as follows. In Sec.~\ref{sec:4dSYM} we study mass-deformed $\N=1$ SYM on $S^1 \times \mathbb{R}^3$, in Sec.~\ref{sec:EK-model} we study the EK reduction of strongly-coupled YM with probe quarks, and in Sec.~\ref{sec:conclusion} we conclude with a summary and outlook for future research. Two appendices contain technical results useful for our study of $\N=1$ SYM.

\section{Weakly-coupled, softly-broken massive ${\cal N}=1$ SYM on $S^1\times{\mathbb R}^3$}
\label{sec:4dSYM}
\hspace{0.51cm}
In this section, we consider ${\cal N}=1$ $SU(N)$ SYM on Euclidean $S^1 \times {\mathbb R}^3$, with a non-zero gluino mass, which breaks SUSY. The action for this theory is
\beq
\label{eq:symaction}
S_{\textrm{SYM}}= \int dx_0 \, d^3\vec{x} \left[ \frac{1}{4g^2}  \mathrm{tr}(F_{\mu\nu}F^{\mu\nu}) - \frac{i\theta}{8\pi^2} \mathrm{tr}(F_{\mu\nu} \tilde{F}^{\mu\nu})
    + \frac{2i}{g^2} \mathrm{tr} (\bar{\Psi}\bar{\sigma}^\mu D_\mu \Psi) + \frac{m}{g^2}(\mathrm{tr}(\Psi \Psi) + \mathrm{c.c.})\right],
\eeq
with $x_0$ the $S^1$ coordinate, $\vec{x}$ the $\mathbb{R}^3$ coordinates, $g$ the gauge coupling, $\theta$ the theta-angle, $F_{\mu\nu}$ the field strength of the $SU(N)$ gauge field $A_{\mu}$, called the gluon, where $\mu,\nu=0,1,2,3$, $\tilde{F}_{\mu\nu}$ the Hodge dual of $F_{\mu\nu}$, $\Psi$ an adjoint Weyl fermion, called the gluino, where $D_{\mu} \Psi = \partial_{\mu}\Psi + i [A_{\mu},\Psi]$, and $m$ the gluino mass.

If $m=0$ then the theory has $\N=1$ SUSY. Although the fields in the action are then massless, a mass gap is dynamically generated at the scale $\Lambda$, given at two-loop order by
\begin{subequations}
\beq
\label{eq:lambdadef}
\Lambda^3 \equiv \frac{4 \pi}{3N} \, M_{\rm PV}^3 \,\, \textrm{Im}\,\tau(M_{\rm PV})\,\textrm{exp}\left[\frac{2 \pi i \,\tau(M_{\rm PV})}{N}\right],
\eeq
\beq
\label{eq:taudef}
\tau \equiv \left(\frac{4\pi \, i}{g^2}+\frac{\theta}{2\pi} \right),
\eeq
\end{subequations}
with $M_{\rm PV}$ a Pauli-Villars mass and $\tau(M_{PV})$ the running coupling in eq.~\eqref{eq:taudef} evaluated at $M_{\rm PV}$. If $m\neq 0$, then all SUSY is broken, and in particular if $m\to \infty$ then the theory reduces to pure $SU(N)$ YM theory.

The theory in eq.~\eqref{eq:symaction} has a $\mathbb{Z}_N$ centre symmetry for all values of $N$, $g$, $\theta$, and $m$, and has CP symmetry for all $N$, $g$, and $m$, but only for certain values of $\theta$. To see why, recall that the term $\frac{1}{8\pi^2} \mathrm{tr}(F\tilde{F})$ in eq.~\eqref{eq:symaction} is CP-odd and topological, and that the integral of this term takes integer values. As a result, $\theta$ is periodic, $\theta\sim\theta+2\pi$, and transforms under CP as $\theta\to -\theta$. The theory thus has CP symmetry only when $\theta=0$ or $\theta=\pi$ mod $2\pi$. In what follows, for brevity we will restrict to $\theta \in [0,\pi]$, leaving the $2 \pi$ periodicity implicit.

The theory in eq.~\eqref{eq:symaction} has a mixed CP-centre symmetry anomaly. Assuming that the vacuum is gapped and confining for all $\theta$,  when $\theta=\pi$ this anomaly forces the vacuum to break CP symmetry spontaneously~\cite{Gaiotto:2017yup}, an effect known as the Dashen phenomenon~\cite{Dashen:1970et}.

We denote the  $S^1$ circumference as $L$. If we write the generating functional of connected correlation functions as a path integral, we have two options for boundary conditions around the $S^1$. First are ``thermal'' boundary conditions, namely a periodic boundary condition for the gluon and an anti-periodic boundary condition for the gluino. The resulting path integral computes correlators in a thermal state at temperature $T = 1/L$. For example, the path integral with no operator insertions gives the partition function, i.e. the exponential of minus the free energy over $T$. If we could compute such a path integral, then we could identify phase transitions from the free energy. However, such a path integral is prohibitively difficult to calculate: perturbative methods are unreliable because deconfinement occurs when $T \sim \Lambda$ and so the theory is strongly coupled, and lattice methods are unreliable because we want $\theta=\pi$, but any $\theta \neq 0$ has a ``sign problem'' due to the $i$ multiplying $\theta$ in eq.~\eqref{eq:symaction}, which comes from the Wick rotation to Euclidean signature.

Following ref.~\cite{Chen:2020syd}, we thus choose the second option for boundary conditions, namely periodic boundary conditions around the $S^1$ for \textit{both} the gluon and gluino. These boundary conditions preserve SUSY, which at first may not seem helpful. If SUSY is preserved then the path integral with no operator insertions is a Witten index~\cite{Witten:1982df,Witten:1982im}, which counts SUSY ground states. For gauge group $SU(N)$, this theory's Witten index is simply $SU(N)$'s dual Coxeter number, $N$. Moreover, the Witten index is invariant under any continuous deformation that preserves SUSY, including continuous changes in $1/L$. As a result, no phase transitions can occur as a function of $1/L$: the theory is always in a confined phase, with centre symmetry preserved, and CP symmetry preserved when $\theta=0$, explicitly broken when $\theta \in (0,\pi)$, and spontaneously broken when $\theta=\pi$.

However, following ref.~\cite{Chen:2020syd}, we will break SUSY softly by introducing a small gluino mass, $m \ll \Lambda$, so that the path integral is no longer a Witten index, and in particular can vary as a function of $1/L$, allowing for the possibility of phase transitions. Actually, strictly speaking, given our boundary conditions, any changes in symmetries will be \textit{quantum} phase transitions, occurring at $T=0$, and hence arising from quantum rather than thermal fluctuations. Nevertheless, again following ref.~\cite{Chen:2020syd}, to guide our intuition we will think of $1/L$ as temperature, treat the path integral as a measure of ``free energy'', and refer to any symmetry changes simply as ``phase transitions''.

Crucially, when $m\ll \Lambda$, SUSY still provides enough control to compute the path integral, at least in combination with approximations that we will describe in detail in sec.~\ref{sec:R3S1-eff-action}, including $L$ small enough to justify perturbation theory in $g(1/L)\ll 1$, and energy small enough to justify an effective description, i.e. to integrate out all but the lightest modes. In these limits, for any $N$ and $\theta$ a deconfinement transition occurs as $L$ shrinks, so that centre symmetry breaks spontaneously at some critical $L$, i.e. some critical ``temperature''. The deconfinement transition is first order when $N>2$ and second order when $N=2$~\cite{Poppitz:2012sw}. When $\theta=0$ or $\theta=(0,\pi)$, CP symmetry is unchanged in the deconfinement transition, remaining preserved or explicitly broken, respectively. However, when $\theta=\pi$, CP symmetry is restored at some critical $L$, which for $N>2$ is the same critical $L$ as deconfinement, but when $N=2$ is a \textit{smaller} critical $L$. Indeed, the main result of ref.~\cite{Chen:2020syd} is that when $N=2$, as $L$ shrinks deconfinement occurs before CP restoration, producing, for some range of $L$, a phase with spontaneously broken centre \textit{and} CP symmetries. From our perspective, this is a partial phase.

In sec.~\ref{sec:R3S1-eff-action} we will work in the limits mentioned above: $m\ll \Lambda$, small $L$, and low energy. In particular, following refs.~\cite{Davies:1999uw,Davies_2003,Poppitz:2012nz,Poppitz:2012sw,Anber_2014,Poppitz:2021cxe} we will perform a dimensional reduction on the $S^1$, retaining sufficiently light modes, and obtaining an effective theory on $\mathbb{R}^3$. If we then take the so-called Abelian large-$N$ limit, explained below, then the extrema of that effective theory's potential give the free energy of the original theory on $S^1 \times \mathbb{R}^3$. In sec.~\ref{sec:minima}, we will review known solutions for locally stable minima describing the confined phase for any $N$, and the deconfined phase in the Abelian large-$N$ limit. In sec.~\ref{sec:PDC-phase} we will present novel solutions for locally unstable maxima describing the partial phase in the Abelian large-$N$ limit. Our main result will be that in the partial phase, $\mathrm{tr}(F_{\mu\nu} \tilde{F}^{\mu\nu})$ has a non-zero expectation value, $\langle\mathrm{tr}(F_{\mu\nu} \tilde{F}^{\mu\nu})\rangle \neq 0$, indicating spontaneous breaking of CP symmetry. As a result, $\langle\mathrm{tr}(F_{\mu\nu} \tilde{F}^{\mu\nu})\rangle$ provides a gauge-invariant order parameter distinguishing the deconfined and partial phases.

\subsection{The Effective Theory on $\mathbb{R}^3$}\label{sec:R3S1-eff-action}
\hspace{0.51cm}

In this subsection we will briefly review the derivation of the effective theory of refs.~\cite{Poppitz:2012sw,Davies_2003,Poppitz:2012nz,Anber_2014}, reviewed recently in ref.~\cite{Poppitz:2021cxe}. The derivation consists of Kaluza-Klein (KK) reducing all fields on the $S^1$ and integrating out all modes except the lightest modes, thus obtaining an effective theory of these light modes in $\mathbb{R}^3$. Our goal is to derive this effective theory's (bosonic) potential, $\mathcal{V}$, whose extrema give the free energy of the original theory on $S^1 \times \mathbb{R}^3$. We will begin with $m=0$, and exploit the full power of SUSY to derive the effective theory in $\mathbb{R}^3$. We will then break SUSY softly, by introducing $m \ll \Lambda$, and compute $\mathcal{V}$ to order $m$, and in the limit of weak coupling, as mentioned above.

In the KK reduction from $S^1 \times \mathbb{R}^3$ to $\mathbb{R}^3$, the gluon's $S^1$ component, $A_0$, reduces to a KK tower of adjoint scalars in $\mathbb{R}^3$, including a massless adjoint scalar. In fact, we can gauge away the massive adjoint scalars: via gauge transformations we make $A_0$ constant over the $S^1$. We can also gauge transform $A_0$ to be diagonal in $SU(N)$. Doing so makes the Polyakov loop operator diagonal, with each diagonal entry being pure phase,
\begin{align}
    \langle &\mathcal{P}e^{i\oint dx_0 \bm{A}_0} \rangle
    =
    \langle e^{iL\bm{A}_0} \rangle =
    \begin{pmatrix}
    e^{i\phi_0} &  &   \\
     & \ddots & \\
     &        & e^{i\phi_{N-1}}
    \end{pmatrix}, \label{eq:full-SYM-action}
\end{align}
where $\mathcal{P}$ denotes path-ordering, each phase $\phi_i$ is $2\pi$ periodic, $\phi_i\sim\phi_i+2\pi$ for $i=0,\ldots,N-1$, and because the gauge group is $SU(N)$, $\sum_{i=0}^{N-1}\phi_i\equiv 0\ {\rm mod}\ 2\pi$. The only remaining gauge invariance is the Weyl group, which for $SU(N)$ is the permutation group, $S_N$. We use $S_N$ permutations to order the phases,
\begin{align}
0\le \phi_{N-1}\le\phi_{N-2}\le\cdots\le\phi_{0}\le 2\pi,
\end{align}
modulo their periodic identifications. With these choices we have fixed the gauge completely. We next collect these phases into an $N$-component vector,
\beq
\bm{\phi} = (\phi_0,\phi_1,\cdots,\phi_{N-1}),
\eeq
which in the effective theory on $\mathbb{R}^3$ is an adjoint Higgs field.

In the KK reduction, the gluon's spatial components reduce to a KK tower of vector fields in $\mathbb{R}^3$, including massless $SU(N)$ gauge fields and their massive KK partners. The gluino reduces to a KK tower of fermionic fields, which when $m=0$ are the superpartners of the vector fields and the adjoint Higgs field, $\bm{\phi}$. We integrate out all the massive KK fields, obtaining an effective theory of the massless fields alone, namely the massless $SU(N)$ vector, a massless three-dimensional Dirac fermion, and $\bm{\phi}$, which together comprise a three-dimensional $\N=2$ vector multiplet.

A non-trivial Higgs field $\bm{\phi}$ breaks the gauge group $SU(N)$ to a subgroup. Following refs.~\cite{Poppitz:2012sw,Davies_2003,Poppitz:2012nz,Anber_2014,Poppitz:2021cxe}, we assume that all the phases $\phi_i$ are distinct, so that the gauge group breaks to the maximal Abelian subgroup, $SU(N)\rightarrow U(1)^{N-1}$. Such a vacuum is generic, compared to vacua where some $\phi_i$ are the same, and hence $SU(N)$ breaks to a subgroup with at least one non-Abelian factor. In the maximal Abelian case, the off-diagonal entries of the $SU(N)$ gauge field, also known as the W-bosons, acquire masses $\geq\frac{2\pi}{NL}$, whereas the diagonal gauge field entries comprise a number $N$ of massless $U(1)$ gauge fields in $\mathbb{R}^3$. We integrate out the W-bosons, obtaining an effective theory of the massless fields alone, namely the $N$ $U(1)$ gauge fields, the adjoint Higgs field $\bm{\phi}$, and their fermionic superpartners. These fields interact with coupling strength $g(\frac{2\pi}{NL})$, that is, the running coupling ``frozen'' at the scale of the lightest W-boson. As mentioned above, we assume $\frac{2\pi}{NL} \gg \Lambda$, so that $g(\frac{2\pi}{NL})\ll 1$. Henceforth, $g$ will always denote $g(\frac{2\pi}{NL})$.

Again following refs.~\cite{Poppitz:2012sw,Davies_2003,Poppitz:2012nz,Anber_2014,Poppitz:2021cxe}, we collect the $N$ $U(1)$ gauge fields into an $N$-component vector of Abelian gauge fields, $\bm{A}$. We then Hodge dualise in $\mathbb{R}^3$ to obtain an $N$-component scalar field, $\bm{\sigma}$, the ``scalar photon''. Explicitly, we define $\bm{\sigma}$ via $d\bm{\sigma}\equiv\frac{4\pi L}{g^2} (\ast d\bm{A})$. Both $\bm{\phi}$ and $\bm{\sigma}$ come from a compact gauge group, so the components of each are $2\pi$ periodic.

The massless sector now consists of $\bm{\sigma}$, $\bm{\phi}$, and their fermionic superpartners. Together, these comprise a number $N$ of three-dimensional $\N=2$ chiral superfields. We collect these into an $N$-component $\N=2$ chiral superfield, ${\bf{X}}=(X_0,X_1,\ldots,X_{N-1})$, whose lowest component is an $N$-component complex scalar, $\bm{z}$, built from $\bm{\phi}$ and $\bm{\sigma}$:
\beq
\label{eq:zdef}
\bm{z} \equiv i \left[ \tau (\bm{\phi} - \bm{\phi}_W) + \bm{\sigma} \right],
\eeq
where $\bm{\phi}_W$ is proportional to the $SU(N)$ Weyl vector: $\left(\bm{\phi}_W\right)_j=\frac{2\pi}{N} \, (N-1-j)$ for $j=0,1,\ldots,N-1$. SUSY requires the bosonic part of the effective action to take the form~\cite{Davies_2003} 
\begin{align}
    S_{\textit{eff}} =\int d^3\vec{x} \left[ \mathcal{K}(\bm{X}, \bm{X}^\dagger)|_{\Theta \Theta \bar{\Theta} \bar{\Theta}} + \mathcal{W}(\bm{X})|_{\Theta \Theta} + \bar{\mathcal{W}}(\bm{X}^\dagger)|_{\bar{ \Theta} \bar{\Theta}}\right],
\end{align}
where $\Theta$ and $\bar{\Theta}$ are the fermionic coordinates of superspace, $\mathcal{K}(\bm{X}, \bm{X}^\dagger)$ is the K\"ahler potential, and $\mathcal{W}(\bm{X})$ is the superpotential. The bosonic potential is then
\beq
\label{eq:susyv}
   {\cal V} = \left(\frac{\partial^2\mathcal{K}}{\partial d\bm{z} \, \partial d\bm{z}^\dagger}\right)^{-1} \left| \frac{\partial \mathcal{W}}{\partial \bm{z}} \right|^2.
\eeq

In the effective theory of refs.~\cite{Poppitz:2012sw,Poppitz:2012nz,Anber_2014,Poppitz:2021cxe}, the K\"ahler potential includes a trivial, classical contribution, plus perturbative loop corrections. In what follows, we will need only the trivial, classical contribution, for which the bosonic fields' kinetic terms are simply
\begin{align}
\label{eq:kdef}
\mathcal{K}(\bm{X}, \bm{X}^\dagger)|_{\Theta \Theta \bar{\Theta} \bar{\Theta}} = \frac{g^2}{16\pi^2 L}|d\bm{z}|^2=\frac{1}{g^2L} \lvert d\bm{\phi} \rvert ^2 + \frac{g^2}{16\pi^2L}|d\bm{\sigma} + \frac{\theta}{2\pi} d\bm{\phi}|^2.
\end{align}

Generically, in any SUSY theory the superpotential, $\mathcal{W}$, is a sum of two contributions, one perturbative and one non-perturbative. In this case, the classical superpotential vanishes, and SUSY non-renormalisation theorems subsequently guarantee that all perturbative contributions vanish.

The non-perturbative contribution to $\mathcal{W}$ is non-zero due to so-called monopole-instantons, which come from the KK reduction of instantons. Monopole-instantons have non-zero Chern number, and hence non-zero topological charge, and an index theorem then implies that they have two fermionic zero modes~\cite{Poppitz:2008hr}. They can therefore contribute to $\mathcal{W}$. The resulting form of $\mathcal{W}$ is determined, up to an overall constant, by holomorphy in $\bm{X}$, single-valuedness under the periodic shifts in $\sigma$, and the R-symmetry. A calculation of $\Psi$'s two-point correlator then fixes the overall constant. Detailed discussions of the monopole-instantons and their contribution to $\mathcal{W}$, appear in refs.~\cite{Poppitz:2012sw,Davies_2003,Poppitz:2012nz,Anber_2014,Poppitz:2008hr,Poppitz:2021cxe}. We only need the final result for $\mathcal{W}$, which is
\begin{align}
\label{eq:wdef}
    \mathcal{W}(\bm{X}) = \frac{L M_{\rm PV}}{g^2}\left( \sum_{j=0}^{N-1}  e^{(X_j-X_{j+1})} +  e^{2\pi i \tau + (X_0-X_1)}  \right).
\end{align}
The monopole-instanton operators are
\begin{align}
\label{eq:monopole-instanton-operator}
    M_j \equiv \exp\left((z_j-z_{j+1}) + i \frac{\theta}{N}\right),
\end{align}
where for $j=N-1$ we define $z_{j+1}=z_N \equiv z_0$. Crucially, the $M_j$ are not all independent: their definition in eq.~\eqref{eq:monopole-instanton-operator} implies the constraint,
\begin{eqnarray} \label{eq:constraint}
M_0\times M_1\times\cdots \times M_{N-1}=e^{i\theta}. 
\end{eqnarray}
Evaluating $\mathcal{W}$ in eq.~\eqref{eq:wdef} on the lowest component of $\bm{X}$ shows explicitly how the monopole-instantons contribute:
\begin{align}
\label{eq:wdef2}
    \mathcal{W}(\bm{z}) = \frac{L M_{\rm PV}}{g^2}\,e^{-i\frac{\theta}{N}} \left[\sum_{j=0}^{N-1} M_j +  e^{2\pi i \tau} M_0  \right].
\end{align}

Plugging $\mathcal{K}$ from eq.~\eqref{eq:kdef} and $\mathcal{W}$ from eq.~\eqref{eq:wdef2} into eq.~\eqref{eq:susyv} gives us the bosonic potential,
\begin{subequations}
 \label{eq:SUSYpotential}
\begin{align}
    {\cal V} & = {\cal V}_0 \, \sum_{j=0}^{N-1} |M_j - M_{j-1}|^2, \\ & = {\cal V}_0 \, \sum_{i=0}^{N-1} \left [ M_j M_j^* - M_j M_{j-1}^* - M_{j-1} M_j^* + M_{j-1}M_{j-1}^*\right],
\end{align}
\end{subequations}
where the overall constant is most compactly written in terms of $\Lambda$ in eq.~\eqref{eq:lambdadef},
\begin{align}
    {\cal V}_0 \equiv \frac{9\, N^2}{\left(4 \pi\right)^2} \frac{L^3 \, \Lambda^6}{g^2}.
\end{align}
Although the superpotential $\mathcal{W}$ in eq.~\eqref{eq:wdef2} received contributions from individual monopole-instantons, $M_j$, the bosonic potential $\mathcal{V}$ in eq.~\eqref{eq:SUSYpotential} receives contributions only from bound states of monopole-instantons with anti-monopole-instantons, such as $M_jM_j^*$. These so-called ``bions'' have zero net topological charge, and hence have no fermionic zero modes. A microscopic analysis shows that these bions are held together by fermion exchange~\cite{Unsal_2009}.

Using eqs.~\eqref{eq:zdef} and~\eqref{eq:monopole-instanton-operator} we can determine how the monopole-instantons $M_j$ transform under centre and CP symmetries. A centre symmetry transformation simply permutes the Polyakov line phases, $\phi_j \to \phi_{j+1}$, with $\phi_{N+1} \equiv \phi_0$. A CP symmetry transformation sends $\theta \to -\theta$, leaving everything else in $M_j$ unchanged. We thus find
\begin{center}
\label{eq:symmaction}
\begin{tabular}{ccc}
centre:  & $M_j \rightarrow M_{j+1}$, & \\
CP: & $M_j \rightarrow M_j^\ast$.  & \\
\end{tabular}
\end{center}
As a result, centre symmetry is preserved only when all the $M_j$ take the same value, and CP symmetry is preserved only when all the $M_j$ are real-valued. The bion contributions to $\cal{V}$ in eq.~\eqref{eq:SUSYpotential} provide a repulsive interaction for the $\phi_j$, and hence ultimate produce centre symmetry and confinement. The potential in eq.~\eqref{eq:SUSYpotential} also clearly preserves CP symmetry.

We now introduce a small gaugino mass, $m \ll \Lambda$, to break SUSY softly. In this case, we can write $\mathcal{V}$ as a sum of two contributions, one perturbative and one non-perturbative, where each contribution consists of the SUSY result plus terms scaling with powers of $m$.

When $m \ll \Lambda$, the perturbative contribution to $\mathcal{V}$ (sometimes called the Gross-Pisarski-Yaffe potential~\cite{Gross:1980br}) becomes non-zero, but turns out to be $\mathcal{O}(m^2\mathcal{V}_0\,g^6 \, N^2)$~\cite{Chen:2020syd,Poppitz:2012sw,Poppitz:2012nz}. In our limit $g \ll 1$, this will be sub-leading compared to the non-perturbative contribution described below, so we will henceforth ignore the perturbative contribution to $\mathcal{V}$.

When SUSY is broken, the non-perturbative contribution to $\mathcal{V}$ again comes from monopole-instantons with no fermion zero modes. However, $m$ lifts the fermion zero modes of the monopole-instantons and anti-monopole instantons, $M_j$ and $M_j^*$, so that these can now contribute individually, rather than only via bions. In other words, (anti-)monopole-instantons with non-zero topological charge can now contribute. Indeed, to leading non-trivial order in $m$, the result of refs.~\cite{Poppitz:2012sw,Poppitz:2012nz,Anber_2014} for the bosonic potential is
\beq
\label{effective-action-finite-alpha}
    {\cal V}  = {\cal V}_0\sum_{j=0}^{N-1} |M_j-M_{j-1}|^2 - {\cal V}_0\,\frac{\gamma}{2}\left[1 - \frac{g^2\,N}{(4\pi)^2}\,\log(M_j^\ast M_j)\right]\left(M_j+M_j^\ast\right),
\eeq
where we have defined
\begin{align}
\label{eq:gammadef}
    \gamma \equiv \frac{32\pi^2}{3N^2}\frac{m}{L^2\Lambda^3},
\end{align}
which is a measure of the inverse circumference, $L$. In what follows, we will treat $\gamma$ as a proxy for temperature, or more precisely, because $\gamma \propto L^{-2}$, as a proxy for $T^2$. We will also re-scale the bosonic potential by ${\cal V}_0$,
\beq
\label{eq:vdef}
V \equiv {\cal V}/{\cal V}_0,
\eeq
and henceforth we will work only with the re-scaled bosonic potential, $V$, rather than ${\cal V}$.

While the bion contributions to $V$ provide a repulsive interaction for the $\phi_j$, thus encouraging centre symmetry and confinement, in contrast, the individual monopole-instanton contributions provide an attractive interaction for the $\phi_j$, encouraging them to bunch up and break centre symmetry, signaling deconfinement. As the ``temperature'' $\gamma$ increases, the latter contributions increase until a deconfinement transition occurs that is first order when $N>2$ and second order when $N=2$, as mentioned above~\cite{Chen:2020syd,Poppitz:2012sw,Poppitz:2012nz,Anber_2014,Poppitz:2021cxe}.

We will work with coupling $g$ sufficiently weak that the kinetic terms, including those for the bosons in eq.~\eqref{eq:kdef}, are negligible, and the potential in eqs.~\eqref{effective-action-finite-alpha} and~\eqref{eq:vdef} reduces to
\begin{eqnarray}
V=\sum_{j=0}^{N-1}|M_j-M_{j-1}|^2-\frac{\gamma}{2}\sum_{j=0}^{N-1}\left(M_j+M_j^\ast\right).
\label{effective-action}
\end{eqnarray}
Eq.~\eqref{effective-action} is the effective action that we will use in all that follows. The weak-coupling limit also justifies a saddle-point approximation, where for us the extrema of the effective action are the extrema of $V$ in eq.~\eqref{effective-action}, subject to the constraint in eq.~\eqref{eq:constraint}. These extrema coincide with the ``free energy'' of the mass-deformed ${\cal N}=1$ SU($N$) super Yang-Mills on ${\mathbb R}^3\times$S$^1$, so we will use the symbol $V$ to denote this free energy as well.

This effective action is valid for any $N$, however for a precise definition of the partial phase we need a continuous eigenvalue spectrum, which requires $N \to \infty$. Crucially, as observed in ref.~\cite{Poppitz:2012nz}, this effective theory breaks down in the standard 't Hooft large-$N$ limit, $N \to \infty$ with $g^2 N$ is fixed, because the W-boson masses, which are $\propto 1/(NL)$, approach zero. In the 't Hooft large-$N$ limit we would thus need to modify the effective action to include these ``extra'' massless degrees of freedom.

Instead of the 't Hooft large-$N$ limit, following refs.~\cite{Poppitz:2012nz,Chen:2020syd} we will take the so-called \textit{Abelian} large-$N$ limit: $N \to \infty$ with the W-boson masses fixed, so that in particular $L \propto 1/N \to 0$, that is, we shrink the $S^1$. The hierarchy of scales thus remains $m \ll \Lambda \ll 2\pi/(NL)$. To emphasise how the 't Hooft and Abelian large-$N$ limits are different: the former has $g \propto 1/N^2 \to 0$ with any $L$, while the latter has any $g$ with $L \propto 1/N \to 0$. In other words, the 't Hooft large-$N$ limit says nothing about $L$, while the Abelian large-$N$ limit says nothing about $g$.

\subsection{Review: Confined and Deconfined Phases}
\label{sec:minima}
\hspace{0.51cm}

In this subsection we will discuss the two extrema of the effective theory with action $V$ in eq.~\eqref{effective-action} that were discovered in ref.~\cite{Chen:2020syd}. These extrema are minima describing the confined and deconfined phases. In the next subsection we will present our new results for the maximum that connects these minima, and which describes the partial phase.

A central challenge in extremising $V$ in eq.~\eqref{effective-action} is implementing the constraint in eq.~\eqref{eq:constraint}. We can deal with this constraint in several ways. A simple approach is to treat $M_0$ as a function of $M_1,M_2,\cdots,M_{N-1}$, as $M_0=\frac{e^{i\theta}}{M_1\times\cdots \times M_{N-1}}$. In that case, for $j>0$, we easily find $\frac{\partial M_0}{\partial M_j}=-\frac{M_0}{M_j}$. 
The saddle-point equation, $\frac{\partial V}{\partial M_j}=0$, can then be written as 
\begin{eqnarray}
2M_j^\ast-M_{j+1}^\ast-M_{j-1}^\ast
+
\frac{M_0}{M_j}\left(M_1^\ast-M_0^\ast\right)
+
\frac{M_0}{M_j}\left(M_{N-1}^\ast-M_0^\ast\right)
-
\frac{\gamma}{2}
+
\frac{\gamma}{2}\frac{M_0}{M_j}=0.
\label{saddle-point-equation}
\end{eqnarray}

\subsubsection{Confined phase}
\label{sec:conf}
\hspace{0.51cm}
Ref.~\cite{Chen:2020syd} found the saddle point solution of eq.~\eqref{saddle-point-equation} describing the confined phase, valid for any $N$ (not just large $N$):
\begin{eqnarray}
\label{eq:confsol}
M_0=M_1=M_2=\cdots=M_{N-1}=e^{i\theta/N}.
\end{eqnarray}
The free energy of the confined phase is then 
\begin{eqnarray}
\label{eq:confonshell}
\left . V \right|_{\rm conf} = -N\gamma\cos(\theta/N). 
\end{eqnarray}

To see that the solution in eq.~\eqref{eq:confsol} describes the confined phase, we use the definitions of $z_j$ and $M_j$ in eqs.~\eqref{eq:zdef} and~\eqref{eq:monopole-instanton-operator}, respectively, to find an expression for the Polyakov loop phases $\phi_j$ in terms of the $M_j$,
\begin{eqnarray}
\phi_{j}-\phi_{j+1}=\frac{2\pi}{N} - \frac{g^2}{4\pi} \log|M_j|.
\label{eq:M-vs-phi}
\end{eqnarray}
In eq.~\eqref{eq:confsol}, $|M_j|=1$ for all $j$, hence from eq.~\eqref{eq:M-vs-phi} we have $\phi_{j}-\phi_{j+1}=\frac{2\pi}{N}$. This saddle-point solution therefore possesses a uniform Polyakov phase distribution, and thus unbroken centre symmetry, indicating that this is indeed the confined phase. The distribution is normalised, having height $\frac{1}{2\pi}$.

As an order parameter for the CP symmetry we will use $\frac{\partial V}{\partial \theta}$, which in the effective theory is proportional to the expectation value of a CP-odd operator, namely the instanton density: $\frac{\partial V}{\partial \theta}\propto \langle \mathrm{tr}(F_{\mu\nu} \tilde{F}^{\mu\nu})\rangle$. From eqs.~\eqref{eq:confsol} and~\eqref{eq:confonshell} we find
\beq
\left . \frac{\partial V}{\partial \theta} \right|_{\rm conf} = \gamma \sin\left(\theta/N\right).
\eeq
When $\theta=0$, the CP symmetry is not explicitly broken, and the $M_j$ in eq.~\eqref{eq:confsol} are purely real, hence CP symmetry is not spontaneously broken. Correspondingly, $\left . \frac{\partial V}{\partial \theta} \right|_{\rm con}=0$. When $\theta \in (0,\pi)$, CP symmetry is explicitly broken, and unsurprisingly, all the $M_j$ in eq.~\eqref{eq:confsol} are complex, and $\left . \frac{\partial V}{\partial \theta} \right|_{\rm con}\neq0$. When $\theta=\pi$, the CP symmetry is not explicitly broken, but all the $M_j$'s in eq.~\eqref{eq:confsol} are complex, hence CP symmetry is spontaneously broken. Correspondingly, $\left . \frac{\partial V}{\partial \theta} \right|_{\rm con}=\gamma \sin(\pi/N)\neq0$.

The results above are valid for any $N$. If we take the Abelian large-$N$ limit, then remembering from eq.~\eqref{eq:gammadef} that $\gamma \propto N^{-2}$, we find that $\left . V \right|_{\rm conf}\sim N^{-1}$, while $\left . \frac{\partial V}{\partial \theta} \right|_{\rm conf}\sim N^{-3}$. These large-$N$ scalings are in fact generic, as we will see below.

\subsubsection{Deconfined phase}
\label{sec:deconf}
\hspace{0.51cm}

Following ref.~\cite{Chen:2020syd}, we find the saddle point solution describing the deconfined phase only in the Abelian large $N$ limit, as follows.

To deal with the constraint in eq.~\eqref{eq:constraint}, we will assume that $M_0$ is much smaller than all the other $M_j$, that is, $M_0\ll M_j$ for all $j>0$. As we will see later, this is self-consistent only for sufficiently large temperature $\gamma$. With this assumption, the saddle-point equation as written in eq.~\eqref{saddle-point-equation} simplifies to 
\begin{eqnarray}
2M_j^\ast-M_{j+1}^\ast-M_{j-1}^\ast -
\frac{\gamma}{2}
=
0. 
\label{saddle-point-equation-2}
\end{eqnarray}
To solve eq.~\eqref{saddle-point-equation-2}, we take all the $M_j$'s to be real-valued, which among other things ensures that the solution will preserve CP symmetry. We also take the Abelian large-$N$ limit with $N$ sufficiently large to justify a continuum approximation, and retain only contributions at leading order in $N$. Specifically, we treat $t\equiv \frac{j}{N}-\frac{1}{2}$ as a continuous ``time'' parameter, and we use a notation $X(t)\equiv M_j$. Using the periodicity $X(t) = X(t+1)$ we can take $t \in [-1/2,1/2]$. Furthermore, since our temperature parameter $\gamma \propto N^{-2}$ from eq.~\eqref{eq:gammadef}, we also define for convenience a re-scaled temperature parameter that will remain order $N^0$ in the Abelian large-$N$ limit,
\beq
\label{eq:gammatildedef}
\tilde{\gamma} \equiv N^2\gamma.
\eeq
With these approximations and definitions, eq.~\eqref{saddle-point-equation-2} becomes a simple second-order, linear, ordinary differential equation,
\begin{eqnarray}
X''(t)
=
-\frac{\tilde{\gamma}}{2},
\label{saddle-point-equation-3}
\end{eqnarray}
where $X'(t) \equiv \frac{dX}{dt}$. The solution of eq.~\eqref{saddle-point-equation-3} is quadratic in $t$,
\begin{eqnarray}
\label{eq:deconfsol}
X(t)
=
A+Bt-\frac{\tilde{\gamma}}{4}t^2,
\end{eqnarray}
with integration constants $A$ and $B$. Plugging the solution in eq.~\eqref{eq:deconfsol} and our approximations and definitions into the effective action $V$ in eq.~\eqref{effective-action}, we find
\begin{eqnarray}
\label{eq:deconfv1}
V
=
\frac{1}{N}\int_{-1/2}^{+1/2}|X'|^2dt
-
\frac{\tilde{\gamma}}{N}\int_{-1/2}^{+1/2}Xdt
+
X(-1/2)^2
+
X(1/2)^2. 
\end{eqnarray}
The last two terms come from $|M_1-M_0|^2$ and $|M_{N-1}-M_0|^2$, which have to be treated separately because we assumed $M_0$ to be an outlier. Assuming that $X$ is of order $N^0$, the first two terms in eq.~\eqref{eq:deconfv1}, i.e. the terms involving integrals over $t$, are actually sub-leading in $N$, while the last two terms are dominant. We already minimised the former by solving eq.~\eqref{saddle-point-equation-3}, so to be consistent we must also minimise the latter, meaning we take
\begin{eqnarray}
X(-1/2)
=
X(1/2)
=
0.
\end{eqnarray}
These boundary conditions fix the integration constants $A$ and $B$, so that the saddle-point solution in eq.~\eqref{eq:deconfsol} becomes
\begin{eqnarray}
\label{eq:deconfsol2}
X(t)
=
\frac{\tilde{\gamma}}{16}
\left(
1-4t^2
\right). 
\end{eqnarray}
The corresponding free energy is 
\begin{eqnarray}
\label{eq:deconfv2}
\left . V \right|_{\rm deconf}
=
-\frac{1}{48}\frac{\tilde{\gamma}^2}{N}. 
\end{eqnarray}
This conforms to the generic behaviour $V \sim N^{-1}$ mentioned at the end of sec.~\ref{sec:conf}.

We now need to check when $M_0\ll M_j$ for $j>0$, and hence the solution is self-consistent. To do so, we follow ref.~\cite{Chen:2020syd}, and estimate $M_0$ by integrating $\log(X(t))$ over $t\in[-1/2,1/2]$, which gives the continuum version of $\sum_{j>0}\ln M_j$, and then exponentiating and using $M_0 = e^{i\theta}/(M_1\ldots M_N)$ to obtain
\begin{eqnarray}
\label{eq:deconfM0approx}
M_0
\simeq
\frac{e^{i\theta}}{N}
\left(
\frac{4e^2}{\tilde{\gamma}}
\right)^{N-1},
\end{eqnarray}
where because we assumed all the $M_j$ are real, including $M_0$, in eq.~\eqref{eq:deconfM0approx} $\theta=0$ or $\pi$. For $M_0$ to be vanishingly small, we demand $\frac{4e^2}{\tilde{\gamma}}<1$, or equivalently $\tilde{\gamma}>4e^2=29.556\cdots$. In other words, we need the temperature $\tilde{\gamma}\equiv N^2 \gamma$ to be sufficiently large, as advertised. 

Let us now show that the centre symmetry is spontaneously broken. Eq.~\eqref{eq:deconfM0approx} gives
\begin{eqnarray}
-\frac{g^2}{4\pi} \ln|M_0|
\sim
-\frac{g^2}{4\pi} N\log\left(\frac{4e^2}{\tilde{\gamma}}\right),
\end{eqnarray}
which indeed can be much bigger than that of all the other $M_j$. Defining for convenience
\beq
\epsilon \equiv N\,\frac{g^2}{4\pi},
\eeq
and using eq.~\eqref{eq:M-vs-phi}, we find a finite spacing between the Polyakov loop phases $\phi_1$ and $\phi_0$,
\begin{eqnarray}
\label{eq:spacing}
\phi_0-\phi_1
\simeq
-\epsilon\log\left(\frac{4e^2}{\tilde{\gamma}}\right), 
\end{eqnarray}
which is $\geq 0$ in our limit $\frac{4e^2}{\tilde{\gamma}}<1$. As a result, the saddle point solution in eq.~\eqref{eq:deconfsol2} describes a distribution of Polyakov line phases with a gap, and thus the centre symmetry is broken, indicating that this is indeed the deconfined phase. The gap closes as $4e^2/\tilde{\gamma}\to 1$. As we will see in subsection~\ref{sec:PDC-phase}, $4e^2/\tilde{\gamma}=1$ can be regarded as the GWW transition point. 

The rest of the distribution behaves as follows. The solution $X(t)$ starts at zero when $t=-\frac{1}{2}$, goes all the way up to $\frac{\tilde{\gamma}}{16}>1$ when $t=0$, and then comes down to zero when $t=\frac{1}{2}$. As a result, as $t$ increases $\ln (X(t))$ is initially negative, then at some point turns positive, and then becomes negative again. If $X(t)< 1$ then $\phi_{j}-\phi_{j+1}<\frac{2\pi}{N}$, namely the distribution is more dense. If $X(t)> 1$ then $\phi_{j}-\phi_{j+1}>\frac{2\pi}{N}$, namely the distribution is more sparse. In short, the Polyakov phases grow sparser as $j$ increases.

As mentioned below eq.~\eqref{saddle-point-equation-2}, this saddle-point solution assumed all the $M_j$ are real, and hence CP symmetry is not spontaneously broken. Correspondingly, the $V$ in eq.~\eqref{eq:deconfv1} or~\eqref{eq:deconfv2} does not depend on $\theta$ at all. As a result, $\frac{\partial V}{\partial \theta}=0$, and hence our order parameter for CP symmetry breaking vanishes, $\langle \mathrm{tr}(F_{\mu\nu} \tilde{F}^{\mu\nu})\rangle=0$.

\subsection{Partial phase}\label{sec:PDC-phase}
\hspace{0.51cm}

In this subsection, we will find new solutions of the effective theory with action $V$ in eq.~\eqref{effective-action} that describe a partial phase. We will begin with the Abelian large-$N$ limit, and with $\theta=0$, where we will find an unstable partial phase connecting the confined and deconfined phases. Of course, when $\theta=0$ in this theory, CP symmetry is unbroken in both the confined and deconfined phases. We will find that CP symmetry is unbroken in the partial phase as well. We then consider $\theta=\pi$, where CP symmetry is broken in the confined phase and restored in the deconfined phase. We will find that when $\theta=\pi$, CP symmetry is broken in the partial phase.

We will then perform numerical computations with finite $N$, although we restrict to numerically large values $N\geq 30$. Our finite-$N$ results are similar to those of the Abelian large-$N$ limit. In particular, we will identify a partial phase that connects the confined and deconfined phases and, when $\theta=\pi$, exhibits spontaneous breaking of CP symmetry. Our numerical results thus support our conjecture that the partial phase can be distinguished from both confined and deconfined phases by global symmetries, even at finite $N$.

\subsubsection{Abelian large-$N$ limit}
\label{sec:abelianlargen}
\hspace{0.51cm}

We begin with the Abelian large-$N$ limit, where we will use the continuous time parameter $t$ and $M_j \to X(t)$, as defined in sec.~\ref{sec:deconf}. We will begin with $\theta=0$, and will assume the $M_j$ are all real, so that CP symmetry is unbroken. The large-$N$ continuum limit of eq.~\eqref{saddle-point-equation} can then be written as
\begin{align}
X''(t) = \frac{X(-1/2)}{X(t)}\left(X''(-1/2) + \frac{\tilde{\gamma}}{2}\right) - \frac{\tilde{\gamma}}{2}, \label{eq:continuumDiffEqn}
\end{align}
where for convenience we choose $X(-1/2)$ to be the minimum of the configuration. We must find solutions of eq.~\eqref{eq:continuumDiffEqn} that are periodic, $X(t)=X(t+1)$, and that obey the continuum version of the constraint in eq.~\eqref{eq:constraint},
\begin{align}
    \int_{-1/2}^{+1/2} dt \log X(t) = i\theta,
    \label{eq:continuumConstraint}
\end{align}
with $\theta=0$.

For given initial values $X(-1/2)$ and $X''(-1/2)$, we solve eq.~\eqref{eq:continuumDiffEqn} numerically by the second-order Taylor method, reviewed in appendix~\ref{Appendix:2ndOrderTaylorMethod}. Actually, we numerically solve an equivalent equation, obtained by multiplying eq.~\eqref{eq:continuumDiffEqn} by $2X'(t)$, integrating from time $0$ to $t$, and then taking the square root:
\begin{align}
    X'(t) = \pm\sqrt{X(-1/2)\log \Big(\frac{X(t)}{X(-1/2)}\Big)(2X''(-1/2) + \tilde{\gamma}) - \tilde{\gamma} (X(t)-X(-1/2))},
    \label{eq:dXdt}
\end{align}
where we used $X'(-1/2)=0$ to set the integration constant. This is a consequence of our convention $M_j=M_{N-j}$ and the smoothness of the solution. Since we chose $X(-1/2)$ as the minimum, we use the $+$ branch of the root in eq.~\eqref{eq:dXdt} in all of our computations.

For a particular choice of $X(-1/2)$ and $X''(-1/2)$ we check for the periodicity condition by integrating eq.~\eqref{eq:dXdt} numerically to obtain the half-period $\tau$,
\begin{align}
    \tau = \int_{X(-1/2)}^{X_{\rm max}} \frac{dX}{\sqrt{X(-1/2)\log \Big(\frac{X}{X(-1/2)}\Big)(2X''(-1/2) + \tilde{\gamma}) - \tilde{\gamma} (X-X(-1/2))}}. \label{eq:tau}
\end{align}
A solution with the required periodicity will have $\tau=\frac{1}{2}$. For fixed $X(-1/2)$, we use Newton's method to find $X''(-1/2)$ such that  $\tau=\frac{1}{2}$ is satisfied to high precision.

For a given value of $X(-1/2)$, therefore, we finally need to satisfy the constraint in eq.~\eqref{eq:continuumConstraint}. To do so, we use the bisection method, varying upper and lower bounds on $X(-1/2)$ until the corresponding solution $X(t)$ satisfies eq.~\eqref{eq:continuumConstraint} to high precision. By choosing the initial upper and lower bounds on $X(-1/2)$ to be between 0 and 1, we are able to avoid the confined and deconfined solutions reviewed in subsection~\ref{sec:minima}.

\begin{figure}[htbp]
  \begin{center}
   \includegraphics[width=100mm]{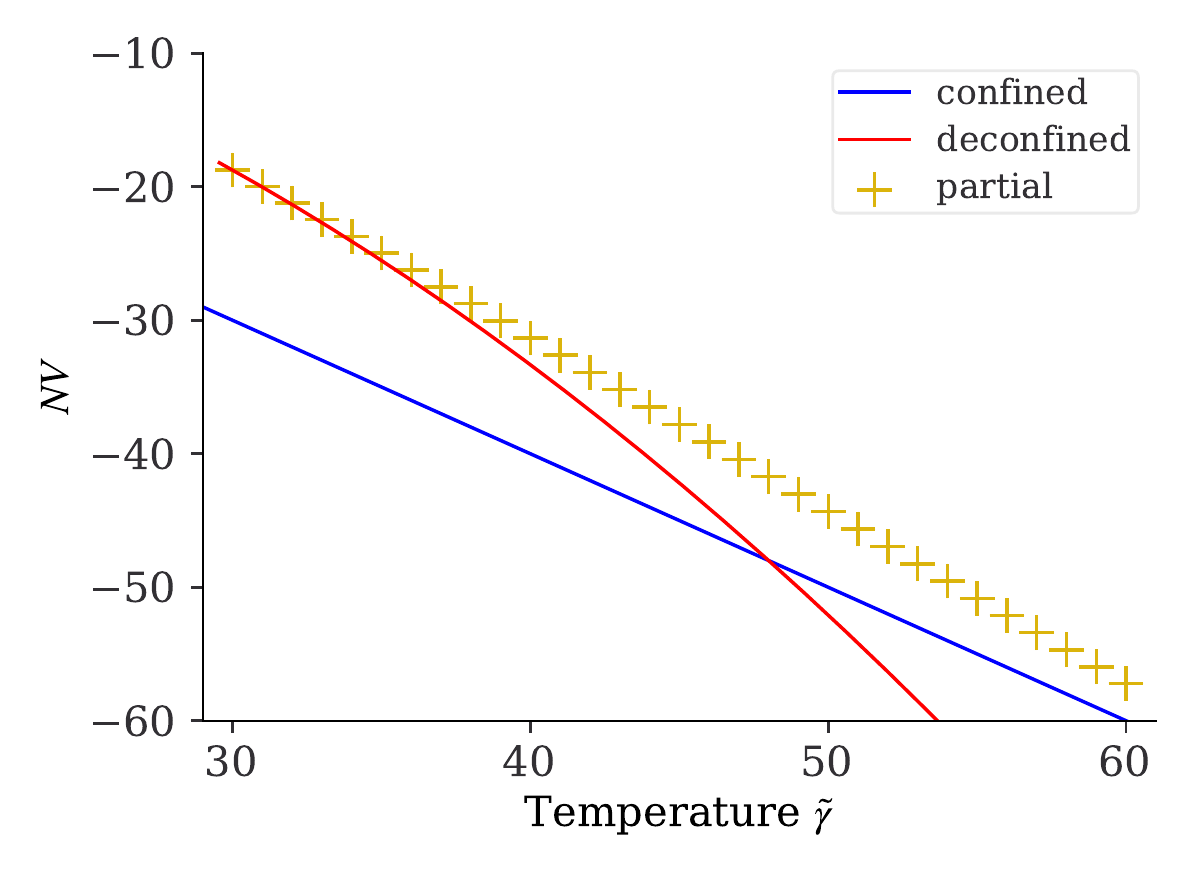}
  \end{center}
  \caption{Free energy $V$ times $N$ versus ``temperature'' $\tilde{\gamma}$ in the Abelian $N=\infty$ limit, for any $\theta$. We show $NV$ of the confined phase from eq.~\eqref{eq:confonshell} (blue line), the deconfined phase from eq.~\eqref{eq:deconfv2} (red line), and our numerical results for the partial phase (gold plus signs). We find the ``swallow-tail'' shape characteristic of a first-order phase transition, with the partial phase as the unstable branch connecting the confined and deconfined branches.}\label{fig:V-vs-gt-largeN}
\end{figure}

Given a numerical solution for $X(t)$, we can straightforwardly compute the on-shell action $V$. Fig.~\ref{fig:V-vs-gt-largeN} shows our numerical results for $NV$ as a function of the ``temperature'' $\tilde{\gamma}$, together with the results for $NV$ in the confined phase eq.~\eqref{eq:confonshell} and in the deconfined phase from eq.~\eqref{eq:deconfv2}. (Appendix~\ref{Appendix:2ndOrderTaylorMethod} describes in detail how we determine the numerical uncertainties in $V$ and $X(-1/2)$.) In fig.~\ref{fig:V-vs-gt-largeN} we observe the ``swallow-tail'' shape characteristic of a first-order phase transition, with the partial phase as the unstable branch connecting the confined and deconfined branches.

When $\theta=0$, CP symmetry is preserved in all three phases. To study the spontaneous breaking of CP symmetry we need to set $\theta = \pi$, and compute our order parameter, $\frac{\partial V}{\partial \theta}\propto \langle \mathrm{tr}(F_{\mu\nu} \tilde{F}^{\mu\nu})\rangle$. Crucially, however, in the Abelian large-$N$ limit, generically $V \sim N^{-1}$, while $\frac{\partial V}{\partial\theta}\sim N^{-3}$, as we saw for example in the confined phase of sec.~\ref{sec:conf}, and as we will find for the partial phase below. In other words, our results for $NV$ in fig.~\ref{fig:V-vs-gt-largeN} are in fact the leading contribution in the Abelian large-$N$ limit for \textit{any} value of $\theta$. We therefore do not need to re-compute $NV$ for $\theta=\pi$: the result is the same as in fig.~\ref{fig:V-vs-gt-largeN}.

For $\theta=0$ we assumed the $M_j$ were real-valued, hence $\frac{\partial V}{\partial \theta}=0$. If $\theta\neq 0$, then we expect $\frac{\partial V}{\partial\theta} \sim N^{-3}$. Generically, the magnitudes $|M_j|\sim N^0$, hence their leading contribution to $\frac{\partial V}{\partial\theta}$ will be unchanged in going from $\theta=0$ to $\theta=\pi$. In other words, to compute $\frac{\partial V}{\partial\theta}$ when $\theta\neq 0$, we can use our existing solutions for the magnitudes $|M_j|$. By extension, if $\theta \neq 0$, then $\frac{\partial V}{\partial\theta}$ can be non-zero only due to the \textit{phases} of the $M_j$. We thus need to compute the phases of the $M_j$, in the Abelian large-$N$ limit.

Writing $M_j = |M_j|e^{i \varphi_j}$, from $V$ in eq.~\eqref{effective-action} we find
\begin{align}
    \frac{\partial V}{\partial \varphi_j} = |M_j|\left(2|M_{j+1}|\sin(\varphi_j - \varphi_{j+1}) + 2|M_{j-1}|\sin(\varphi_j - \varphi_{j-1}) + \gamma \sin(\varphi_j)\right),
    \label{dVdPhi_expanded}
\end{align}
while the constraint in eq.~\eqref{eq:constraint} becomes
\begin{align}
 \sum_{j=0}^{N-1} \varphi_j = \theta.
 \label{eq:phi-constraint}
\end{align}
The confined phase satisfies this constraint with $\varphi_j=\theta\,N^{-1}$ for all $j$, while in the deconfined phase all of the $\theta$ dependence is in $\varphi_0=\theta$. Generically, the partial phase satisfies this constraint with all $\varphi_j$ being different, but with $\varphi_j\sim N^{-1}$ for all $j$. Using the constraint, we can write the equation of motion of each $\varphi_j$ as
\begin{align}
    \frac{\partial V}{\partial \varphi_j} = \frac{\partial V}{\partial \varphi_0},
    \label{eq:phi-eom}
\end{align}
that is, because of the constraint, all $\partial V/\partial \varphi_j$ take the same value, namely $\partial V/\partial \varphi_0$. Large-$N$ counting in this equation of motion provides another way to see $\varphi_j\sim N^{-1}$ for all $j$, assuming $\tilde{\gamma}$ is an order $N^0$ distance from the GWW transition, so that $|M_j|\sim N^0$.

As mentioned above, when $\theta \neq 0$, to compute $\frac{\partial V}{\partial\theta}$ we can use the solutions for the $|M_j|$ at $\theta=0$, so the only new contribution is from the $\varphi_j$. We can thus write
\begin{align}
    \frac{\partial V}{\partial \theta} = \sum_j \frac{\partial{\varphi_j}}{\partial\theta}\frac{\partial V}{\partial \varphi_j}. \label{eq:dVdTheta_phiDecomposition}
\end{align}
Taking $\partial/\partial \theta$ of the constraint in eq.~\eqref{eq:phi-constraint} gives us
\begin{align}
    \sum_j \frac{\partial \varphi_j}{\partial \theta} = 1. 
    \label{eq:deriv-theta-phi-relation}
\end{align}
Using eqs.~\eqref{eq:phi-eom} and~\eqref{eq:deriv-theta-phi-relation}, we can re-write eq.~\eqref{eq:dVdTheta_phiDecomposition} as
\begin{align}
    \frac{\partial V}{\partial \theta} = \frac{\partial V}{\partial \varphi_j},
    \label{eq:dVdTheta-as-dVdPhi_k}
\end{align}
that is, to obtain $\frac{\partial V}{\partial \theta}$ we can just compute $\partial V/\partial \varphi_j$ for any one value of $j$. If we choose $j=0$, then
\begin{align}
    \frac{\partial V}{\partial \theta} = |M_0|\left(2|M_{1}|\sin(\varphi_0 - \varphi_{1}) + 2|M_{N-1}|\sin(\varphi_0 - \varphi_{N-1}) + \gamma \sin(\varphi_0) \right).
    \label{eq:dvdthetajzero}
\end{align}
From eq.~\eqref{eq:dvdthetajzero} we can argue that the order parameter will jump discontinuously at the GWW transition, as we expect for a first-order transition. In general, as we approach the GWW transition we expect $\phi_0\to\pi$ and $\phi_{\pm 1}\to 0$, so that the partial phase matches onto the deconfined phase. In eq.~\eqref{eq:dvdthetajzero}, as we approach the GWW transition, na\"ively we expect each sine function to approach zero, and hence $\frac{\partial V}{\partial \theta} \to 0$, indicating restoration of CP symmetry, as expected. Additionally, as we approach the GWW transition, we expect $|M_0|\to0$ tends towards zero, further suppressing $\frac{\partial V}{\partial \theta}$. Our finite-$N$ results in the next subsection will indeed exhibit such behaviour, although $|M_0|$ will never precisely reach zero (but will decrease at the GWW transition as $N$ increases). However, as mentioned above, in the Abelian large-$N$ limit, $\varphi_j \sim N^{-1}$ for all $j$, including $j=0$, so as we approach the GWW transition we will not see $\varphi_0 \to \pi$ smoothly. Instead, in the GWW transition, $\varphi_0$ will jump discontinuously from $\varphi_0\sim N^{-1}$ in the partial phase to $\varphi_0 = \pi$ in the deconfined phase, and correspondingly $\frac{\partial V}{\partial \theta}$ will jump discontinuously. In the Abelian large-$N$ limit, the CP symmetry restoration transition at the GWW point will therefore be first order.

To solve for the $\varphi_j$ in the Abelian large-$N$ limit, we expand the sine functions in eq.~\eqref{dVdPhi_expanded} to linear order, and then take the continuum limit and introduce the time parameter $t$ from sec.~\ref{sec:deconf}, with $\varphi_j \to \varphi(t)$, so that eq.~\eqref{eq:phi-eom} becomes a differential equation for $\varphi(t)$. In this equation, $X(t)$ and its derivatives appear. To leading order in $N$ we can use our existing $\theta=0$ solutions for these, which allows us to write $\varphi(t)$'s equation of motion as
\begin{align}
    \varphi''(t) = \frac{-4XX'\varphi' + 2 X(-1/2)^2 \varphi''(-1/2) + \tilde{\gamma} (X \varphi - X(-1/2) \varphi(-1/2))}{ 2X^2}.
    \label{eq:phicontinuumeom}
\end{align}
A solution $\varphi(t)$ must also be periodic, $\varphi(t+1)=\varphi(t)$, and obey the continuum version of the constraint in eq.~\eqref{eq:phi-constraint}
\begin{align}
\int_{-1/2}^{1/2} dt \, \varphi(t) = \theta.
\label{eq:phicontinuumconstraint}
\end{align}

For given values of $\varphi(-1/2)$ and $\varphi''(-1/2)$, we solve eq.~\eqref{eq:phicontinuumeom} numerically using the second-order Taylor method, in analogy to what we did for $X(t)$ above (and reviewed in Appendix~\ref{Appendix:2ndOrderTaylorMethod}). To guarantee that $\varphi(t)$ is periodic in $t$ and obeys the constraint in eq.~\eqref{eq:phicontinuumconstraint}, we fix values of $\varphi(-1/2)$ and $\varphi''(-1/2)$, as follows. We first choose $\varphi(-1/2)=1$, and use the bisection method to determine $\varphi''(-1/2)$ such that the extreme values of $\varphi(t)$ occur at the terminating values of $t$, i.e. $\varphi'(\pm\frac{1}{2}) = 0$. This guarantees that the solution $\varphi(t)$ will be periodic. To guarantee that the solution obeys the constraint in eq.~\eqref{eq:phicontinuumconstraint}, we observe that eq.~\eqref{eq:phicontinuumeom} is linear in $\varphi(t)$, and hence remains unchanged under a re-scaling of $\varphi(t)$. We thus simply re-scale $\varphi(-1/2)$ and $\varphi''(-1/2)$ until the constraint in eq.~\eqref{eq:phicontinuumconstraint} is satisfied. Explicitly, if $c \equiv \int dt \, \varphi(t)$, then we re-scale $\varphi(-1/2) \rightarrow \varphi(-1/2) \frac{\theta}{c}$ and $\varphi''(-1/2) \rightarrow \varphi''(-1/2) \frac{\theta}{c}$.

For a given solution $\varphi(t)$, we computed $\frac{\partial V}{\partial \theta}$ in two different ways, as a cross-check. The first way was using the continuum version of eq.~\eqref{eq:dVdTheta-as-dVdPhi_k}. The second way begins by observing that if we re-scale $\theta \to \xi \, \theta$ then we can obtain the new solution for $\varphi(t)$ simply by re-scaling $\varphi(t) \rightarrow \xi \varphi(t)$. Furthermore, we observe that in the continuum version of $V$, in the terms involving an integral over $t$, $\varphi$ and $\varphi'$ enter quadratically,
\begin{align}
     V \ni V_{\varphi} \equiv \int dt \left( X^2 \varphi'^2 + \frac{1}{2}\,\tilde{\gamma} X \varphi^2 \right).
\end{align}
We can thus evaluate $\partial V/\partial\theta$ at any $\theta$ by extracting the coefficient of the $\varphi$ terms, $V_\varphi$, evaluated at $\theta$, as
\begin{align}
    \frac{\partial V}{\partial \theta} = \frac{2}{\theta} \left . V_{\varphi}\right |_{\theta}.
    \label{eq:largendVdtheta2}
\end{align}
We have checked that numerical results obtained from eq.~\eqref{eq:dVdTheta-as-dVdPhi_k} or eq.~\eqref{eq:largendVdtheta2} agree to great numerical accuracy.

\begin{figure}[htbp]
  \begin{center}
   \includegraphics[width=80mm]{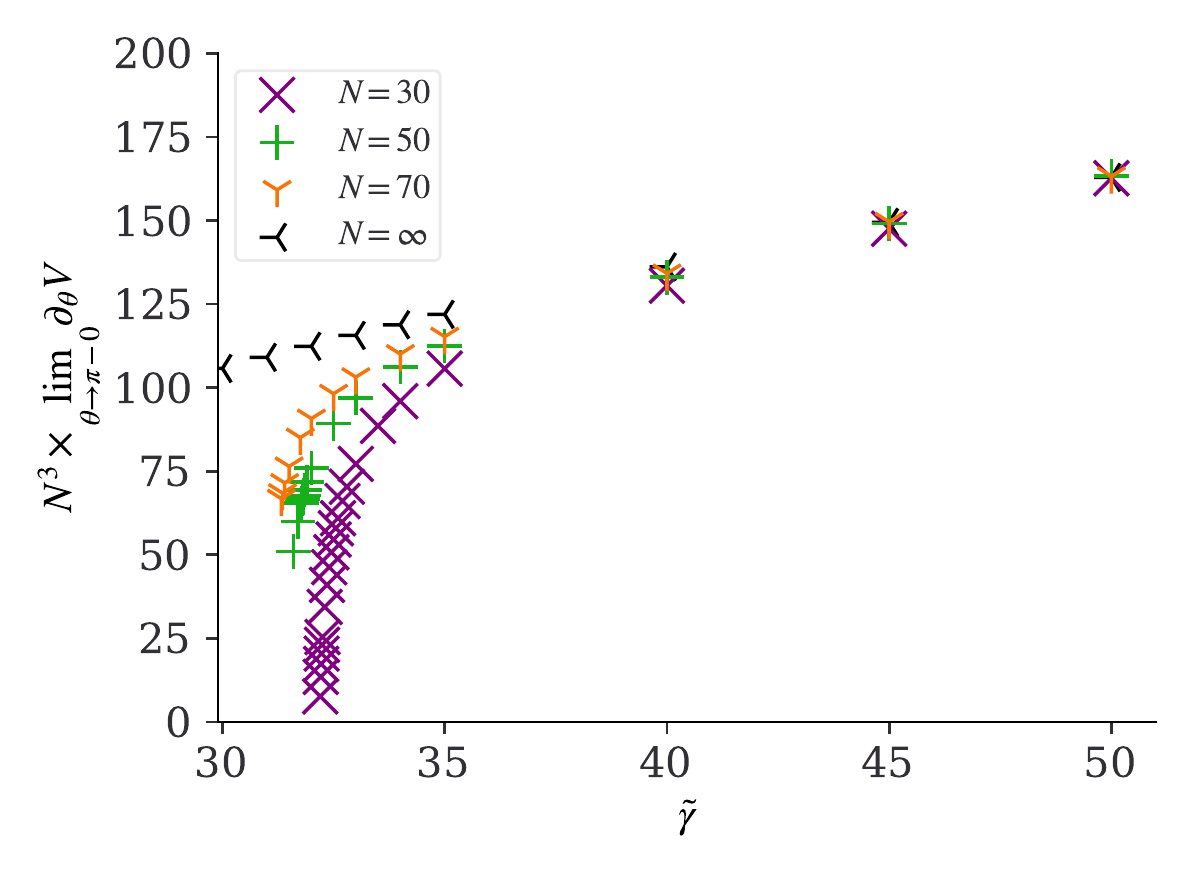}
  \end{center}
  \caption{Numerical results in the partial phase for 
  $\frac{\partial V}{\partial\theta}\propto\langle \mathrm{tr}(F_{\mu\nu} \tilde{F}^{\mu\nu})\rangle$ times $N^3$ in the limit $\theta \to \pi-0$ as a function of ``temperature'' $\tilde{\gamma}$, for $N=30$ (purple crosses), $50$ (green plus signs), $70$ (orange 3-point star), and $\infty$ (black 3-point star). The error bars are invisibly small on the scale of the plot. In all cases, we see clearly that $\langle \mathrm{tr}(F_{\mu\nu} \tilde{F}^{\mu\nu})\rangle\neq0$ and hence CP symmetry is spontaneously broken in the partial phase. We also see that the finite-$N$ results smoothly approach the $N=\infty$ result as $\tilde{\gamma}$ increases.
  } 
  \label{fig:TrFF_N70N50N30}
\end{figure}

Fig.~\ref{fig:TrFF_N70N50N30} shows our numerical results for $\frac{\partial V}{\partial \theta}$ times $N^3$ at $\theta=\pi$ as a function of the ``temperature'' $\tilde{\gamma}$ in the partial phase. We see clearly that $\partial V/\partial\theta\propto\langle \mathrm{tr}(F_{\mu\nu} \tilde{F}^{\mu\nu})\rangle\neq 0$, indicating spontaneous CP symmetry breaking. We also see that as we approach the GWW transition point, $\tilde{\gamma} \to 4 e^2 \approx 29.556\ldots$, the order parameter $\frac{\partial V}{\partial \theta}$ does not approach zero, indicating a first-order CP symmetry restoring transition, as mentioned above. Fig.~\ref{fig:TrFF_N70N50N30} also shows our finite-$N$ numerical results, which we describe in the next subsection.

\subsubsection{Finite $N$ numerics}
\label{sec:nummethods}
\hspace{0.51cm}

In this subsection we will perform numerical calculations in the effective theory with action $V$ in eq.~\eqref{effective-action}, with finite $N\geq 30$. Our goal will be to identify a partial phase. However, as mentioned in sec.~\ref{sec:intro}, our definition of the partial phase is valid in the large-$N$ limit ('t Hooft, Abelian, or otherwise), where we have a continuous distribution of Polyakov phases. Namely, we define the partial phase from the breaking of centre symmetry plus no gap in the distribution. Our numerics have finite $N$, and hence a discrete distribution. How then do we define the partial phase? Our working definition is the following. Eq.~\eqref{eq:M-vs-phi} allows us to write the eigenvalue density around any given $\phi_j$ as $\left(2\pi- N g^2/(4\pi) \log|M_j|\right)^{-1}$. A gap occurs if the minimum of this eigenvalue density is zero. Using centre symmetry permutations we will always make $M_0$ the smallest of the $M_j$, so any gap will occur at $\left(2\pi- Ng^2/(4\pi)^2 \log|M_0|\right)^{-1}$. This can vanish only if $|M_0|$ becomes exponentially small. In our new solutions we will explicitly find that $|M_0|<1$, but that $|M_0|$ is not exponentially small at a finite distance from the GWW transition point. As a result, the distribution of the Polyakov line phases will not be gapped, which we use to identify a partial phase.

As further evidence that our numerical solutions describe a partial phase, we will also check the eigenvalues of the Hessian of $V$, whose components are $\frac{\partial^2 V}{\partial M_j \partial M_k}$, $\frac{\partial^2 V}{\partial M_j^* \partial M_k}$, and $\frac{\partial^2 V}{\partial M^\ast_j \partial M^\ast_k}$. At minima of $V$, all eigenvalues of the Hessian are positive, as occurs for the confined and deconfined phase solutions. We expect the partial phase solution to be a local maximum of $V$, with one negative eigenvalue in the Hessian, and all others positive. This is because in the partial phase at fixed temperature, when we change the energy we expect the free energy to decrease. In all the solutions we present below, we confirmed this expectation numerically, as we will discuss.

In appendix~\ref{Appendix:numericaldata} we collect some numerical data useful for anyone who wishes to reproduce our results.

\subsubsection*{Finite $N$ numerical methods}

We will use a combination of two numerical methods to find finite-$N$ solutions for the $M_j$. The first is to solve the saddle-point equation in eq.~\eqref{saddle-point-equation}, $\frac{\partial V}{\partial M_j}=0$, via gradient descent. To do so, we define a function $\Phi_1(M_1,\cdots,M_{N-1})$ as 
\begin{eqnarray}
\lefteqn{
\Phi_1(M_1,\cdots,M_{N-1})
\equiv
\sum_{j=1}^{N-1}
\left|\frac{\partial V}{\partial M_j}\right|^2
}
\nonumber\\
&=&
\sum_{j=1}^{N-1}
\left|
2M_j^\ast-M_{j+1}^\ast-M_{j-1}^\ast
+
\frac{M_0}{M_j}\left(M_1^\ast-M_0^\ast\right)
+
\frac{M_0}{M_j}\left(M_{N-1}^\ast-M_0^\ast\right)
-
\frac{\gamma}{2}
+
\frac{\gamma}{2}\frac{M_0}{M_j}\right|^2. 
\nonumber\\
\end{eqnarray}
Solving eq.~\eqref{saddle-point-equation} is then equivalent to solving $\Phi_1(M_1,\cdots,M_{N-1})=0$. To do so, we minimise $\Phi_1(M_1,\ldots,M_{N-1})$ via gradient descent. Specifically, we choose some initial conditions for the $M_j$ and then iteratively update them as $M_j\to M_j-\epsilon\frac{\partial\Phi_1}{\partial M_j^\ast}$, with some small step size $\epsilon$. A drawback of this method is that generic initial conditions lead to the confined or deconfined solutions reviewed in subsection~\ref{sec:minima}. We thus combine this first method with a second method that can generate suitable initial conditions for the first method, to avoid this problem.

The second method starts with $V$ in eq.~\eqref{effective-action}, and implements the constraint in eq.~\eqref{eq:constraint} not by replacing $M_0 = e^{i\theta}/(M_1\dots M_{N-1})$, as we have done so far, but by introducing a Lagrange multiplier $\kappa$, so that instead of extremising $V$ we extremise $V+\kappa\left(\sum_j\log M_j - i\theta\right)$. This is equivalent to solving 
\begin{align}
M_j\frac{\partial V}{\partial M_j} = M_j\left(2M_j^\ast-M_{j-1}^\ast-M_{j+1}^\ast-\frac{\gamma}{2}\right) = -\kappa,
\label{eq:saddle-point-equation-PD}
\end{align}
and simultaneously solving
\begin{align}
\sum_{j=0}^{N-1}\log M_j - i\theta=0.
\end{align}
The confined and deconfined solutions have $\kappa=\frac{\gamma N}{2}e^{i\theta/N}$ and $\kappa=0$, respectively. Clearly, $\kappa$ measures the size of the confined sector. Intuitively, fixed-$\gamma$ gives the canonical ensemble, while fixed-$\kappa$ gives the microcanonical ensemble. By re-writing eq.~\eqref{eq:saddle-point-equation-PD} as 
\begin{align}
2M_j^\ast-M_{j+1}^\ast-M_{j-1}^\ast-\frac{\gamma}{2}=-\frac{\kappa}{M_j},
\label{eq:meth2eom}
\end{align}
and then taking a sum over $j$ and solving for $\gamma$, we find
\begin{align}
\label{eq:gammameth2}
\gamma = \frac{2\kappa}{N} \sum_{j=0}^{N-1} M_j^{-1}.
\end{align}
We then use the following strategy to find partial-phase solutions. We begin with $\theta=0$, and we choose values for $N$ and $\kappa$. As an initial condition, we use a discretised version of the deconfined solution in eq.~\eqref{eq:deconfsol2}, at the GWW transition point. In that solution, both the $M_j$ and $\kappa$ are real-valued, and together with $N$, determine $\gamma$ via eq.~\eqref{eq:gammameth2}. We then solve eq.~\eqref{eq:meth2eom} via gradient descent. In particular, we define a function
\begin{align}
\Phi_2(M_1,\cdots,M_N)
&\equiv
\frac{1}{2}
\sum_j\left(
2M_j-M_{j+1}-M_{j-1}-\frac{\gamma}{2}
+\frac{\kappa}{M_j}
\right)^2
+
\frac{1}{2}
\left(
\sum_{j=1}^N\log M_j
\right)^2
\nonumber\\
&=
\frac{1}{2}
\sum_j\left(
2M_j-M_{j+1}-M_{j-1}
-
\frac{\kappa}{N}
\sum_{k=1}^N M_k^{-1}
+\frac{\kappa}{M_j}
\right)^2
+
\frac{1}{2}
\left(
\sum_{j=1}^N\log M_j
\right)^2
\nonumber\\
&=
\frac{1}{2}
\sum_j\left(
2M_j-M_{j+1}-M_{j-1}
+\frac{\kappa}{M_j}
\right)^2
\nonumber\\
&\qquad
-
\frac{\kappa^2}{2N}
\left(
\sum_{j=1}^N M_j^{-1}
\right)^2
+
\frac{1}{2}
\left(
\sum_{j=1}^N\log M_j
\right)^2,
\end{align}
so that eq.~\eqref{eq:meth2eom} is equivalent to $\Phi_2(M_1,\cdots,M_N)=0$, and we solve this latter equation via gradient descent. We found that if our initial $\kappa$ is to large, then the gradient descent converges to the confined solution in eq.~\eqref{eq:confsol}, but if we start with sufficiently small $\kappa$ we find partial-phase solutions. With one solution for the partial phase, we can increase $\kappa$ slightly, and use the solution as an initial condition for a new gradient descent. By iterating this procedure, we obtain solutions for a range of $\kappa$ values. We then turn on a small $\theta$ for fixed $N$ and $\gamma$, and use the $\theta=0$ partial-phase solution as an initial condition for the first method. By combining the two methods in this way, we found partial-phase solutions for a range of $N$, $\gamma$, and $\theta$, including $\theta=0$ and $\theta=\pi$.

As mentioned above, to provide additional evidence that our solutions describe a partial phase, we also numerically computed the Hessian of $V$, whose explicit components are
\begin{subequations}
\begin{align}
&\frac{\partial^2 V}{\partial M_j \partial M_k} =
- \frac{M_0}{M_k} 
\left(M_1^* + M_{N-1}^* - 2M_0^* + \frac{\gamma}{2}\right) 
\left(\frac{\delta_{jk}}{M_k} + \frac{1}{M_j}\right) \label{Hessian1},\\
&\frac{\partial^2 V}{\partial M_j^* \partial M_k} =  
2\delta_{k j} - \delta_{k, j+1} - \delta_{k, j-1}
+ \frac{M_0}{M_k} (\delta_{1 j} + \delta_{N-1, j})
+ \frac{M_0^*}{M_j^*} (\delta_{1 k} + \delta_{N-1, k})
+ 2 \frac{M_0 M_0^*}{M_k M_j^*},
\label{Hessian2}
\end{align}
\end{subequations}
and a similar expression for $\frac{\partial^2 V}{\partial M^\ast_j \partial M^\ast_k}$. At minima of the free energy $V$, all eigenvalues of the Hessian are positive, as occurs for the confined and deconfined solutions in eqs.~\eqref{eq:confsol} and~\eqref{eq:deconfsol2}, respectively. We expect the partial phase solution to be a local maximum of $V$, with one negative eigenvalue in the Hessian, and all the other eigenvalues positive. For all numerical solutions used below, we confirmed this expectation explicitly.

\subsubsection*{Finite-$N$ numerical results at $\theta=0$}
\hspace{0.51cm}

\begin{figure}[htbp]
    \begin{center}
    \begin{subfigure}{.5\textwidth}
   \includegraphics[width=80mm]{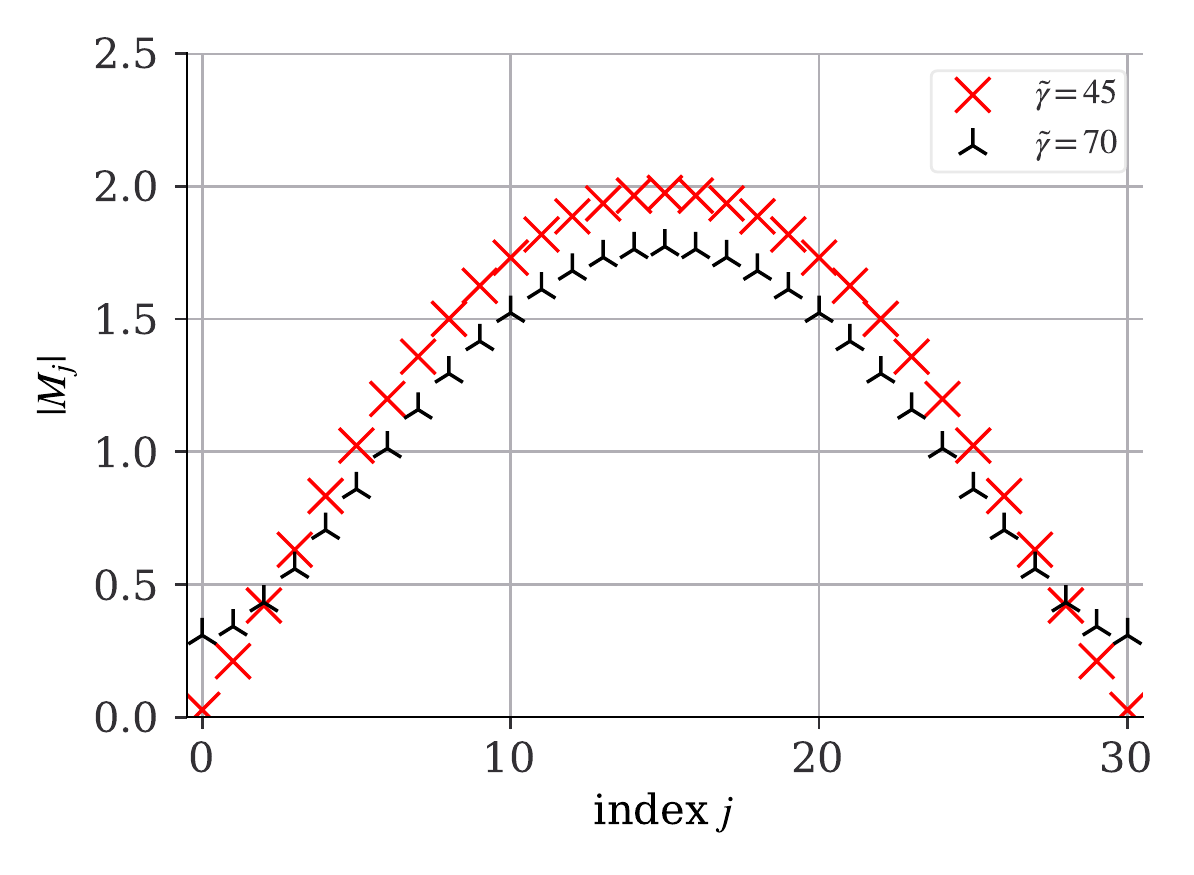}
   \caption{}
   \end{subfigure}%
       \begin{subfigure}{.5\textwidth}
   \includegraphics[width=80mm]{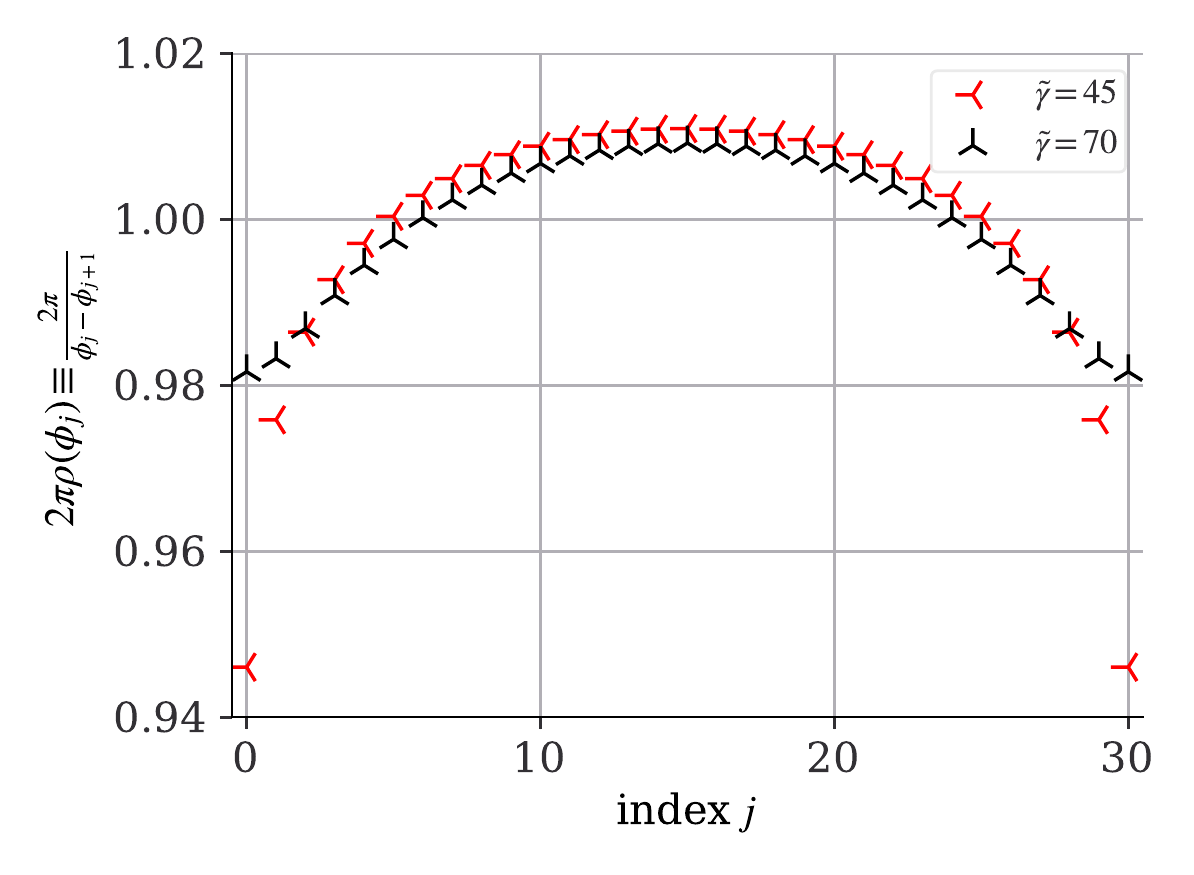}
     \caption{}
   \end{subfigure}
  \end{center}
  \caption{(a) $\theta=0$ numerical results for the $|M_j|$ versus $j$ for $N=30$ in the partial phase, for $\tilde{\gamma}=45$ (red crosses) and $70$ (black 3-point stars). Notice that $|M_0|$ is small but non-zero. (b) Using eq.~\eqref{eq:M-vs-phi}, with $g^2=0.1$, we converted the results of (a) into the Polyakov loop eigenvalue distribution $2 \pi \rho(\phi_j) = \frac{2\pi}{\phi_{j}-\phi_{j+1}}$ versus $j$. The eigenvalue distribution is not flat, indicating spontaneous breaking of centre symmetry, and is not gapped, indicating that the phase is not deconfined. We thus identify these as partial phase solutions.
      }\label{fig:Msandevalstheta0}
\end{figure}

\begin{figure}[htbp]
    \begin{center}
    \begin{subfigure}{.5\textwidth}
   \includegraphics[width=80mm]{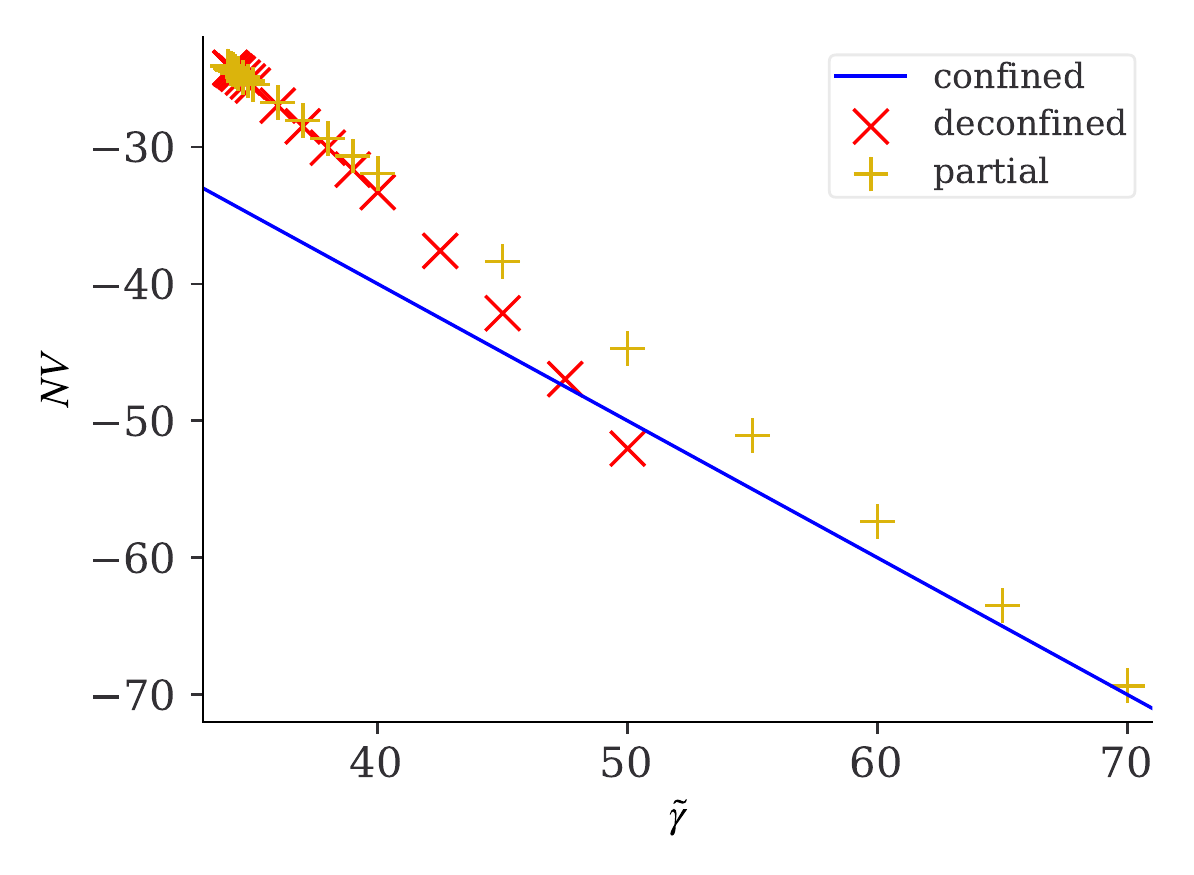}
   \caption{}
   \end{subfigure}%
       \begin{subfigure}{.5\textwidth}
   \includegraphics[width=80mm]{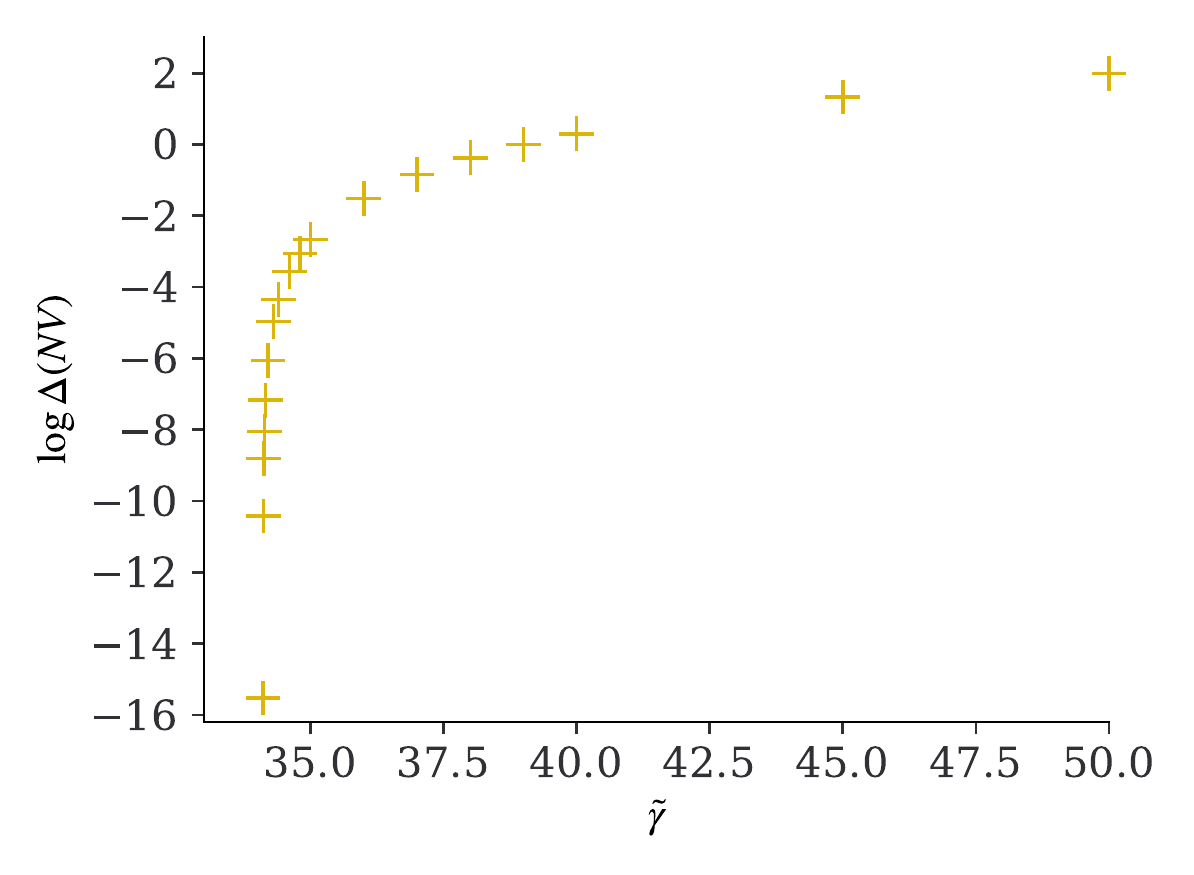}
     \caption{}
   \end{subfigure}
  \end{center}
  \caption{(a) $\theta=0$ numerical results for $N$ times the free energy $V$ as a function of $\tilde{\gamma}$ for $N=30$, showing the confined phase (blue line), deconfined phase (red crosses), and partial phase (gold plus signs). We clearly observe a first-order transition, with the partial phase as the unstable branch connecting the confined and deconfined phases. (b) $\theta=0$ numerical results for $\log (\Delta NV)$ versus $\tilde{\gamma}$ for $N=30$, where $\Delta NV$ is the difference in $NV$ between the partial and deconfined phases, $\Delta NV \equiv N(\left . V \right |_{\textrm{partial}}- \left . V \right|_{\textrm{deconf}})$. Clearly the partial phase has higher free energy than the deconfined phase near the transition between them (which at $N=\infty$ is the GWW transition).
      }\label{fig:N30-theta=0-free-energy}
\end{figure}

\begin{figure}
    \begin{center}
    \includegraphics[width=100mm]{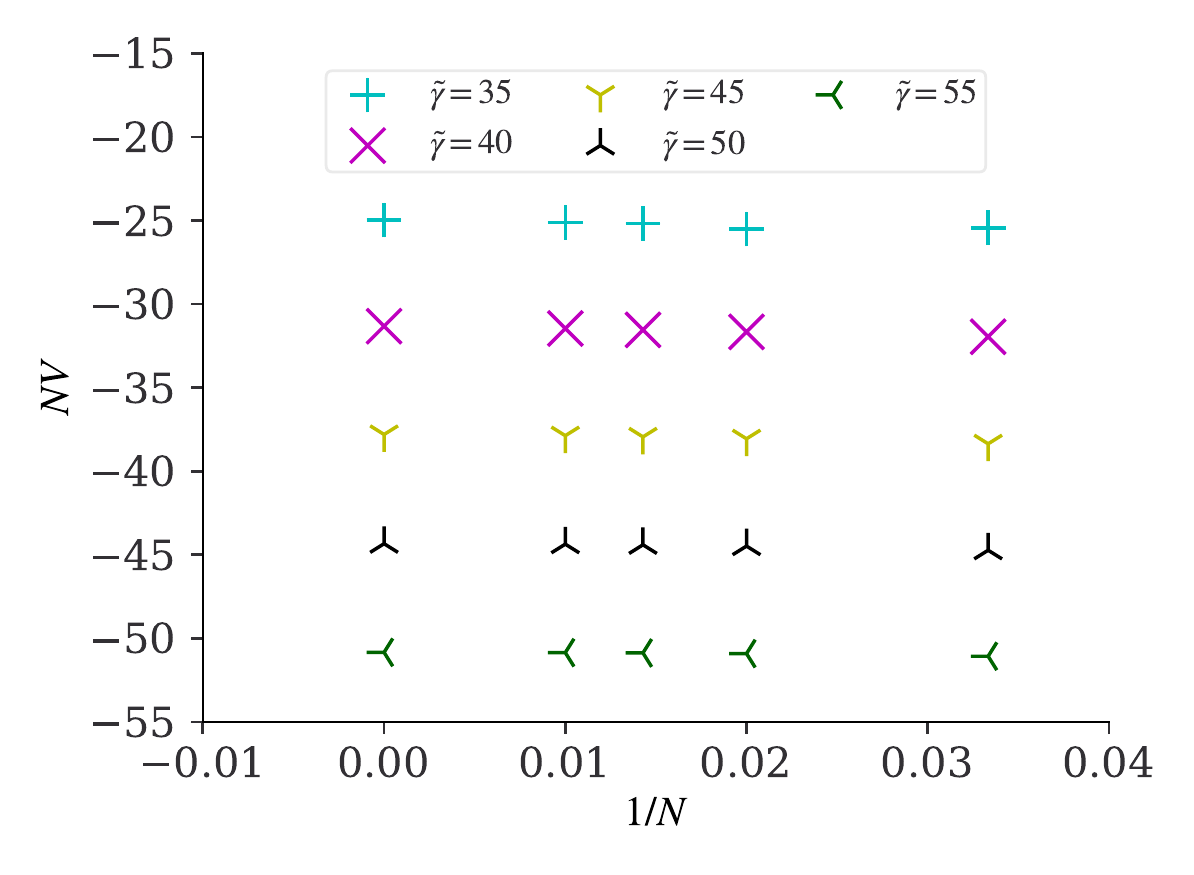}
    \end{center}
    \caption{Our $\theta=0$ numerical results for $N$ times the tree energy $V$ versus $1/N$ for $N=30$, $50$, $70$, $100$, $\infty$, or equivalently $1/N= 0.0333\ldots$, $0.02$, $0.0142\ldots$, $0.01$, $0$, for several values of $\tilde{\gamma}$. These results suggest that the $N \to \infty$ limit of $V$ is smooth at each fixed $\tilde{\gamma}$.}\label{fig:V-vs-Ninv}
\end{figure}

For $\theta=0$, fig.~\ref{fig:Msandevalstheta0} shows typical numerical solutions for the $M_j$ and the corresponding distribution of Polyakov loop eigenvalues, $2 \pi \rho(\phi_j) = \frac{2\pi}{\phi_{j}-\phi_{j+1}}$, in the partial phase, with $\tilde{\gamma}= 45$ and $70$. For both values of $\tilde{\gamma}$, the eigenvalue distribution is not flat, indicating spontaneous breaking of the centre symmetry, and is not gapped, indicating that this phase is not deconfined. In particular, at $\tilde{\gamma}=45$, observe that $M_0$ is small, but non-zero. We will explore $|M_0|$ in more detail momentarily.

Fig.~\ref{fig:N30-theta=0-free-energy} shows our numerical results for the free energy $V$ times $N$ as a function of $\tilde{\gamma}$ with $N=30$ and $\theta=0$. We see the swallow-tail shape characteristic of a first-order phase transition, with the partial phase appearing as the unstable branch connecting the confined and deconfined branches. In particular, the partial phase always has the highest free energy, and hence is always thermodynamically dis-favoured.

The phase diagram fig.~\ref{fig:N30-theta=0-free-energy}, which has $N=30$, is qualitatively similar to the phase diagram in fig.~\ref{fig:V-vs-gt-largeN}, which has $N=\infty$, suggesting that as we increase $N$, fig.~\ref{fig:N30-theta=0-free-energy} may evolve continuously into fig.~\ref{fig:V-vs-gt-largeN}. In other words, our numerical results suggest that the $N\to\infty $ limit of $V$ is smooth. We provide additional evidence for this in fig.~\ref{fig:V-vs-Ninv}, which shows $NV$ versus $1/N$ for several values of $\tilde{\gamma}$, and strongly suggests that as $N$ increases, $NV$ smoothly approaches the $N=\infty$ result. This also provides further evidence that our numerical solutions describe a partial phase, since they appear to connect smoothly to our $N=\infty$ partial phase solutions.

\begin{figure}[htbp]
    \begin{center}
       \begin{subfigure}{.5\textwidth}
   \includegraphics[width=80mm]{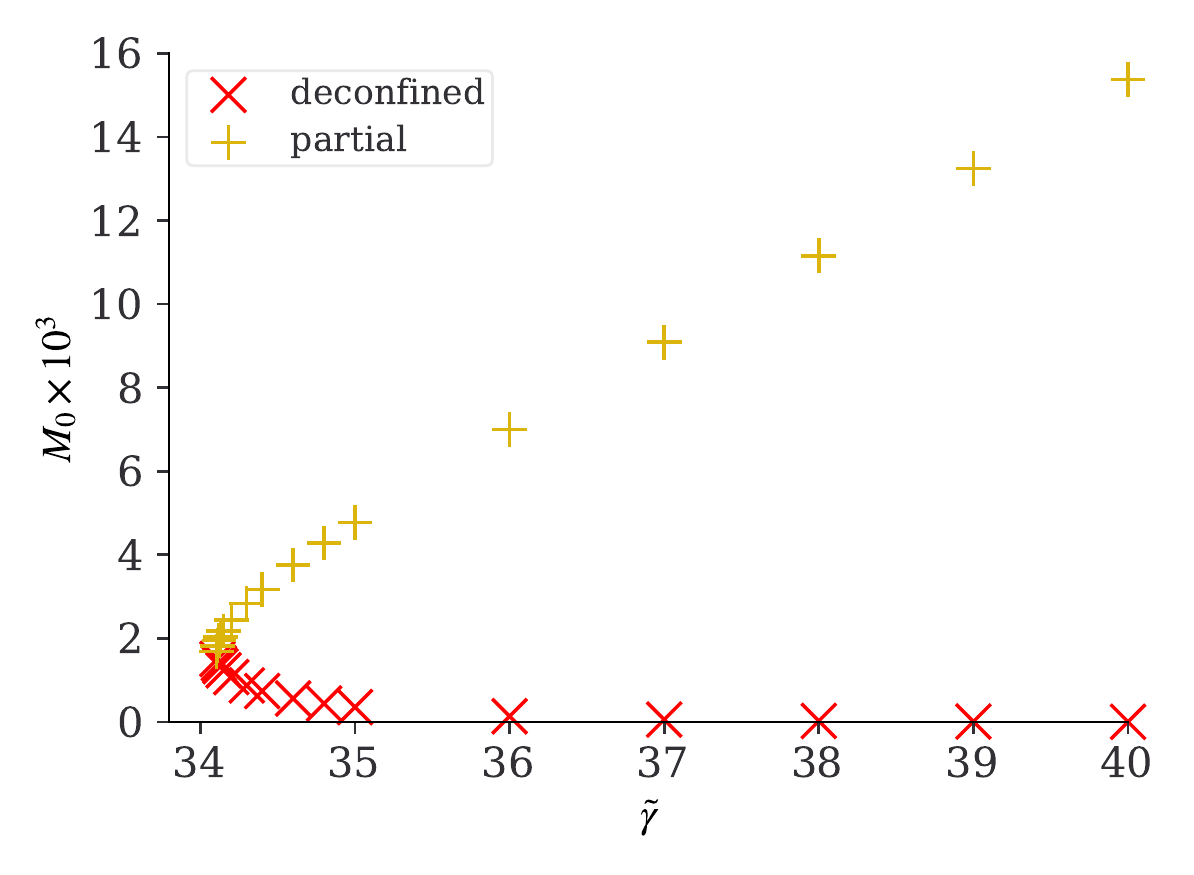}
        \caption{}
   \end{subfigure}%
       \begin{subfigure}{.5\textwidth}
   \includegraphics[width=80mm]{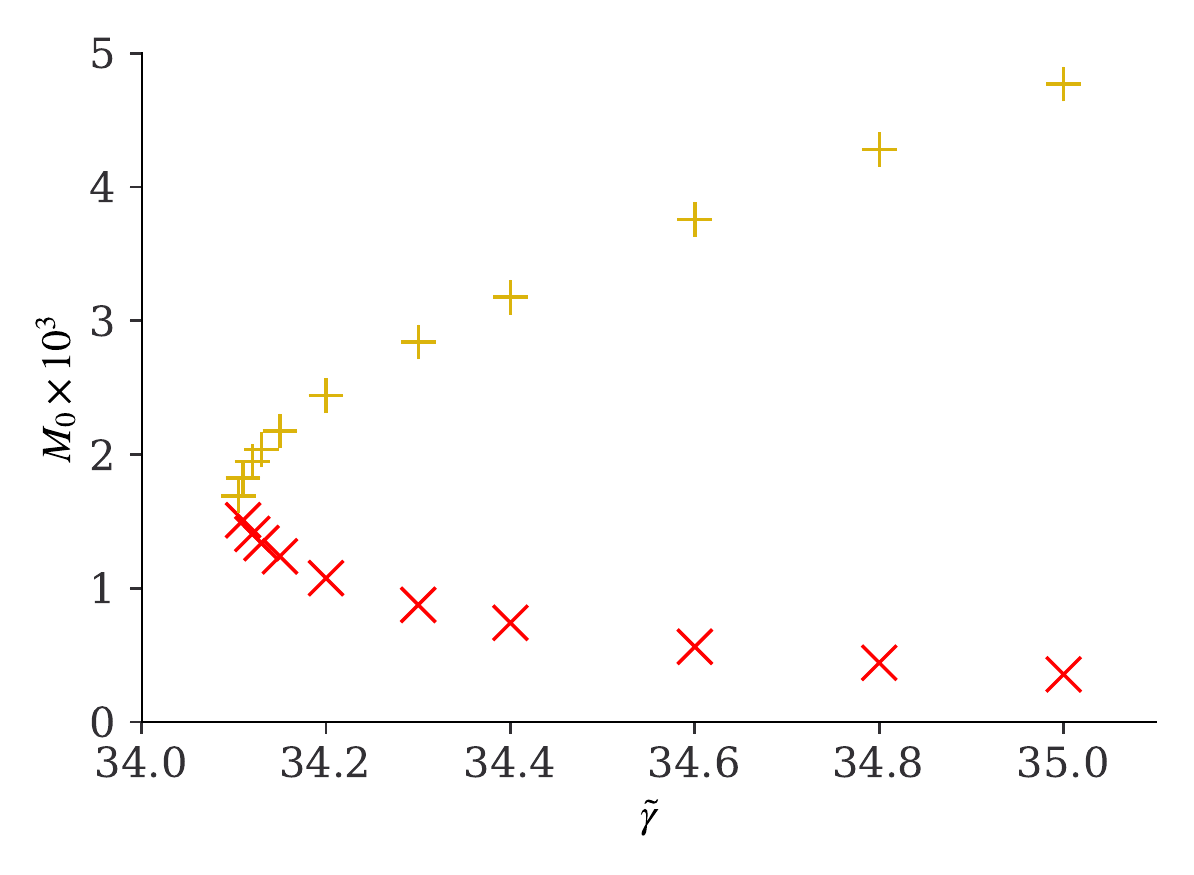}
           \caption{}
   \end{subfigure}%
  \end{center}
  \caption{(a) $\theta=0$ numerical results for $|M_0|$, magnified by a factor of $10^3$ for clarity, as a function of $\tilde{\gamma}$, for $N=30$, in the deconfined phase (red crosses) and partial phase (gold plus signs). $|M_0|$ in the deconfined phase becomes zero in the large-$N$ limit. The gold plus signs have larger values, showing that our solutions indeed describe a partial phase. (b) Close-up of (a) near the transition point.
  }\label{fig:N30-theta=0-M0}
\end{figure}

Fig.~\ref{fig:N30-theta=0-M0} shows $|M_0|$, magnified by a factor of $10^3$ for clarity, as a function of $\tilde{\gamma}$ near the transition to the deconfined phase (which at $N=\infty$ is the GWW transition). Clearly, $|M_0|$ is small, but not exponentially small, near the transition point. In particular, $|M_0|$ is non-zero, justifying our interpretation of these solutions as the partial phase.

When $\theta=0$ all the $M_j$ are real and CP symmetry is preserved in all phases. As a result, our order parameter for CP symmetry breaking, $\frac{\partial V}{\partial \theta} \propto \langle \mathrm{tr}(F_{\mu\nu} \tilde{F}^{\mu\nu})\rangle$, vanishes in all phases when $\theta=0$. To see the spontaneous breaking of CP symmetry in the partial phase, we turn next to $\theta=\pi$.

\subsubsection*{Finite-$N$ numerical results at $\theta=\pi$}
\hspace{0.51cm}

\begin{figure}[htbp]
    \begin{center}
    \begin{subfigure}{.5\textwidth}
   \includegraphics[width=80mm]{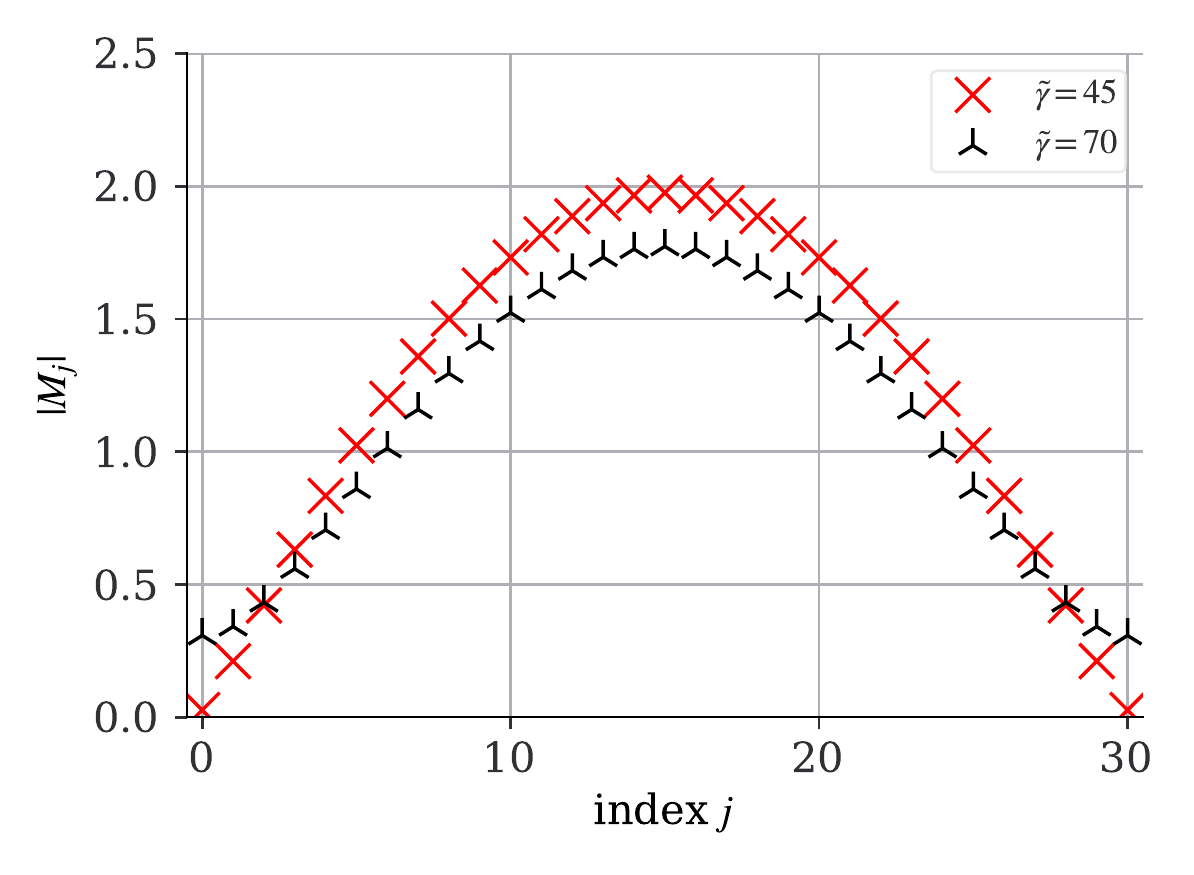}
   \caption{}
   \end{subfigure}%
       \begin{subfigure}{.5\textwidth}
   \includegraphics[width=80mm]{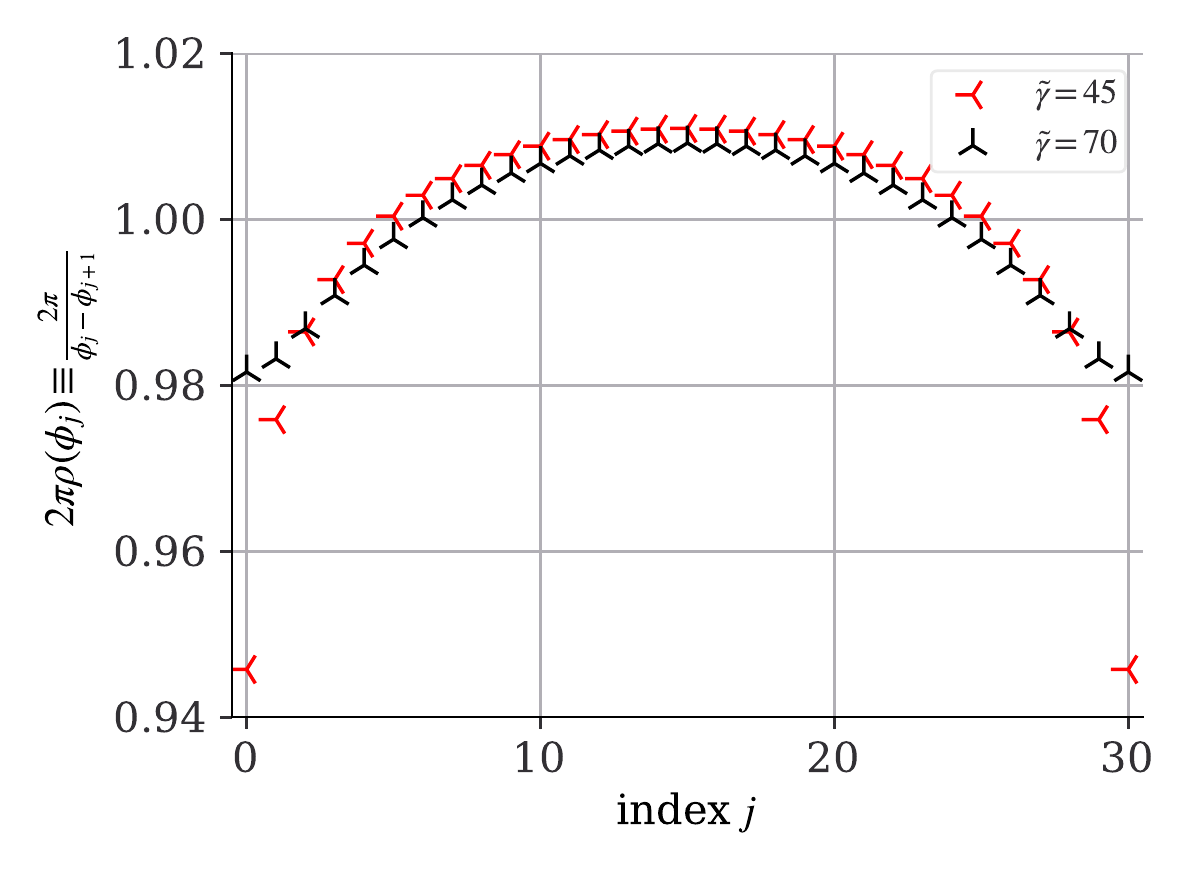}
     \caption{}
   \end{subfigure}
  \end{center}
  \caption{(a) $\theta=\pi$ numerical results for the $|M_j|$ versus $j$ for $N=30$ in the partial phase, for $\tilde{\gamma}=45$ (red crosses) and $70$ (black 3-point stars). Notice that $|M_0|$ is small but non-zero. (b) Using eq.~\eqref{eq:M-vs-phi}, with $g^2=0.1$, we converted the results of (a) into the Polyakov loop eigenvalue distribution $2 \pi \rho(\phi_j) = \frac{2\pi}{\phi_{j}-\phi_{j+1}}$ versus $j$. These results are qualitatively similar to those at $\theta=0$ in fig.~\ref{fig:Msandevalstheta0}.
      }\label{fig:MsandevalsthetaPi}
\end{figure}

For $\theta=\pi$, fig.~\ref{fig:MsandevalsthetaPi} shows typical numerical solutions for the $|M_j|$ and the corresponding distribution of Polyakov loop eigenvalues, in the partial phase, with $\tilde{\gamma}= 45$ and $70$. The results are qualitatively similar to the $\theta=0$ case in fig.~\ref{fig:Msandevalstheta0}, although a key difference is that when $\theta=\pi$, the $M_j$'s can have non-zero phases and hence CP symmetry can be spontaneously broken, as we will discuss in detail momentarily.

We next want to calculate $V$ and $\frac{\partial V}{\partial \theta}$, both at $\theta=\pi$. To do so numerically, we calculated $V$ at three values of $\theta$ near $\pi$, namely $\theta=3.14159$, $3.14155$, and $3.14149$, and then extrapolated to $\theta=\pi$ by fitting to a form $V(\theta)=a\cdot(\theta-\pi)+b$ and numerically extracting the coefficients $a$ and $b$. In this was we obtained $V$ and $\frac{\partial V}{\partial \theta}=a$ at $\theta=\pi$. We performed a similar extrapolation for $|M_0|$ as well.

\begin{figure}[htbp]
  \begin{center}
   \begin{subfigure}{.5\textwidth}
  \includegraphics[width=80mm]{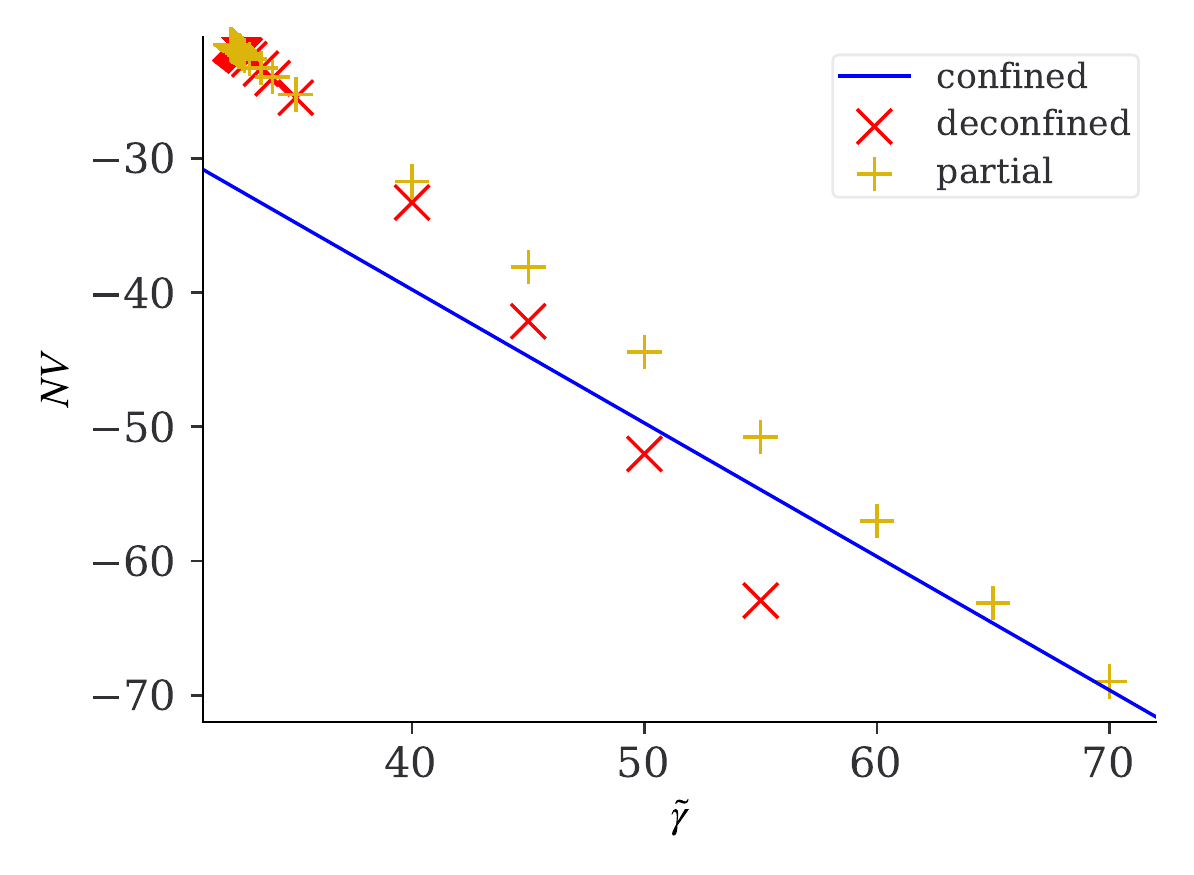}
     \caption{}
   \end{subfigure}%
       \begin{subfigure}{.5\textwidth}
   \includegraphics[width=80mm]{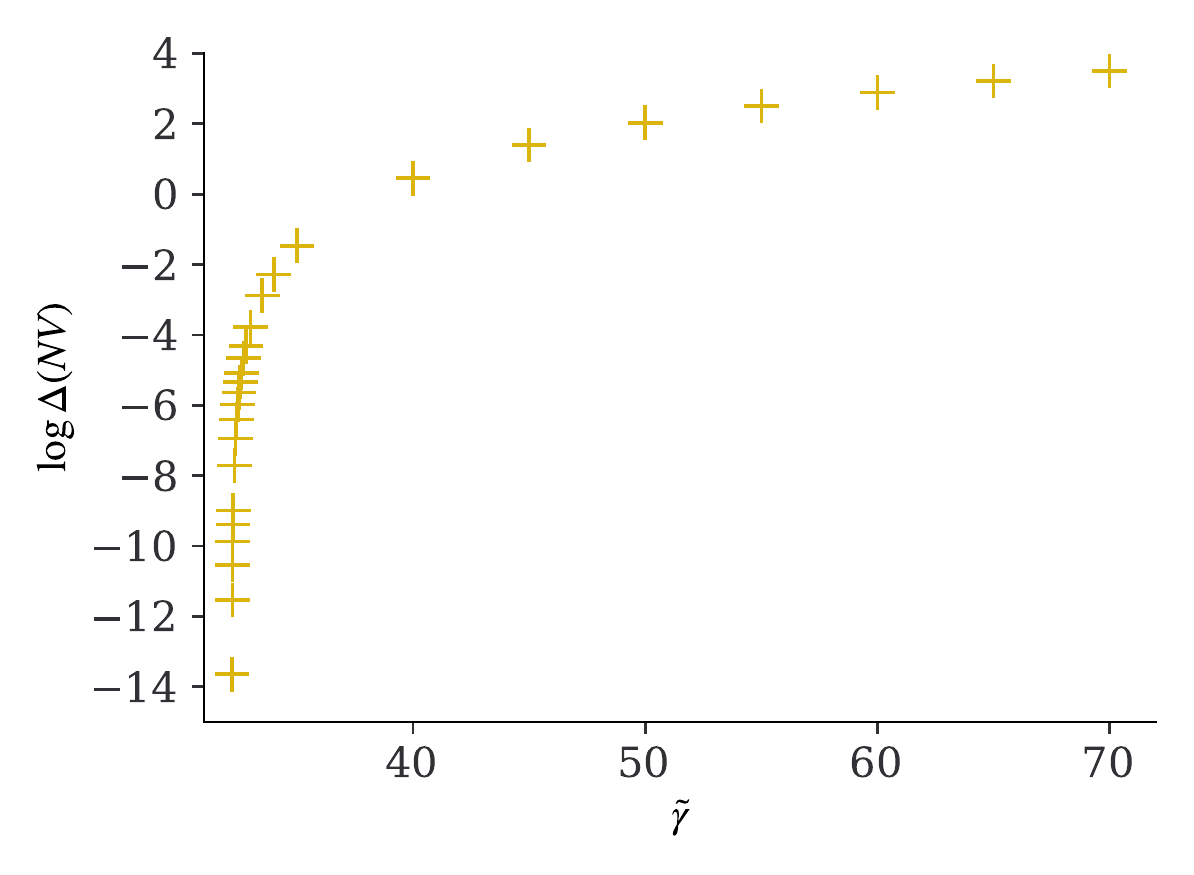}
        \caption{}
   \end{subfigure}
  \end{center}
  \caption{(a) $\theta=\pi$ numerical results for $N$ times the free energy $V$ as a function of $\tilde{\gamma}$ for $N=30$, showing the confined phase (blue line), deconfined phase (red crosses), and partial phase (gold plus signs). We find a first-order transition, similar to the $\theta=0$ case in fig.~\ref{fig:N30-theta=0-free-energy}. (b) Our $\theta=\pi$ numerical results for $\log (\Delta NV)$ versus $\tilde{\gamma}$ for $N=30$, with $\Delta NV$ defined as in fig.~\ref{fig:N30-theta=0-free-energy}. Clearly the partial phase has higher free energy than the deconfined phase near the transition between them (which at $N=\infty$ is the GWW transition).
 }\label{fig:Free-Energy-theta=pi-N30}
\end{figure}

Fig.~\ref{fig:Free-Energy-theta=pi-N30} shows our numerical results for the free energy $V$ times $N$ as a function of $\tilde{\gamma}$ with $N=30$ and $\theta=\pi$. We again see the swallow-tail shape characteristic of a first-order phase transition, with the partial phase appearing as the unstable branch connecting the confined and deconfined branches. Again the partial phase always has the highest free energy, and hence is always thermodynamically dis-favoured.

\begin{figure}[htbp]
  \begin{center}
     \begin{subfigure}{.5\textwidth}
   \includegraphics[width=80mm]{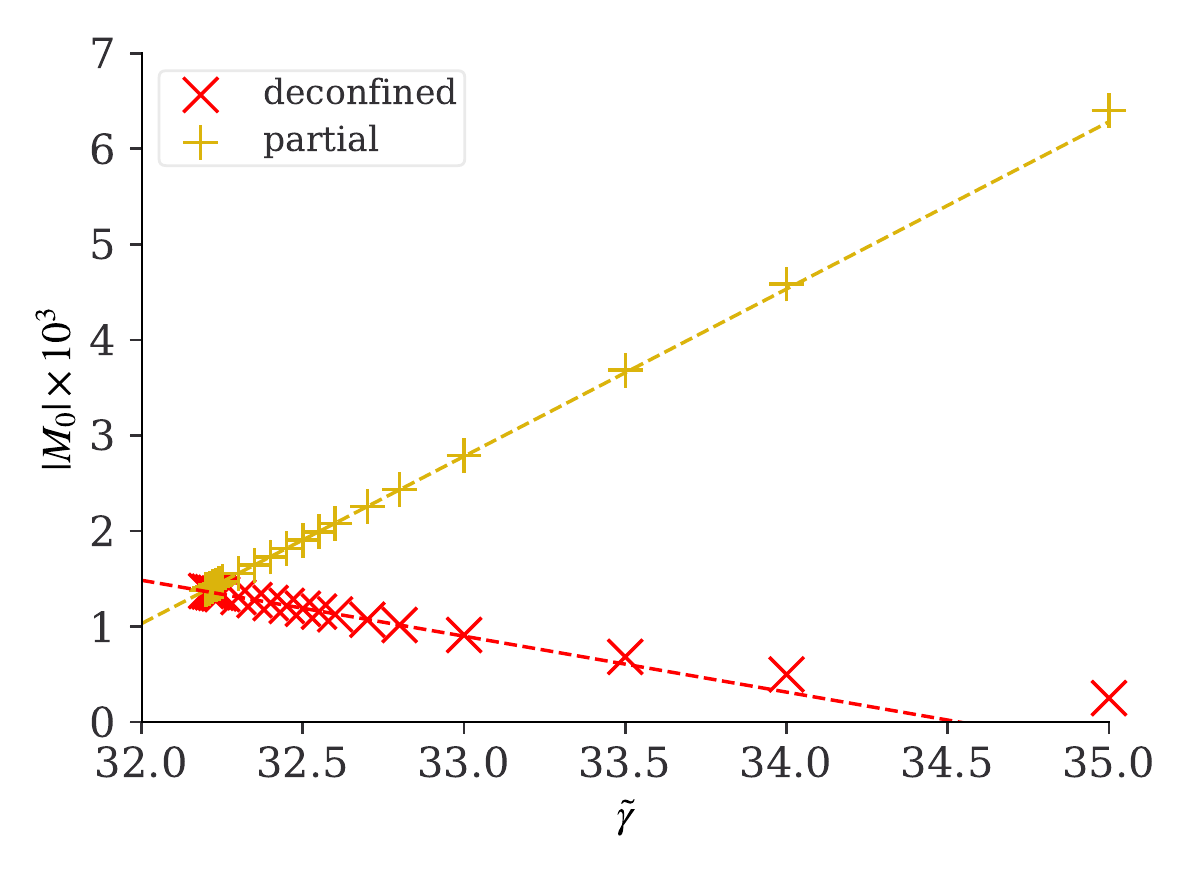}
        \caption{}
   \end{subfigure}%
       \begin{subfigure}{.5\textwidth}
   \includegraphics[width=80mm]{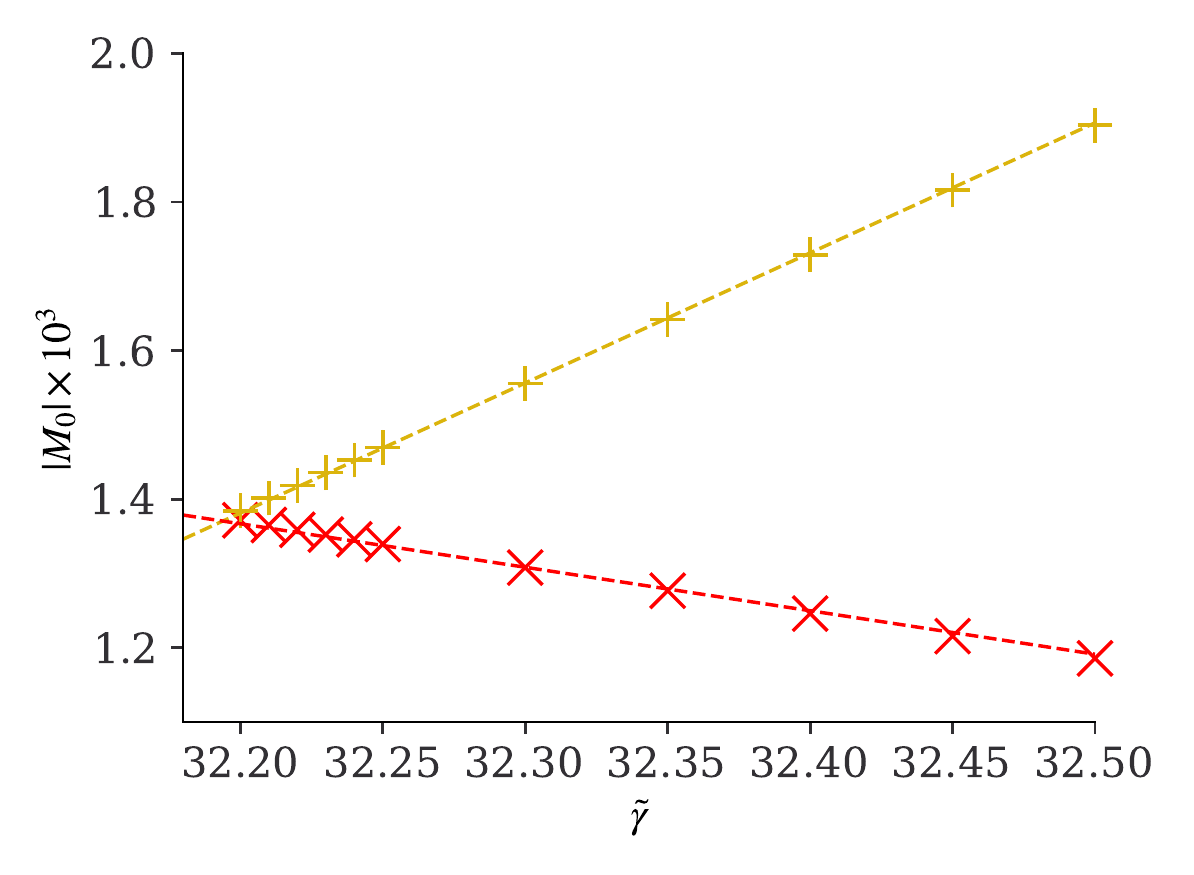}
           \caption{}
   \end{subfigure}%
  \end{center}
  \caption{(a) $\theta=\pi$ numerical results for $|M_0|$, magnified by a factor of $10^3$ for clarity, as a function of $\tilde{\gamma}$, for $N=30$, in the deconfined phase (red crosses) and partial phase (gold plus signs). 
 $|M_0|$ in the deconfined phase becomes zero in the large-$N$ limit. The gold plus signs have larger values, showing that our solutions indeed describe a partial phase. (b) Close-up of (a) near the transition point.
  }\label{fig:M0_N30_comparison}
\end{figure}

Fig.~\ref{fig:M0_N30_comparison} shows $|M_0|$, magnified by a factor of $10^3$ for clarity, as a function of $\tilde{\gamma}$ near the transition to the deconfined phase (which at $N=\infty$ is the GWW transition). Similar to the $\theta=0$ case in fig.~\ref{fig:N30-theta=0-M0}, we again see that $|M_0|$ is small, but not exponentially small, near the transition point, and in particular $|M_0|$ is non-zero, justifying our interpretation of these solutions as the partial phase.

\begin{figure}[htbp]
  \begin{center}
     \begin{subfigure}{.5\textwidth}
   \includegraphics[width=80mm]{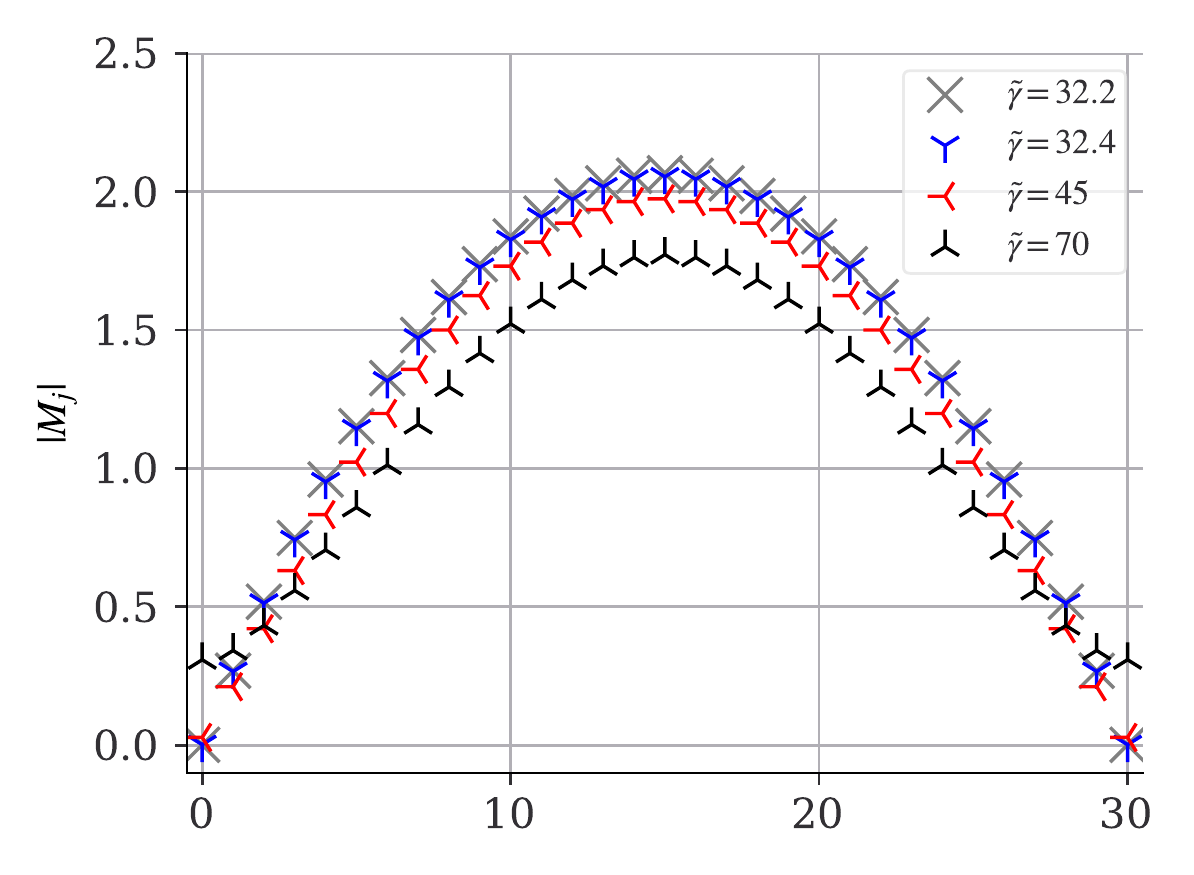}
        \caption{}
   \end{subfigure}%
       \begin{subfigure}{.5\textwidth}
   \includegraphics[width=80mm]{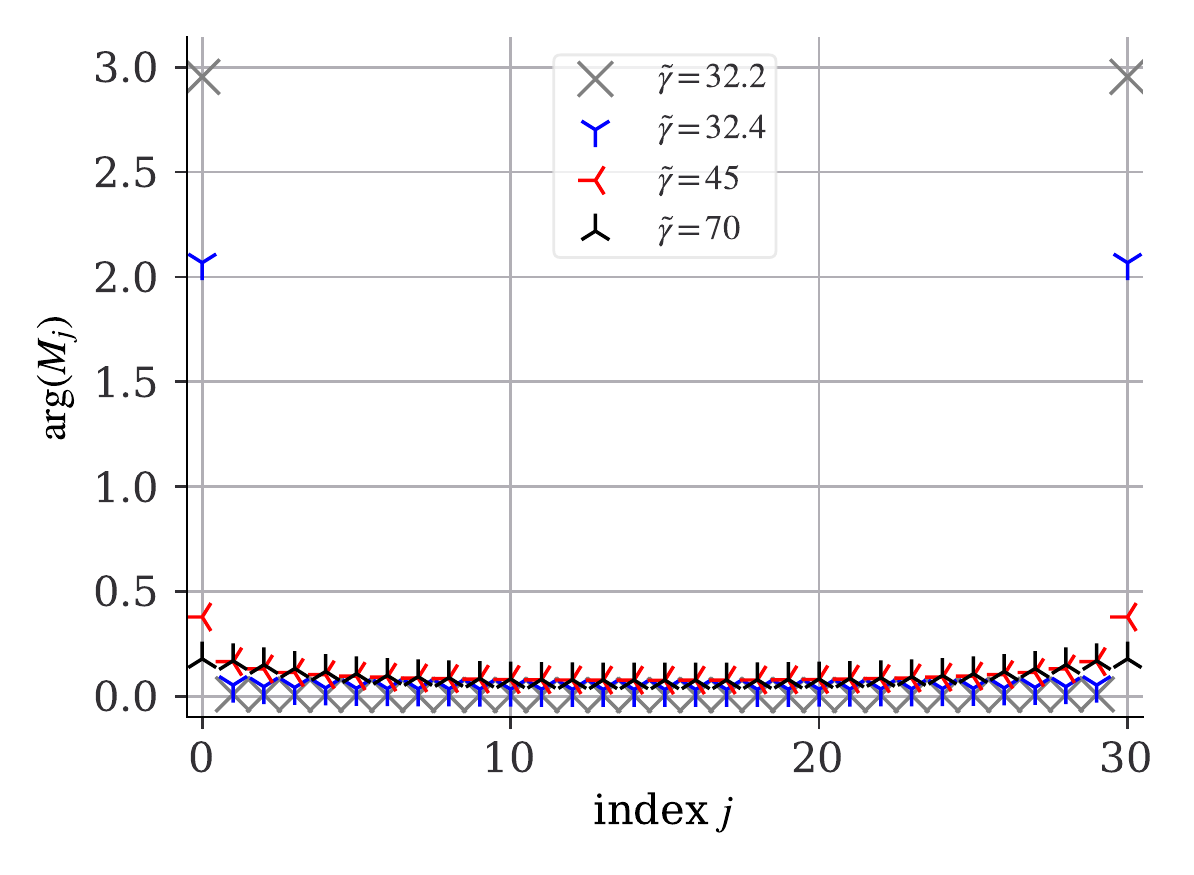}
           \caption{}
   \end{subfigure}
  \end{center}
  \caption{(a) $\theta=\pi$ numerical results for $|M_j|$ as functions of $j$, in the partial phase, at $\tilde{\gamma}=32.2$ (gray crosses) $32.4$ (blue 3-pont star), $45$ (red 3-point star), and $70$ (black 3-point star). (b) $\theta=\pi$ numerical results for $\textrm{arg}(M_j)$ as functions of $j$, in the partial phase, at the same $\tilde{\gamma}$ as in (a). Generically the $M_j$ have non-zero phases, indicating spontaneous breaking of CP symmetry. As we decrease $\tilde{\gamma}$ we see that $\textrm{arg}(M_0)$ grows, to meet the value of the deconfined phase, $\textrm{arg}(M_0)=\theta=\pi$. 
  }\label{fig:partialMcomparison}
\end{figure}

Of course, a key difference between the $\theta=0$ and $\theta=\pi$ cases is that in the former all of the $M_j$ were real-valued while in the latter the $M_j$ acquire non-zero phases, indicating spontaneous breaking of CP symmetry. Fig.~\ref{fig:partialMcomparison} illustrates this difference, showing our numerical results for both $|M_j|$ and $\textrm{arg}(M_j)=\varphi_j$ as a function of $j$ in the partial phase, for $\theta=\pi$ and various $\tilde{\gamma}$. The evolution of the $|M_j|$ is consistent with what we observed in figs.~\ref{fig:Msandevalstheta0} and~\ref{fig:MsandevalsthetaPi}. However, we now see that in the partial phase with $\theta=\pi$, generically all the $\textrm{arg}(M_j)=\varphi_j$ are non-zero. We also see that as we approach the deconfined phase, all of the $\textrm{arg}(M_j)$ approach zero except for $\textrm{arg}(M_0)$, which approaches $\textrm{arg}(M_0)=\theta=\pi$, so that the partial solutions connect to the deconfined solution (recall eq.~\eqref{eq:deconfM0approx}).

Our numerical results for $\frac{\partial V}{\partial \theta} \propto \langle \mathrm{tr}(F_{\mu\nu} \tilde{F}^{\mu\nu})\rangle$ (times $N^3$) with $\theta=\pi$ and $N=30$, $50$, $70$ appear in fig.~\ref{fig:TrFF_N70N50N30}, alongside our $N=\infty$ results from subsection~\ref{sec:abelianlargen}. In all cases we see that $\frac{\partial V}{\partial \theta}\neq 0$ in the partial phase, indicating spontaneous breaking of CP symmetry. This is our main result, that CP symmetry distinguishes the partial and deconfined phases, even at finite $N$. Moreover, at large $N$ we saw that $\frac{\partial V}{\partial \theta}$ was finite at the GWW transition, and hence jumped discontinuously between the partial and deconfined phases, indicating a first-order transition. In contrast, at finite $N$ we see in fig.~\ref{fig:TrFF_N70N50N30} that $\frac{\partial V}{\partial \theta}$ smoothly approaches zero at the transition to the deconfined phase, suggesting a second-order transition. However, as $N$ increases we observe that the slope of $\frac{\partial V}{\partial \theta}$ increases, strongly suggesting that the slope diverges as $N\to \infty$, thus connecting the finite- and large-$N$ calculations.

In summary, we have demonstrated that, in weakly-coupled, softly-broken $\N=1$ SYM on $\mathbb{S}^1\times\mathbb{R}^3$, at both infinite and finite $N$ a partial phase exists that connects the confined and deconfined phases, and has broken centre and CP symmetries. As a result, order parameters exist that can distinguish the partial phase from both the confined and deconfined phases.  In particular, $\frac{\partial V}{\partial \theta} \propto \langle \mathrm{tr}(F_{\mu\nu} \tilde{F}^{\mu\nu})\rangle$ can distinguish the partial and deconfined phases, a phenomenon that to our knowledge is novel.

\section{Chiral symmetry in strongly-coupled lattice gauge theory}
\label{sec:EK-model}
\hspace{0.51cm}
In this section, we consider $U(N)$ YM theory on an anisotropic lattice, with $d=3$ spatial dimensions with lattice spacing $a_s$, and a compact Euclidean time direction with lattice spacing $a_t$. Defining the dimensionless parameter $a \equiv \lambda_{\textrm{tH}}\,a_t/a_s$, with 't Hooft coupling $\lambda_{\textrm{tH}}$, this theory's plaquette action takes the form
\begin{eqnarray}
S
=
\frac{N}{2a}
\sum_{\vec{x}}
\sum_{\mu=1}^{d}\sum_{t=1}^{n_t}
{\rm Tr}\left(
\textbf{1}_N
-
U_{\mu,t,\vec{x}}V_{t,\vec{x}+\hat{\mu}}U_{\mu,t+1,\vec{x}}^\dagger V_{t,\vec{x}}^\dagger 
\right)
+
{\rm h.c.}
+
{\rm magnetic\ term},
\label{YM-action-lattice}
\end{eqnarray}
with number of temporal lattice sites $n_t$, dimensionless spatial and temporal link variables $U_{\mu,t,\vec{x}}$ and $V_{t,\vec{x}}$, respectively, and ``magnetic terms'' representing terms without $V_{t,\vec{x}}$. As temperature, we will use the dimensionless combination  $T=\frac{1}{a n_t}$. 

We will study theory in eq.~\eqref{YM-action-lattice} using three limits, defined as follows. The first limit is strong coupling, $\lambda_{\textrm{tH}} \gg 1$. In this limit, we drop the magnetic terms because they are suppressed by $1/\lambda_{\textrm{tH}}^2$ relative to the terms shown explicitly in eq.~\eqref{YM-action-lattice}. Of course, doing so produces a theory that, in the continuum limit, is no longer pure YM, and in particular cannot be Lorentz invariant. Nevertheless, we expect the theory to share qualitative features with YM. More generally, this theory will provide a non-trivial example of a partial phase distinguished from both the confined and deconfined phases by symmetries, thus providing additional compelling evidence for our conjecture that such behaviour is generic.

The second limit is the 't Hooft large-$N$ limit. The third limit is then to invoke large-$N$ volume independence, and reduce the spatial directions to a single site. In other words, we invoke the Eguchi-Kawai (EK) equivalence~\cite{Eguchi:1982nm}.\footnote{Actually, in the original paper on the EK equivalence~\cite{Eguchi:1982nm}, both the spatial directions and the Euclidean time direction were reduced to a single site. In contrast, we will reduce only the spatial directions to a single site, leaving the Euclidean time direction untouched.} Crucially, EK equivalence requires not only the 't Hooft large-$N$ limit, but also two other conditions: unbroken discrete spatial translational symmetry in the large-volume theory, and centre symmetry unbroken by a Wilson line in a spatial direction in the single-site theory. Upon reducing to a single site, the action in eq.~\eqref{YM-action-lattice} becomes
\begin{eqnarray}
S
=
\frac{N}{2a}
\sum_{\mu=1}^{d}\sum_{t=1}^{n_t}
{\rm Tr}\left(
\textbf{1}_N
-
U_{\mu,t}V_tU_{\mu,t+1}^\dagger V_t^\dagger 
\right)
+
{\rm h.c.},
\label{EK-action-lattice}
\end{eqnarray}
which is the action we will use in all that follows.

These limits come with some subtleties. For example, since we reduce to a single site, our results are valid only in the 't Hooft large-$N$ limit. However, to do numerics we must use link variables of finite size, hence we must work with finite $N$. We will use large but finite $N$, namely $N=8,\ldots,32$. Notice this is qualitatively different from our finite-$N$ results for softly-broken $\N=1$ SYM in sec.~\ref{sec:nummethods}: those were genuinely finite $N$, while here our starting point, namely the action in eq.~\eqref{EK-action-lattice} is equivalent to the one with the action in eq.~\eqref{YM-action-lattice} only at $N=\infty$. We will discuss other subtleties with our limits in what follows.

Our order parameter for the confinement/deconfinement transition will be the Polyakov loop, $P$, defined on the lattice as $P=\frac{1}{N}{\rm Tr}{\cal P}$, where
\begin{eqnarray}
{\cal P}
=
V_1V_2\cdots V_{n_t}.
\label{eq:latticePdef}
\end{eqnarray}
Although $P$ can be complex, the complex phase can be shifted by a centre symmetry transformation, which acts on the Polyakov loop as $P\to e^{i\alpha}P$, where in the large-$N$ limit, $\alpha$ is an arbitrary real number. In the following, we will use the centre symmetry to make $P$ real-valued and non-negative, so strictly speaking, our $P=\left|\frac{1}{N}{\rm Tr}{\cal P}\right|$. Using this order parameter, we will observe a first-order transition as $T$ increases, from a confined to a deconfined phase. In this transition the centre symmetry will be spontaneously broken. We will also find an unstable partial phase connecting the confined and deconfined phases. In the partial phase the centre symmetry will be spontaneously broken.

We will add to this model fermions in the fundamental representation, i.e. quarks. In general, fundamental-representation fields explicitly break centre symmetry. To suppress this explicit breaking, we will take a fourth limit, namely the probe limit: we will introduce a number $N_f$ of quarks, keep $N_f$ fixed as $N\to\infty$, and work only to leading order in $N_f/N$. In the probe limit, the explicit breaking of centre symmetry is invisible in the pure glue sector, so the confinement/deconfinement transition will be unchanged. However, as in ordinary YM theory, the probe quarks exhibit chiral symmetry breaking in the confined phase, and chiral symmetry restoration in the deconfined phase. The order parameter for this transition is the chiral condensate. We will numerically compute the eigenvalue distribution of the probe Dirac operator and use the Banks-Casher relation~\cite{Banks:1979yr} to extract the value of the chiral condensate. Our main result in this section will be that chiral symmetry is broken in the partial phase, and hence the chiral condensate provides an order parameter that can distinguish the partial and deconfined phases.

In sec.~\ref{sec:conf/deconf} we present our numerical results for the confinement/deconfinement transition in this model, including novel results for the partial phase. In sec.~\ref{sec:chiral-symmetry} we present our numerical results for the probe Dirac operator's eigenvalue spectrum, where our main result is that chiral symmetry is spontaneously broken in the partial phase.

\subsection{Confinement/deconfinement transition}
\label{sec:conf/deconf}
\hspace{0.51cm}

We will apply the Hybrid Monte Carlo method~\cite{Duane:1987de} to the single-site lattice action in eq.~\eqref{EK-action-lattice}. Without probe quarks, the cost for configuration generation is not so large, unless we make $N$ and/or $n_t$ large. We will work with $N=8,\ldots,32$ and $n_t=24$.

As we will show shortly, for the current theory in the canonical ensemble the confinement/deconfinement transition is first order. As a result, the partial phase is the maximum of the free energy, and therefore has the \textit{smallest} weight in the ensemble, and hence is not efficiently sampled using the standard Monte Carlo method. To sample configurations at fixed values of $P$ efficiently, we performed a constrained simulation, in which we fixed the value of the Polyakov loop. Specifically, we modified the action $S$ in eq.~\eqref{EK-action-lattice} by adding a term, $S \to S+\Delta S$, where
\begin{eqnarray}
\label{eq:deltaS}
\Delta S
=
\left\{
\begin{array}{cc}
\frac{g_{\rm P}}{2}\left(
P-(P_{\rm fix}+\delta)
\right)^2 & (P > P_{\rm fix}+\delta)\\
0 & (P_{\rm fix}-\delta \le P \le P_{\rm fix}+\delta)\\
\frac{g_{\rm P}}{2}\left(
P-(P_{\rm fix}-\delta)
\right)^2& (P < P_{\rm fix}- \delta)
\end{array}
\right.,
\end{eqnarray}
where we chose the dimensionless constant $g_{\rm P}$ to be large enough that the value of $P$ is fixed to a small window $P_{\rm fix}-\delta \le P \le P_{\rm fix}+\delta$. 

Ideally, to study the partial phase we should find the maximum of the free energy as a function of $T$. If we fix $P$, then as $P$ in the partial phase decreases, $T$ increases slightly. We instead fixed $T$ near the confinement/deconfinement transition temperature $T_c$, and then dialed the value of $P$. In particular, we studied the distribution of the Polyakov line phases at $T=0.29$, $0.30$, and $0.31$, and did not see significant $T$-dependence in the distribution of Polyakov line phases.

As mentioned above, EK reduction requires both translational symmetry in spatial directions, and unbroken centre symmetry in spatial directions. As also mentioned above, a subtlety arise with EK reduction, in the partial phase. To see why, consider taking the infinite-volume limit of the spatial directions in our theory. In this limit, the confined and deconfined phases preserve spatial translational symmetry, hence they pose no problem for the EK reduction. However, the partial phase is a local maximum of the free energy, hence a state in which the partial phase uniformly fills all of space is unstable, even in the micro-canonical ensemble: such a state should in principle separate into regions of confined or deconfined phase, thus breaking spatial translational symmetry. Since we will use the single-site approximation, we will be blind to this effect, that is, we will study states with fixed energy that, in the continuum and infinite-volume limits, correspond to the partial phase uniformly filling all of space.

In the strong-coupling limit, centre symmetry in spatial directions is not spontaneously broken, hence EK reduction is valid. To confirm this, we numerically computed Wilson loops in spatial directions, and verified that they approach zero as $N$ increases. To be explicit, the action in eq.~\eqref{EK-action-lattice} is invariant under a global (i.e. $t$-independent) centre symmetry transformation in each spatial direction, which at large $N$ is a $U(1)$ transformation for each spatial direction. These leave $V_t$ untouched but act on $U_{\mu,t}$ as $U_{\mu,t}\to e^{i\alpha_\mu}U_{\mu,t}$ with each $\alpha_{\mu}$ an arbitrary real number. If centre symmetry in all spatial directions is unbroken, then the Wilson lines ${\rm Tr} \, U_{\mu,t}$ will be zero in all directions. For EK reduction to be valid, this must be the case at least at large $N$. We numerically calculated
\begin{eqnarray}
W\equiv
\frac{1}{3Nn_t}\sum_{\mu=1}^3\sum_{t=1}^{n_t}|{\rm Tr} U_{\mu,t}|,
\end{eqnarray}
for all the solutions we use below, and verified that $W$ approaches zero as $N$ increases.

\begin{figure}[htbp]
\begin{center}
\begin{subfigure}{.33\textwidth}
\scalebox{0.45}{
\includegraphics{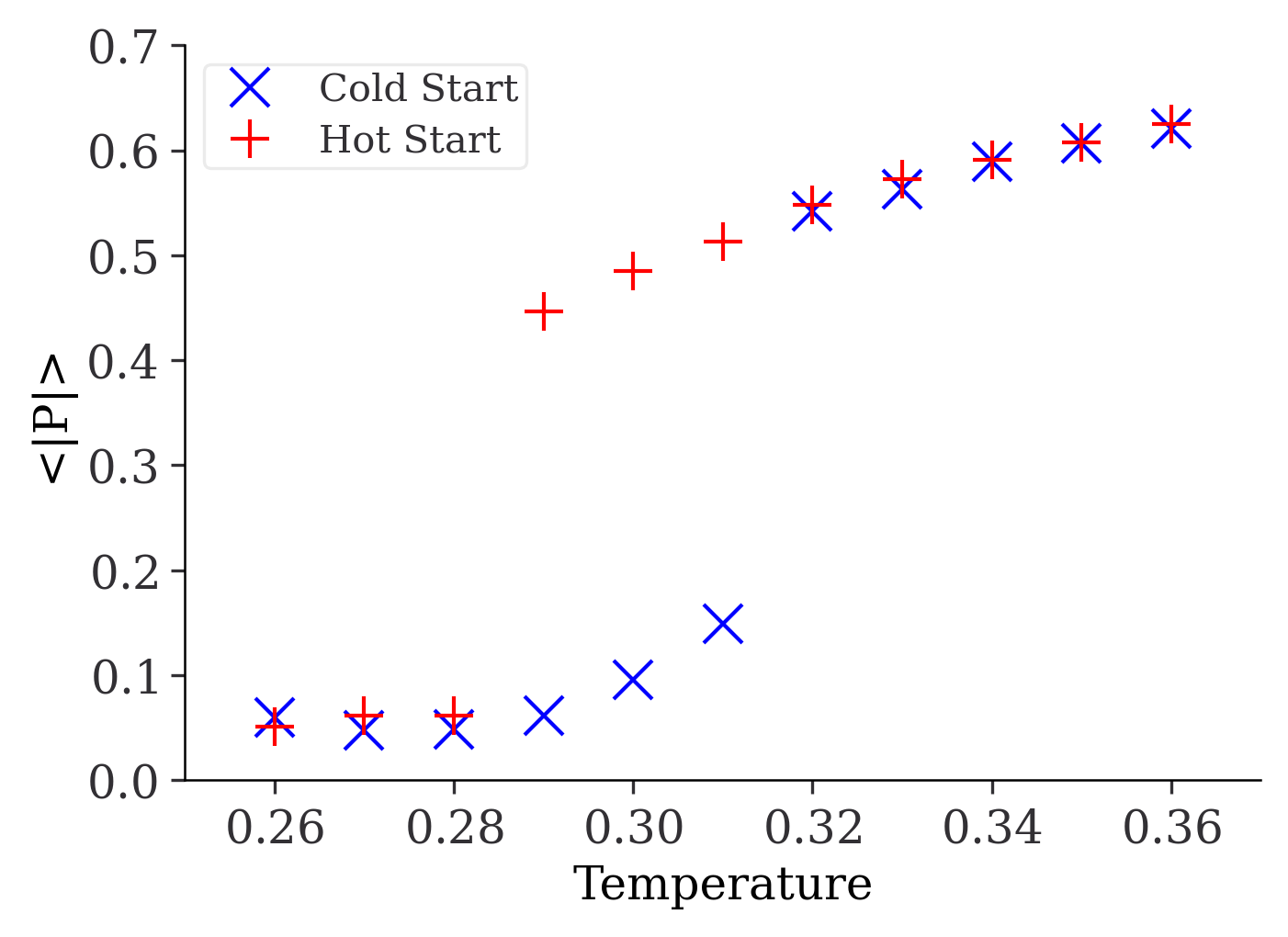}}
   \caption{}
   \end{subfigure}%
       \begin{subfigure}{.33\textwidth}
\scalebox{0.33}{
\includegraphics{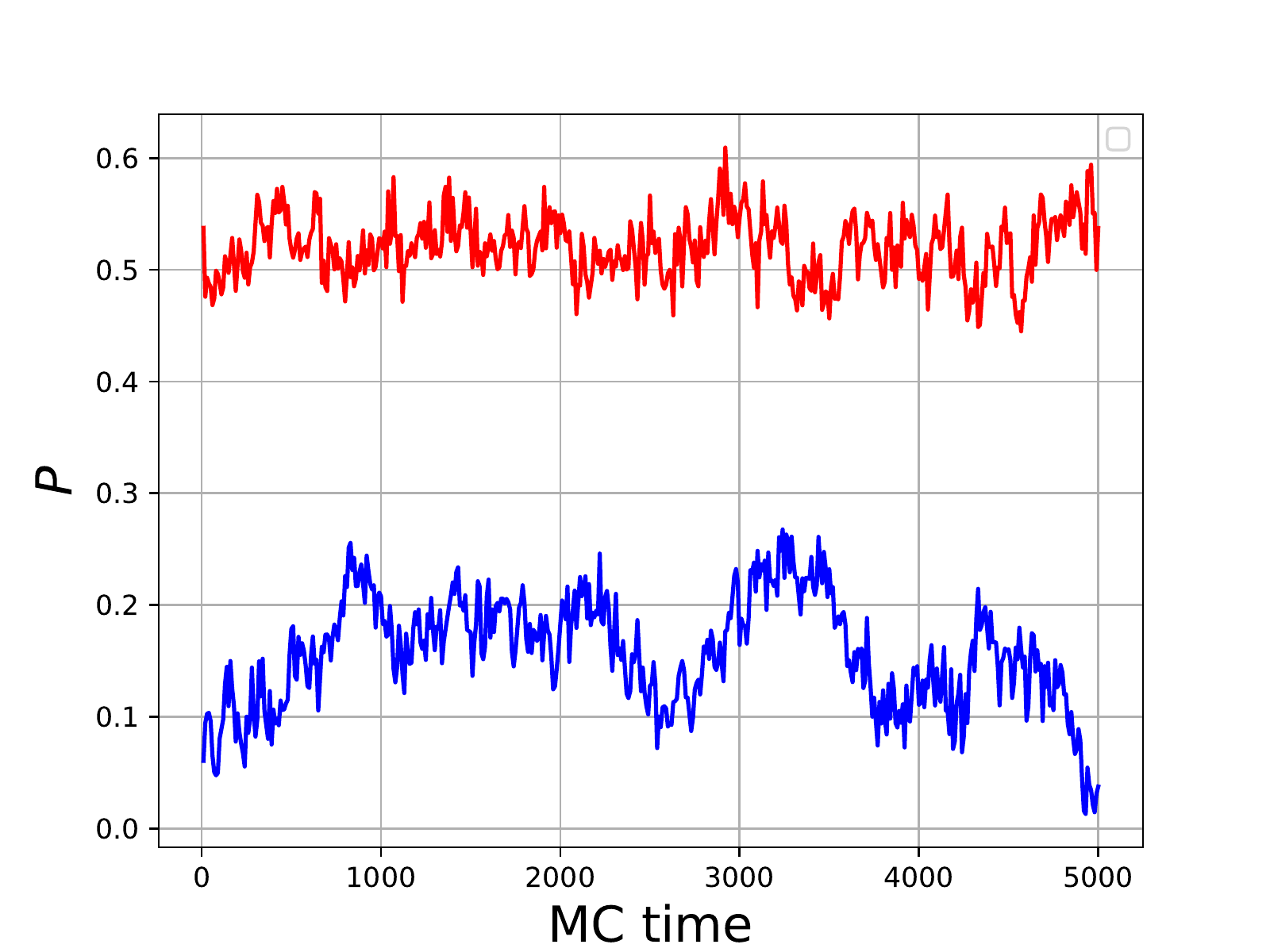}}
   \caption{}
   \end{subfigure}%
       \begin{subfigure}{.33\textwidth}
\scalebox{0.33}{
\includegraphics{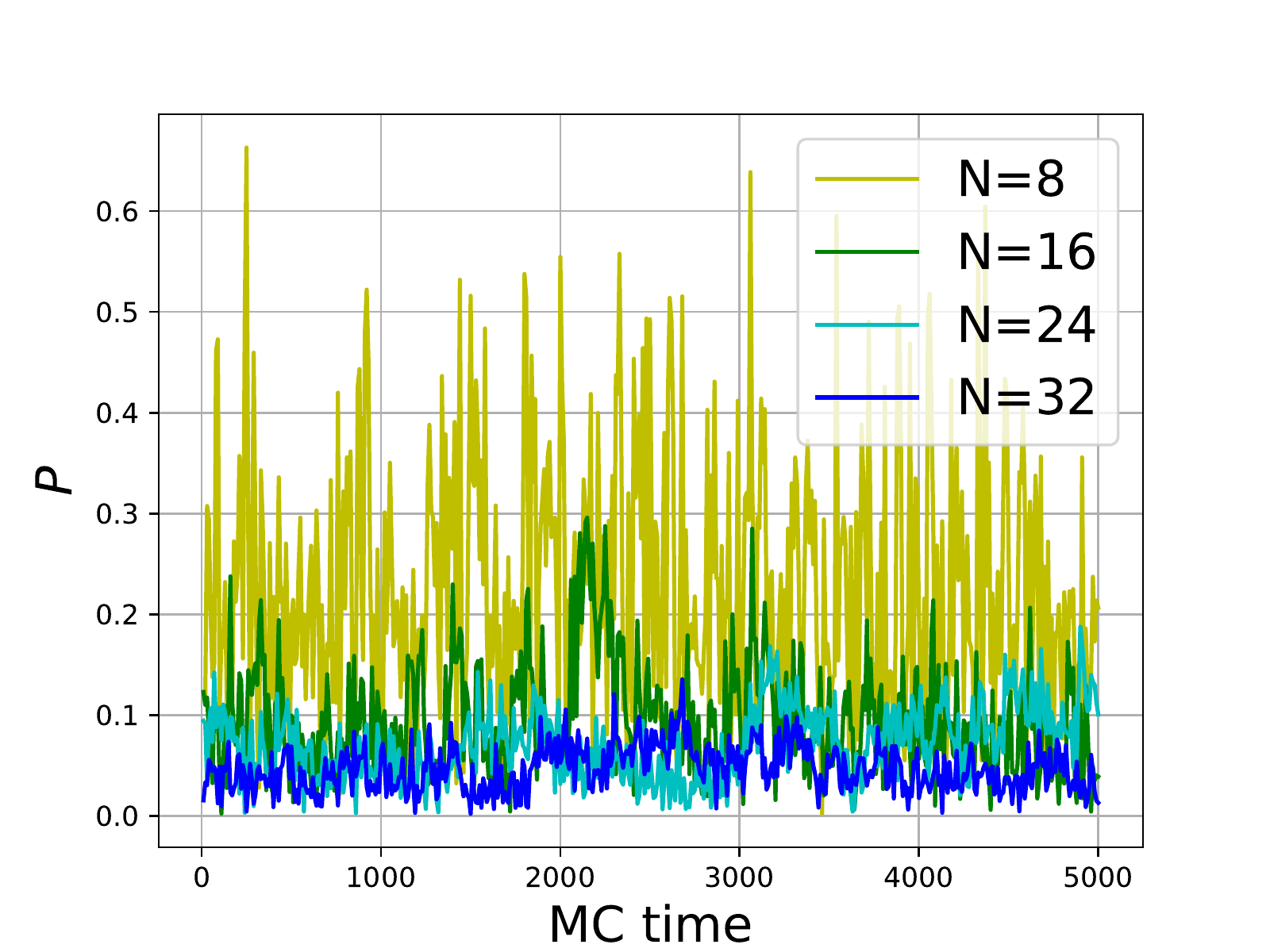}}
   \caption{}
   \end{subfigure}
\end{center}
\caption{Numerical results for the Polyakov loop, $P$, in the theory with action in eq.~\eqref{EK-action-lattice}. As discussed below eq.~\eqref{eq:latticePdef}, our $P$ is always real and non-negative. (a) Mean Polyakov loop versus temperature $T$, for $N=32$ and $n_t=24$. We observe strong hysteresis around $T_c\simeq 0.31$, consistent with a first-order transition. The cold start (blue crosses) and hot start (red plus signs) simulations had initial temperatures $T=0.2$ and $T=0.45$, respectively. The simulations were then thermalised before 200 configurations taken for data collection. (b) The simulation history of $P$ for $N=32$ and $n_t=24$ at $T=0.31$, for a cold start (blue) and hot start (red). The horizontal axis is the Monte  Carlo time. We observe that tunneling between the two phases is strongly suppressed. The fluctuations are larger in the confined phase because $T=0.31$ is close to the endpoint of the confined phase (the Hagedorn temperature). (c) The simulation history of $P$ for $N=8$, $16$, $24$ and $32$, with $n_t=24$, at $T=0.25$. As $N$ becomes larger, $P$ approaches zero. 
}
\label{fig:N16L8-two-state-signal}
\end{figure}

We now present our results for the confinement/deconfinement transition in this theory. As shown in ref.~\cite{Hanada:2014noa}, for this theory with arbitrary spatial dimension $d$, simple state counting reveals a first-order phase transition around $T_c=\frac{1}{2\log(2d-1)}$, which in our $d=3$ case is $T_c=\frac{1}{2\log(5)}\simeq 0.31$. We easily confirmed this numerically. Fig.~\ref{fig:N16L8-two-state-signal} shows our numerical results for the Polyakov loop, $P$. In particular, Fig.~\ref{fig:N16L8-two-state-signal} (a) shows our numerical results for $P$ as a function of $T$, for $N=32$ and $n_t=24$. We observe a strong hysteresis, consistent with a first-order transition at $T_c=\simeq 0.31$. Fig.~\ref{fig:N16L8-two-state-signal} (b) shows the simulation history of $P$ at $T=0.31$, for $N=32$ and $n_t=24$. We see that the confined and deconfined phase both exist rather stably, which also indicates a first-order phase transition. Fig.~\ref{fig:N16L8-two-state-signal} (c) shows $P$ at $T=0.25$, in the confined phase, for $n_t=24$ and $N=8$, $16$, $24$ and $32$. We see that $P$ approaches zero as $N$ increases, as expected for the confined phase.

\begin{figure}[htbp]
\begin{center}
\begin{subfigure}{.5\textwidth}
\scalebox{0.27}{
\includegraphics{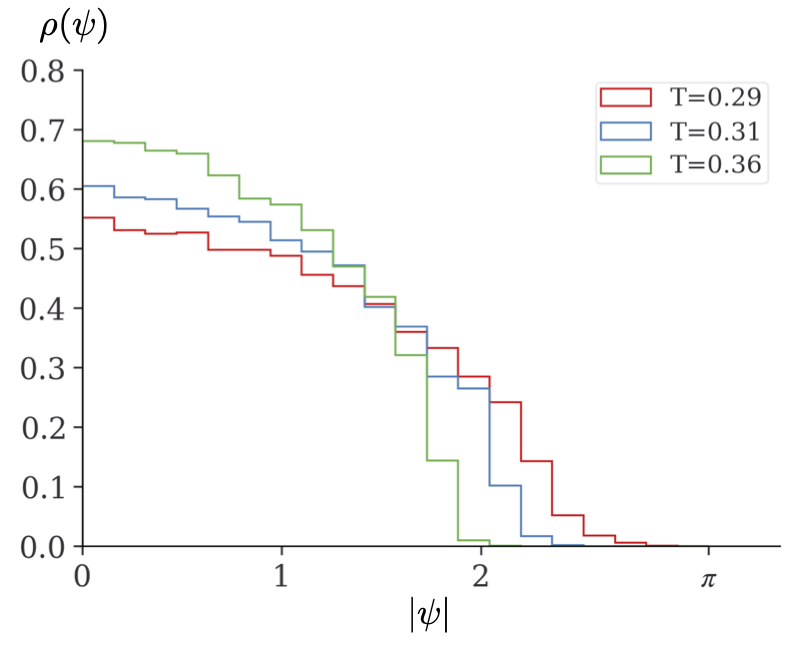}}
   \caption{}
   \end{subfigure}%
\begin{subfigure}{.5\textwidth}
\scalebox{0.24}{
\includegraphics{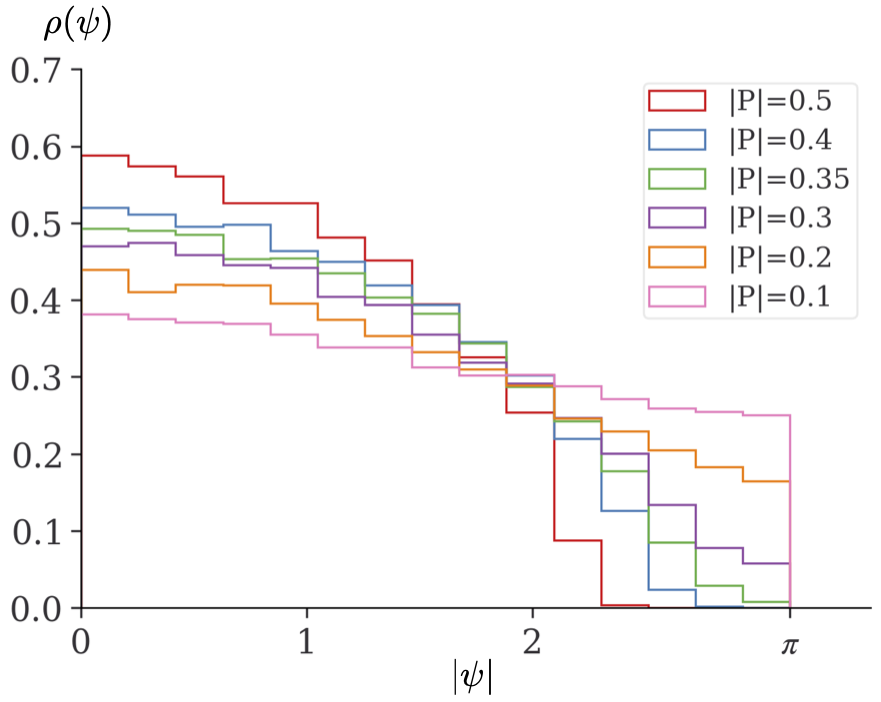}}
   \caption{}
   \end{subfigure}
\end{center}
\caption{Numerical results for the distribution of Polyakov loop phases, $\rho(\psi)$, as a function of $|\psi|$, in the theory with action in eq.~\eqref{EK-action-lattice}, for $N = 32$ and $n_t=24$. We use $|\psi|$ because, if the number of samples is sufficiently large, $\rho(\psi)$ is symmetric about $\psi = 0$. (a) The deconfined phase, as $T$ decreases through the confinement/deconfinement transition temperature $T_c\simeq 0.31$, where the free energies of confined and deconfined phases coincide. We see a gap at $\psi=\pi$ that shrinks as $T$ decreases, as expected. (b) Fixed-$P$ simulations at the lowest $T$ in the deconfined phase, $T=0.29$, i.e. near the GWW transition. Specifically, we fixed $P$ with $\delta = 5 \times 10^{-3}$ in eq.~\eqref{eq:deltaS}, with 400 configurations for each $P$. We see that as $P$ decreases, the gap closes at $P\simeq 0.35$, so we identify $P\leq 0.35$ as the partial phase.
}
\label{fig:Pol-phase-GWW}
\end{figure}

Fig.~\ref{fig:Pol-phase-GWW} shows our numerical results for the distribution of Polyakov line phases, $\rho(\psi)$, versus $\psi$, for $N=32$ and $n_t=24$. Fig.~\ref{fig:Pol-phase-GWW} (a) shows our results for $\rho(\psi)$ in the deconfined phase, as we lower $T$ through $T_c \simeq 0.31$ and towards the GWW transition. We observe a gap that shrinks and approaches zero as $T$ approaches the GWW transition, as expected. We also observed that in the confined phase $\rho(\psi)$ is uniform, up to $1/N$-corrections. Fig.~\ref{fig:Pol-phase-GWW} (b) shows the evolution of $\rho(\psi)$ in fixed-$P$ simulation at $T=0.29$, the lowest $T$ of the deconfined phase, i.e. near the GWW transition. As $P$ decreases, we see that the gap closes at $P\simeq 0.35$, and for $P<0.35$ the gap is gone, but $\rho(\psi)$ is not uniform. We thus identify these $P\leq0.35$ configurations as partial phase solutions.

All of our results above are consistent with a first-order confinement/deconfinement transition in which the partial phase appears as the unstable branch connecting the confined and deconfined branches, similar to what we saw for softly-broken $\N=1$ SYM in sec.~\ref{sec:4dSYM}.

\subsection{Chiral symmetry}
\label{sec:chiral-symmetry}
\hspace{0.51cm}
We now introduce a massless probe quark and study chiral symmetry breaking. We use the na\"ive lattice fermion, with action\footnote{Notice we have no reason to worry about doublers for our current purposes.}
\begin{eqnarray}
S_f
= \frac{1}{2}
\sum_{t=1}^{n_t}
\left[
\frac{1}{a}\left(\bar{\psi}_t \gamma^t V_t\psi_{t+1}-\bar{\psi}_{t+1} \gamma^t V_t^\dagger\psi_t \right)
+
\sum_\mu
\left(\bar{\psi}_t \gamma^\mu U_\mu\psi_{t}-\bar{\psi}_{t} \gamma^\mu U_\mu^\dagger\psi_t\right)
\right]. 
\end{eqnarray}
Let us write this action as $S_f = \bar{\psi}D\psi$, where $D$ is the Dirac operator. The eigenvalues of $D$ are purely imaginary, with two-fold degeneracy. Specifically, they have the form $\pm i \lambda$, where $\lambda$ is real and non-negative. We want to compute the distribution of eigenvalues of the Dirac operator, which we denote $\rho^{({\rm D})}(\lambda)$. We normalise this density as $\int_0^\infty d\lambda\rho^{({\rm D})}(\lambda)=4Nn_t$.

As mentioned above, fields in the fundamental representation explicitly break the centre symmetry, but we suppress this effect by working in the probe limit, i.e. working in the 't Hooft large-$N$ limit with $N_f$ fixed, and keeping the leading order in $N_f/N\ll 1$. In that case, roughly speaking the quark is influenced by the pure glue sector, but the pure glue sector is unaffected by the quark, i.e. the quark's ``back-reaction'' on the pure glue sector is suppressed. As a result, introducing the probe quark does not affect the results of sec.~\ref{sec:conf/deconf}.

Our quarks are massless, hence the quark sector classically has $U(N_f) = U(1) \times SU(N_f)$ chiral symmetry. When $N$ is finite the $U(1)$ factor is anomalous, however the one-loop diagram that produces this anomaly is suppressed in the probe limit. We expect the chiral symmetry to be spontaneously broken in the confined phase and restored in the deconfined phase. The corresponding order parameter is the chiral condensate, $\langle \bar{\psi}\psi\rangle$, which we expect to be non-zero in the confined phase and zero in the deconfined phase.

As discussed in ref.~\cite{Hanada:2019kue}, anomaly-matching arguments suggest that chiral symmetry is spontaneously broken also in the partial phase, and specifically in its confined subsector. Our goal here is to compute $\langle \bar{\psi}\psi\rangle$ in the partial phase solutions of sec.~\ref{sec:conf/deconf}, to determine whether chiral symmetry is indeed spontaneously broken. To do so, we will numerically compute $\rho^{({\rm D})}(\lambda)$, and use the Banks-Casher relation~\cite{Banks:1979yr}, which states that $\rho^{({\rm D})}(0)\propto \langle \bar{\psi}\psi\rangle$.

Computing $D$'s eigenvalues, and hence $\rho^{({\rm D})}(\lambda)$, is the most computationally-demanding part of our numerical calculations for this theory. We calculated $D$'s eigenvalues using the linear algebra package LAPACK.

When the centre symmetry breaks spontaneously, the eigenvalues of $D$ are sensitive to the complex phase of the Polyakov loop. Because we want to interpret our probe quarks as an approximation to dynamical (i.e. non-probe) quarks, we have no reason to pick a specific phase. Instead, we need to perform a centre-symmetry transformation $V_t\to e^{i\alpha}V_t$ and average over $\alpha$. In our simulation, for each configuration we chose ten random values of $\alpha$ from $[0,2\pi)$, and used $e^{i\alpha}V_t$ instead of $V_t$ to calculate $D$'s eigenvalues.\footnote{Because configurations related by the U(1) centre symmetry transformation have the same path-integral weight, such an averaging is, in principle, automatic. In our simulation, because we used the same step size for the $U(1)$ and $SU(N)$ directions, the auto-correlation along the $U(1)$ direction is large. We therefore used such a trick to average over the $U(1)$ direction with a reasonable number of configurations.}

\begin{figure}[htbp]
\begin{center}
\begin{subfigure}{.33\textwidth}
\scalebox{0.3}{
\includegraphics{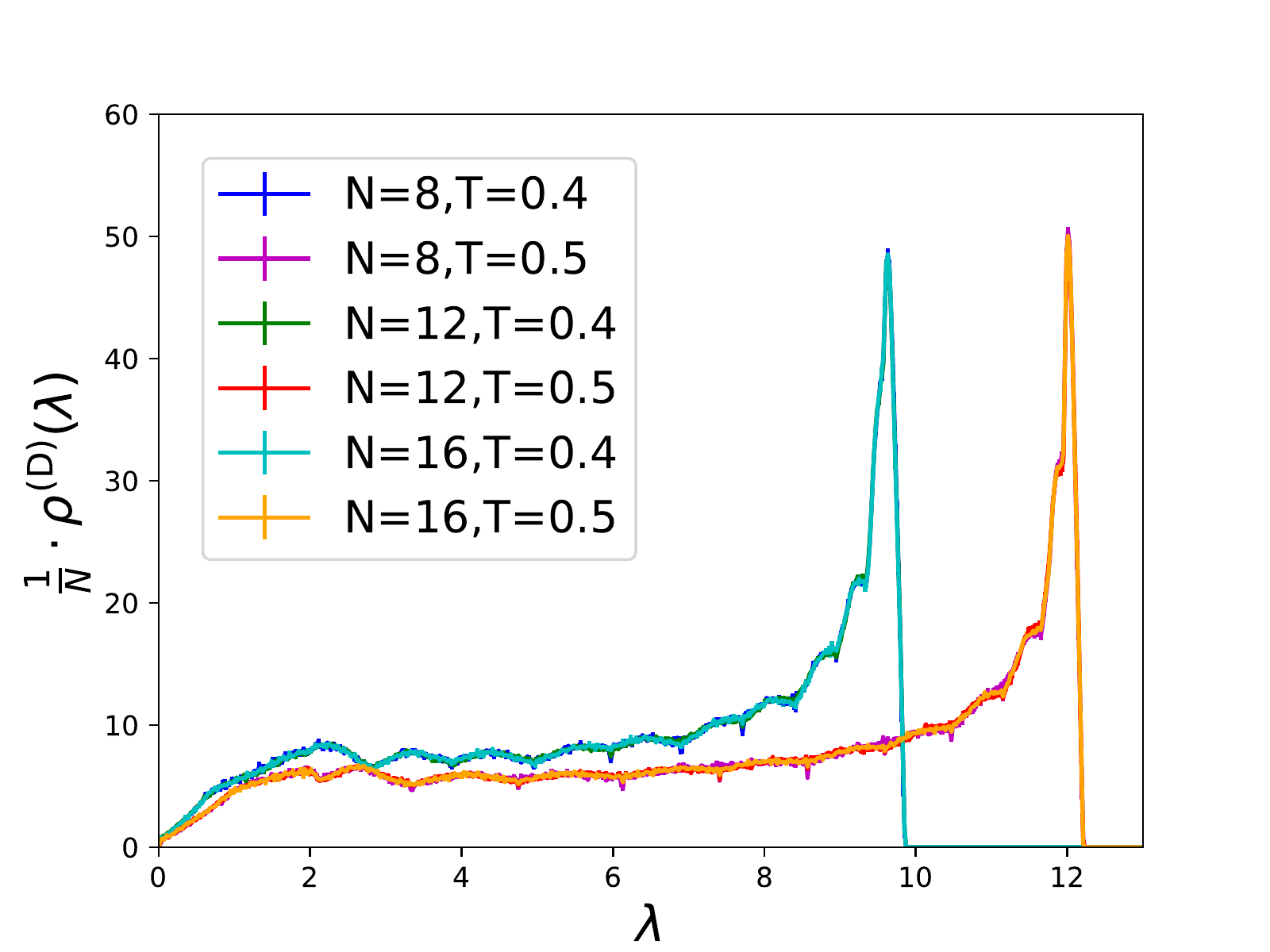}}
   \caption{}
   \end{subfigure}%
\begin{subfigure}{.33\textwidth}
\scalebox{0.3}{
\includegraphics{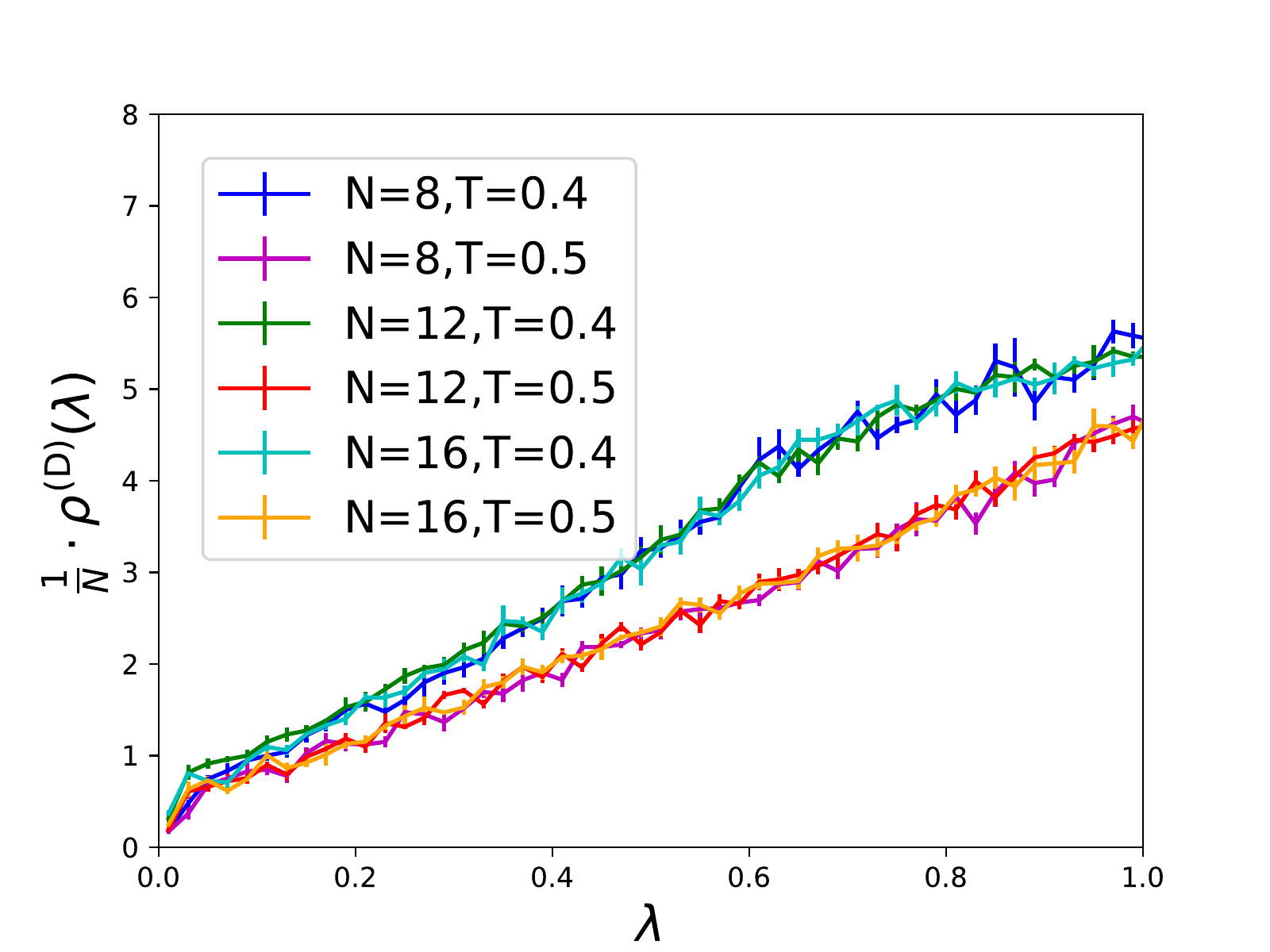}}
   \caption{}
   \end{subfigure}%
   \begin{subfigure}{.33\textwidth}
\scalebox{0.3}{
\includegraphics{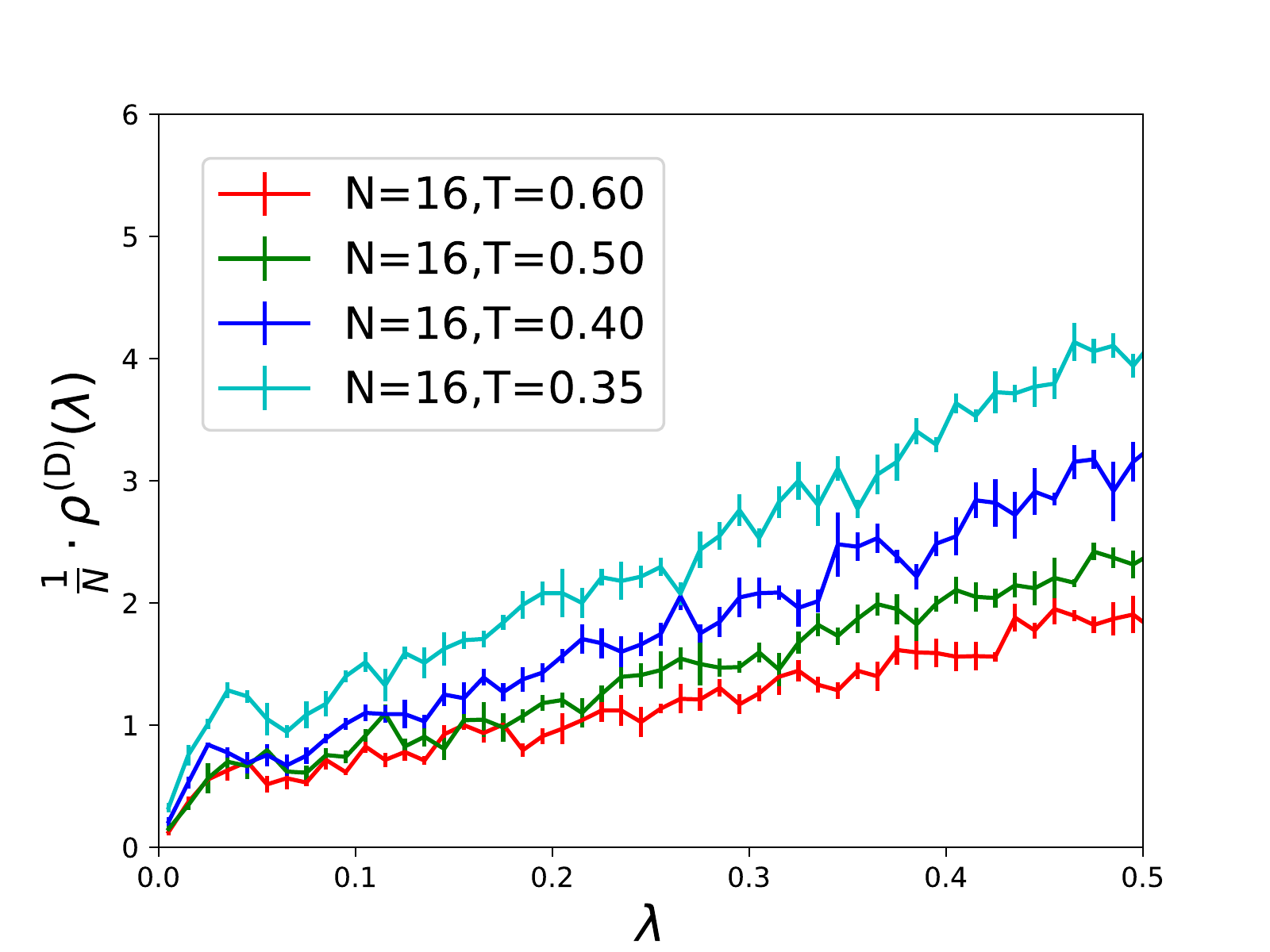}}
   \caption{}
   \end{subfigure}
\end{center}
\caption{The density of Dirac eigenvalues, $\rho^{\rm (D)}(\lambda)$, divided by $N$, for $N=8$, $12$, and $16$, $n_t=24$, and $T=0.4$ and $0.5$, i.e. in the deconfined phase, with 500 configurations (5000 sets of eigenvalues) for each. Error bars are estimated by using 5-bin Jackknife with bin width $0.02$. We see in all cases that $\rho^{\rm (D)}(\lambda)\to0$ as $\lambda\to0$, indicating $\langle \bar{\psi}\psi\rangle=0$ and hence that chiral symmetry is preserved, as expected. (b) Close-up of (a), showing that the slope near $\lambda=0$ increases as $T$ increases. (c) The same, but for $T=0.35$, $0.40$, $0.50$, and $0.60$, showing again that the slope near $\lambda=0$ increases as $T$ increases.
}\label{fig:Dirac_highT}
\end{figure}

\begin{figure}[htbp]
\begin{center}
\begin{subfigure}{.5\textwidth}
\scalebox{0.45}{
\includegraphics{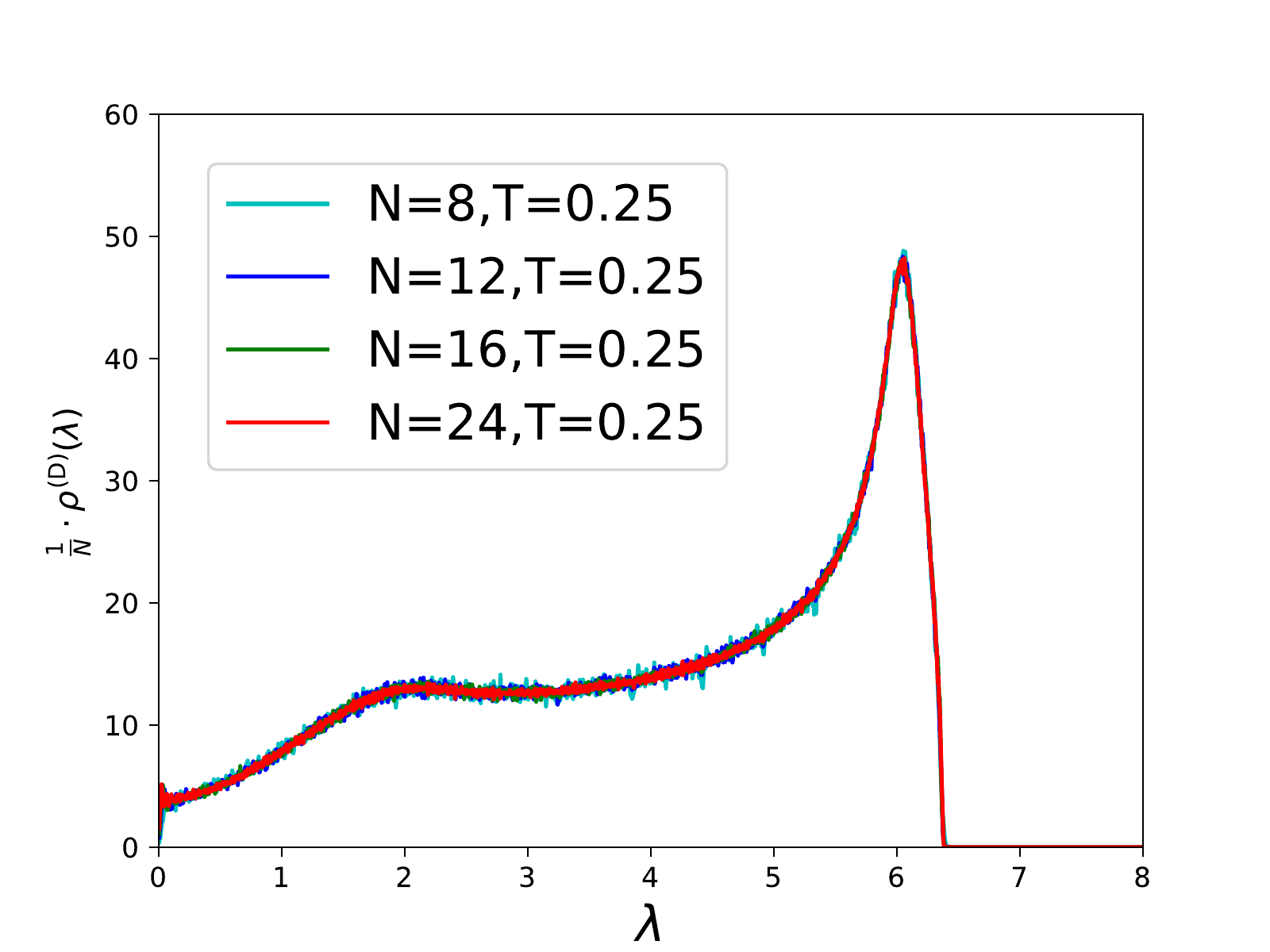}}
   \caption{}
   \end{subfigure}%
   \begin{subfigure}{.5\textwidth}
\scalebox{0.45}{
\includegraphics{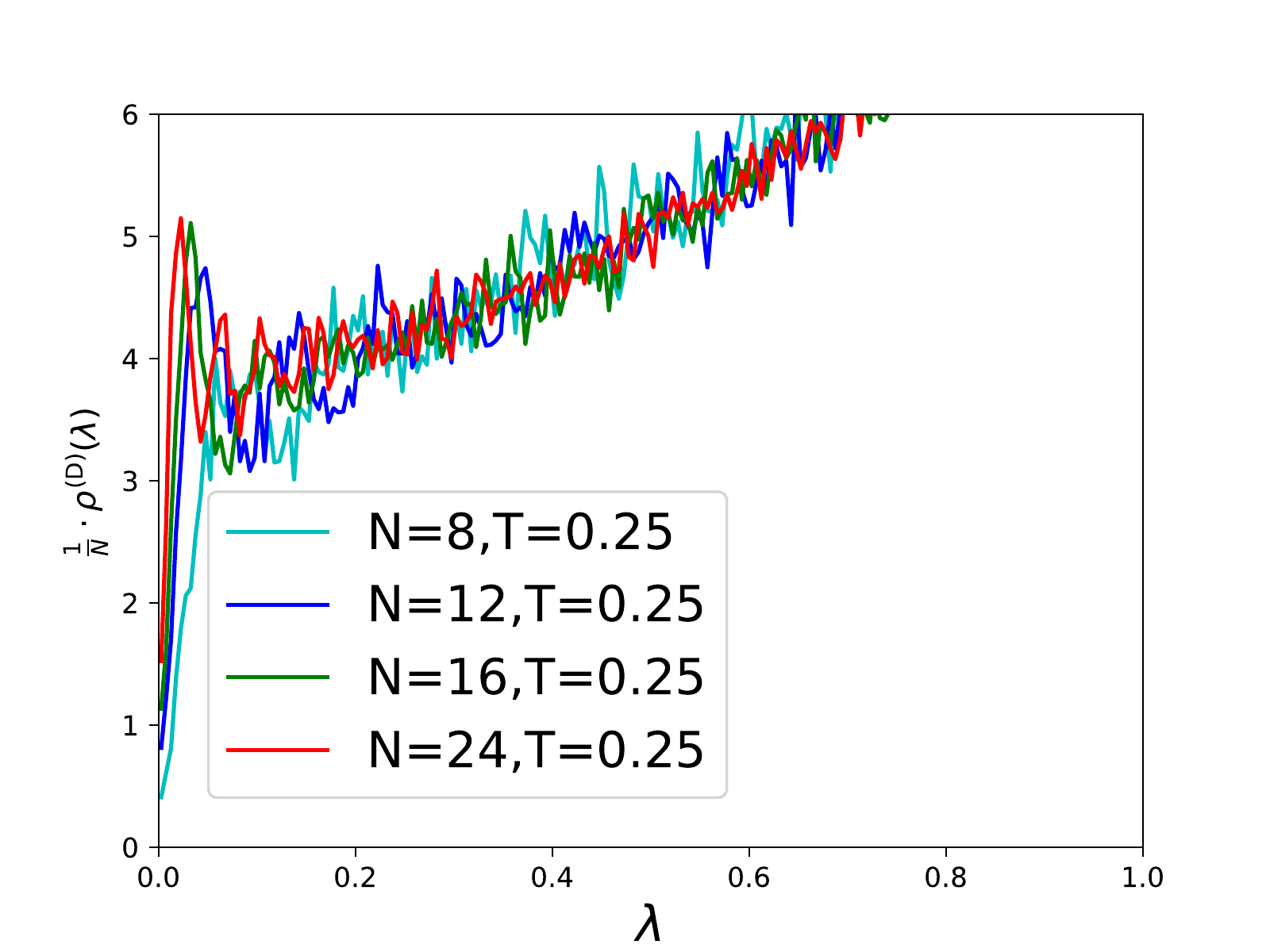}}
   \caption{}
   \end{subfigure}
\end{center}
\caption{(a) The density of Dirac eigenvalues, $\rho^{\rm (D)}(\lambda)$, divided by $N$, for $N=8$, $12$, $16$, and $24$, $n_t=24$, and $T=0.25$, i.e. in the confined phase. We see in all cases that $\rho^{\rm (D)}(\lambda)$ approaches a non-zero value as $\lambda\to0$, indicating $\langle \bar{\psi}\psi\rangle\neq 0$ and hence that chiral symmetry is spontaneously broken, as expected. (b) Close-up of (a).
}\label{fig:Dirac_lowT}
\end{figure}

\begin{figure}[htbp]
\begin{center}
\begin{subfigure}{.5\textwidth}
\scalebox{0.45}{
\includegraphics{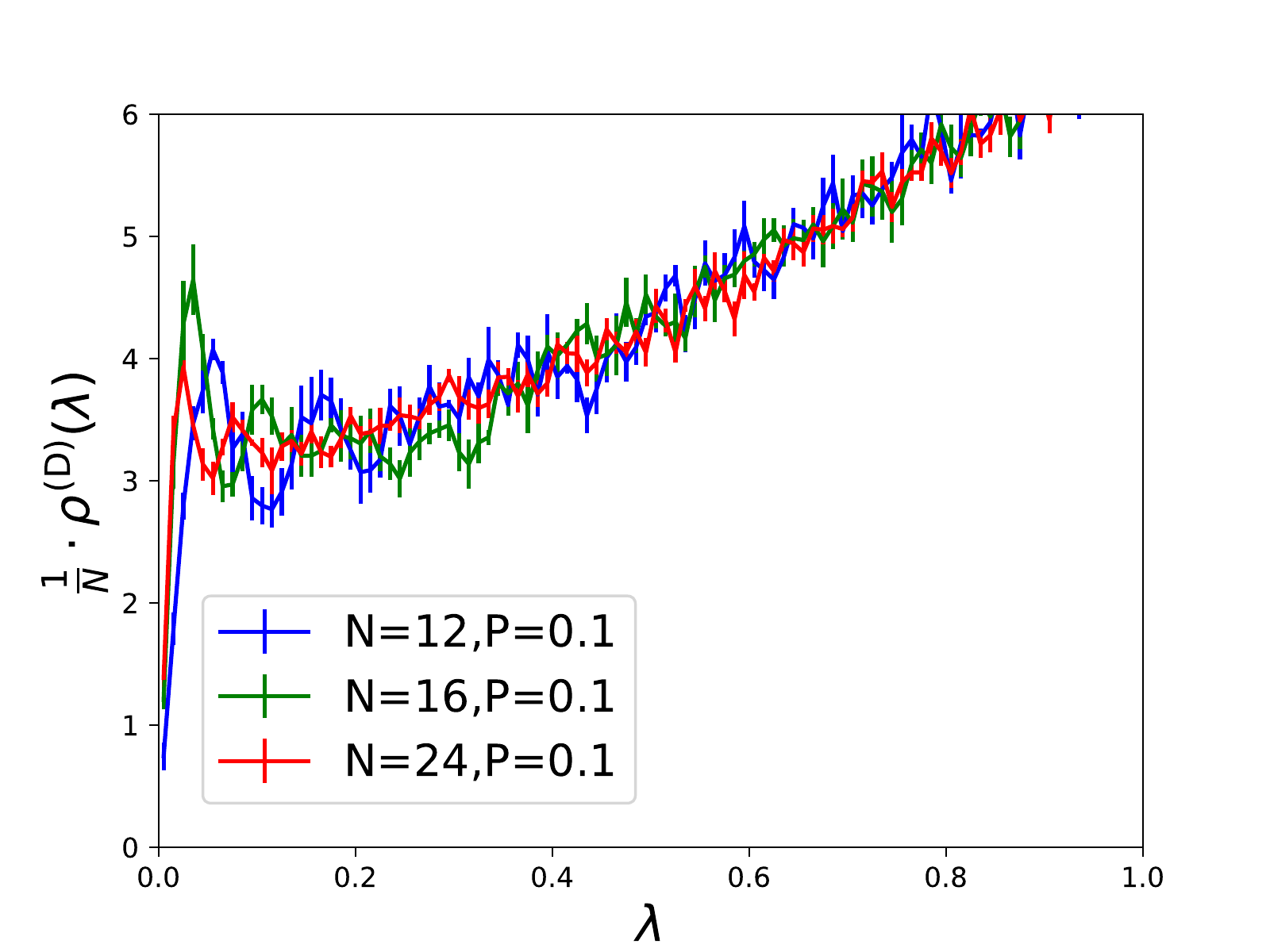}}
   \caption{}
   \end{subfigure}%
\begin{subfigure}{.5\textwidth}
\scalebox{0.45}{
\includegraphics{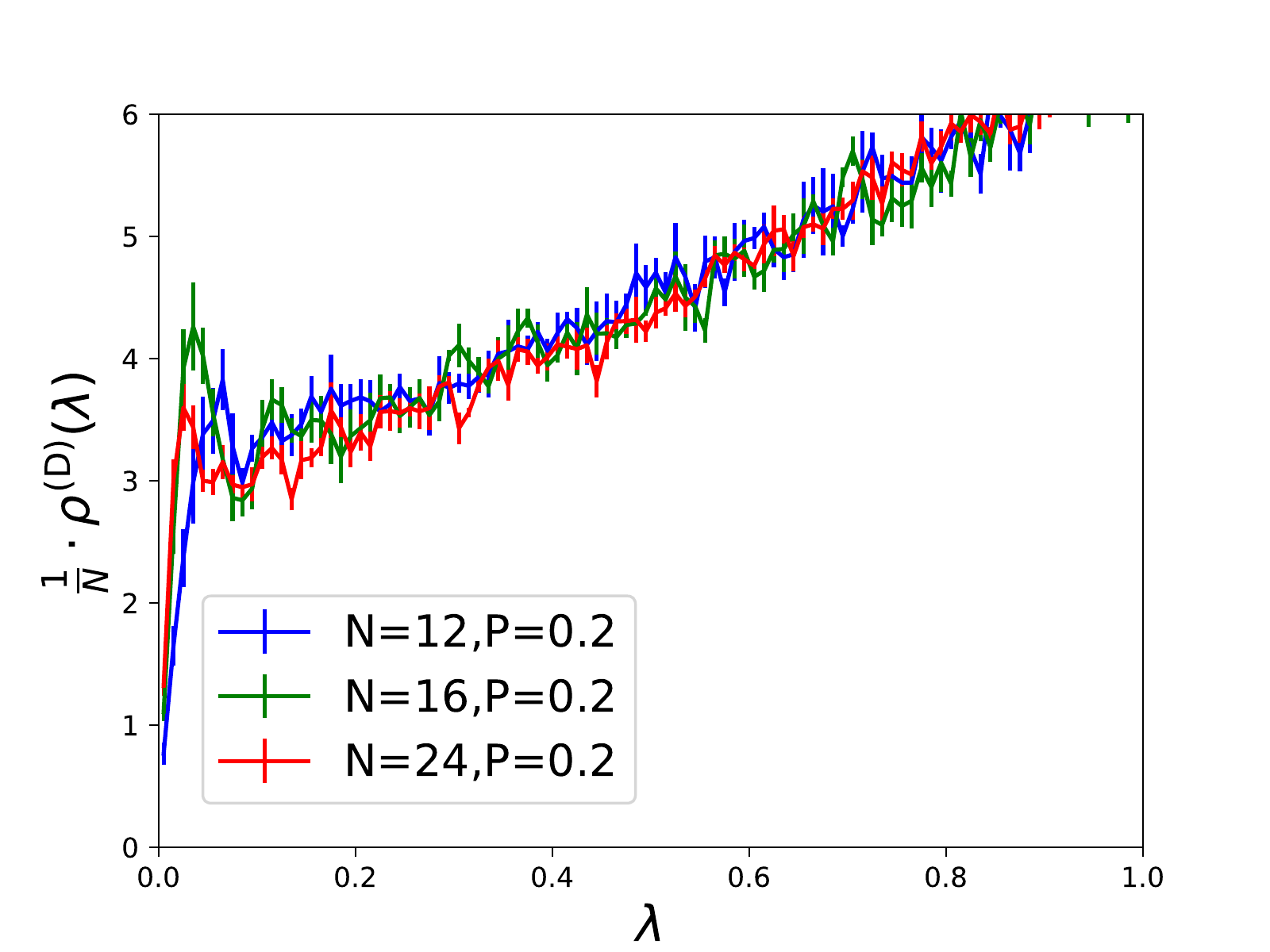}}
   \caption{}
   \end{subfigure}
\begin{subfigure}{.5\textwidth}
\scalebox{0.45}{
\includegraphics{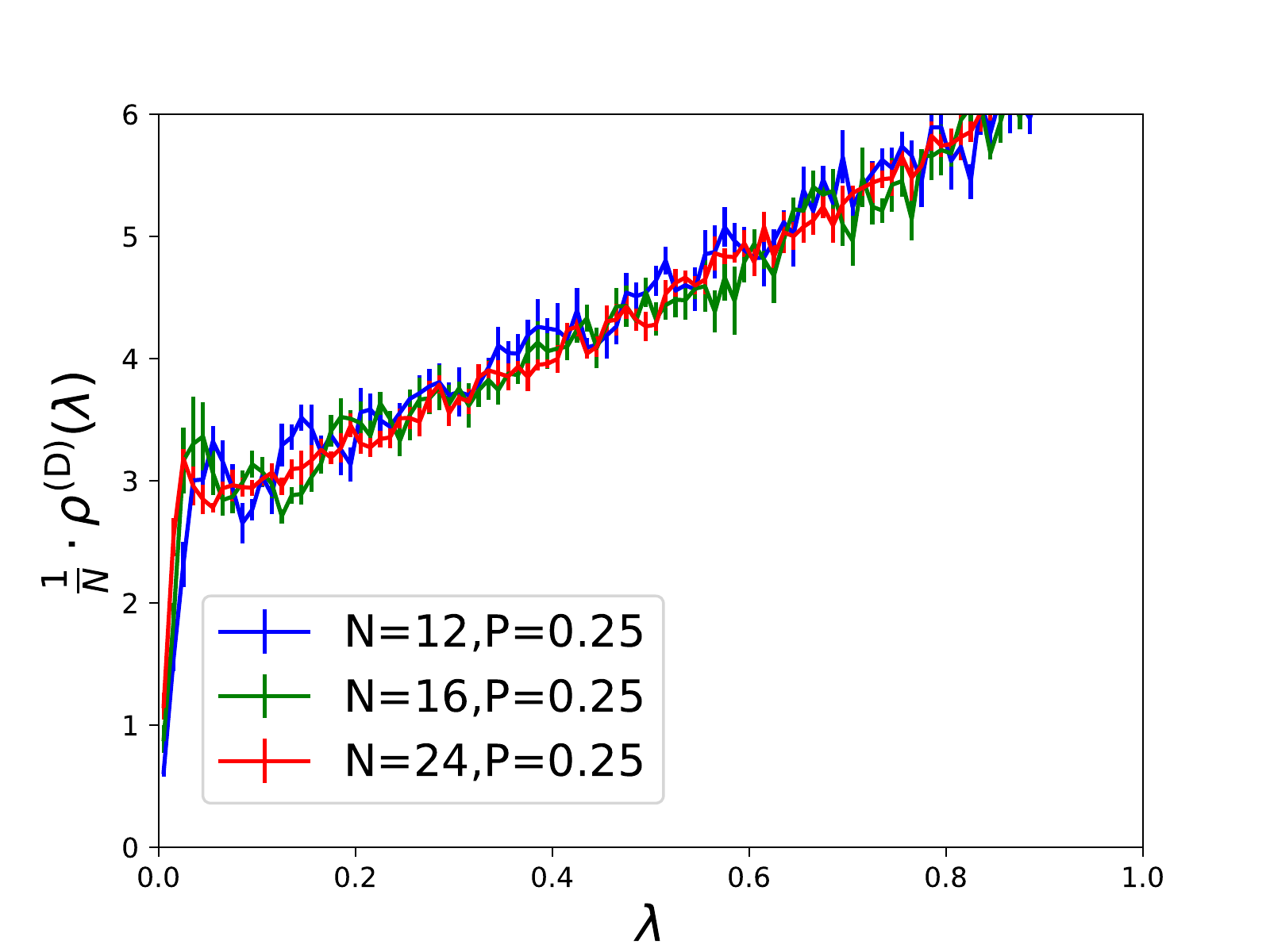}}
   \caption{}
   \end{subfigure}%
\begin{subfigure}{.5\textwidth}
\scalebox{0.45}{
\includegraphics{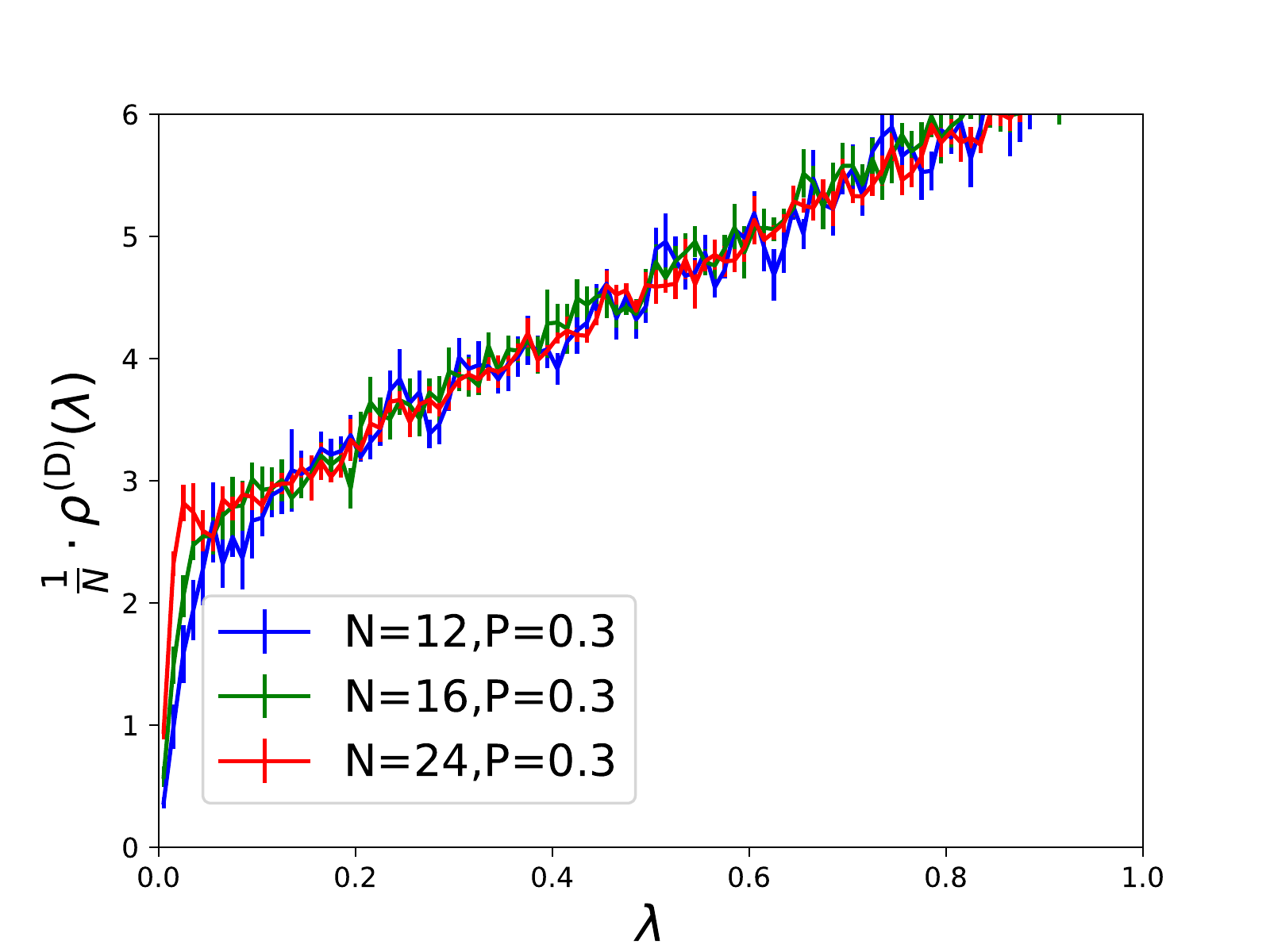}}
   \caption{}
   \end{subfigure}
\begin{subfigure}{.5\textwidth}
\scalebox{0.45}{
\includegraphics{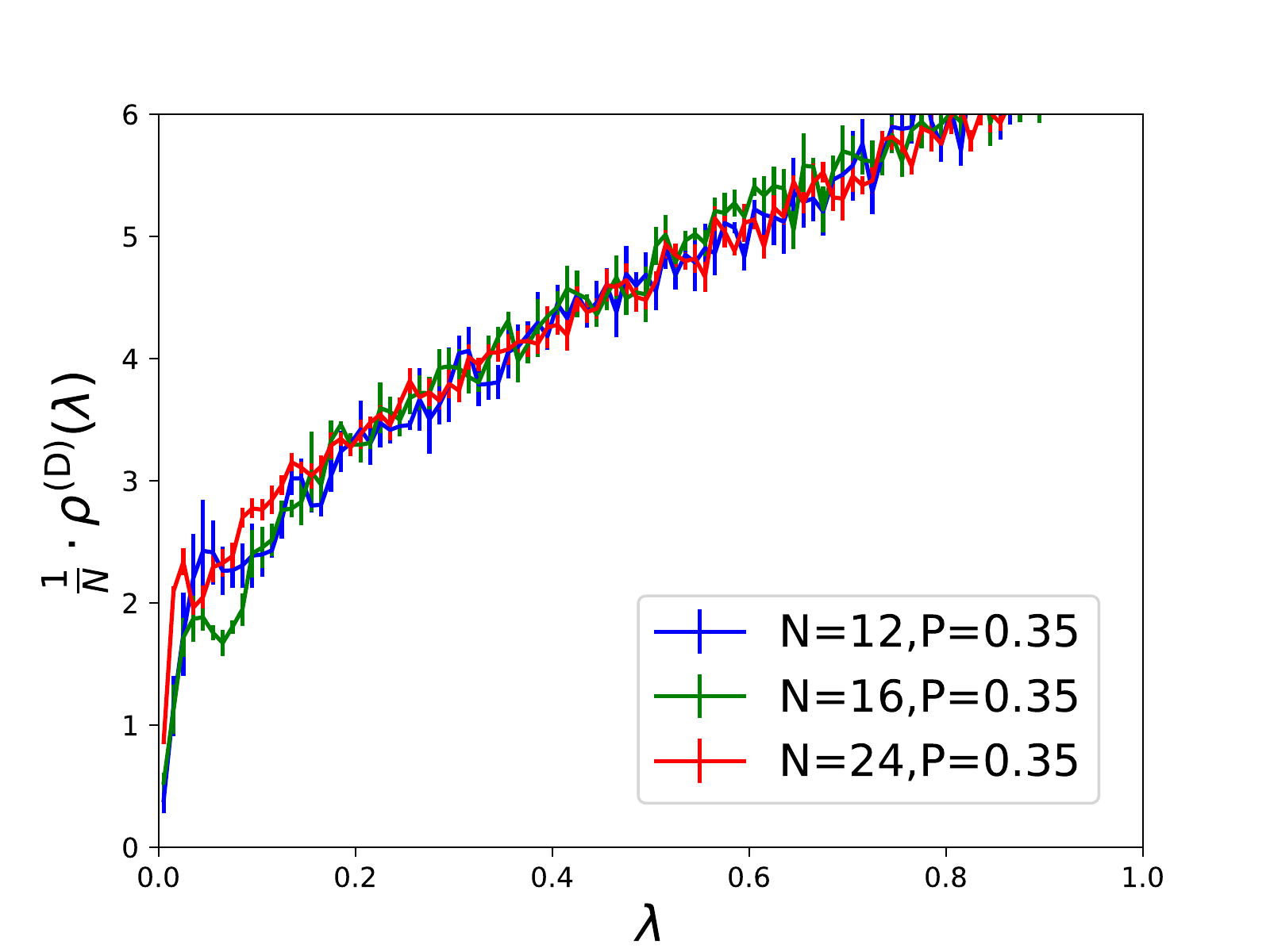}}
   \caption{}
   \end{subfigure}%
\begin{subfigure}{.5\textwidth}
\scalebox{0.45}{
\includegraphics{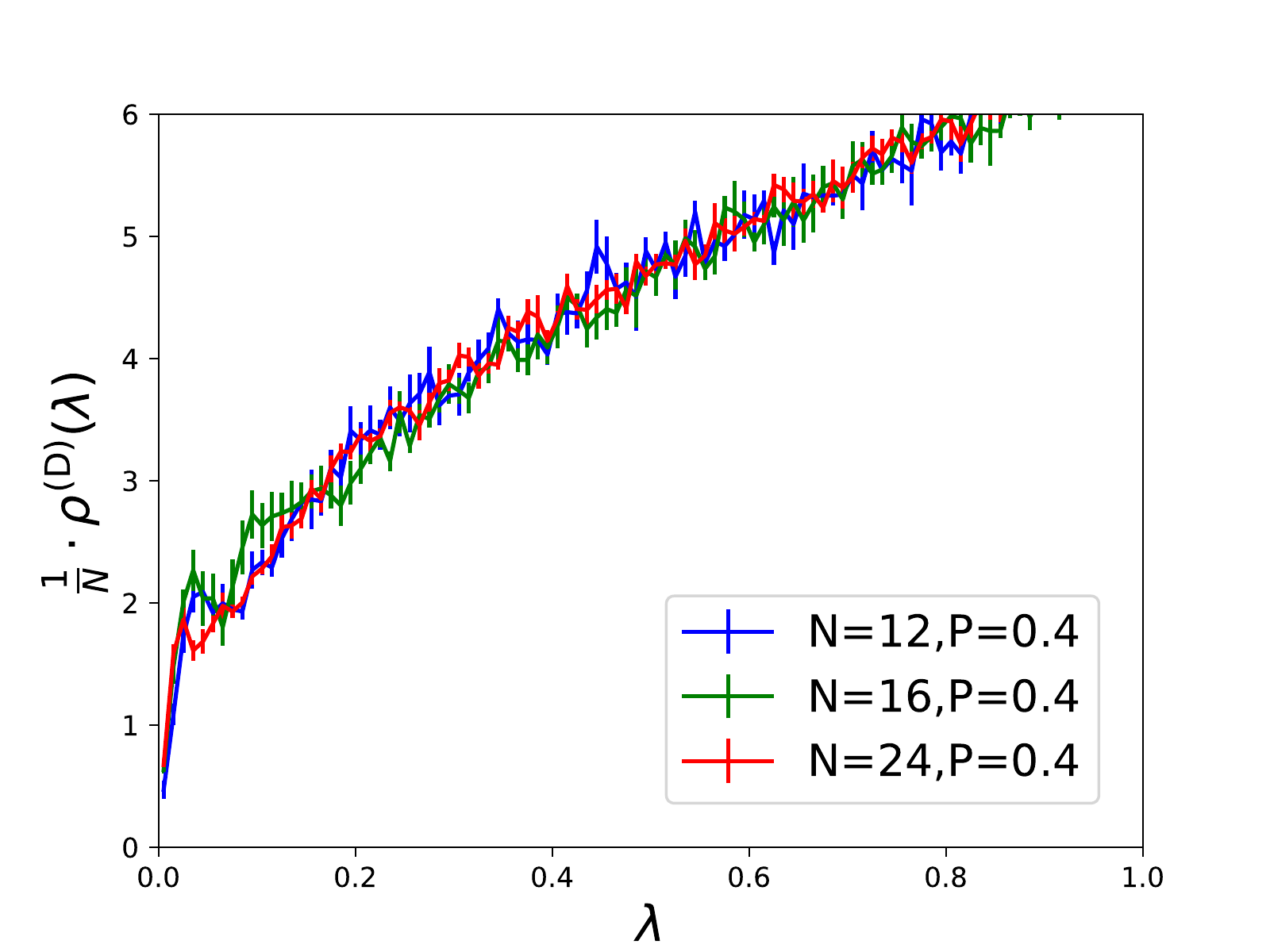}}
   \caption{}
   \end{subfigure}
\end{center}
\caption{The density of Dirac eigenvalues, $\rho^{\rm (D)}(\lambda)$, divided by $N$, for $N=12$, $16$, and $24$, $n_t=24$, and $T=0.29$, for various fixed values of $P$: (a)-(d) have $P=0.1$, $0.2$, $0.25$, $0.3$, respectively, which are all in the partial phase, while (e) and (f) have $P=0.35$ and $0.4$, respectively, which are in the deconfined phase. In the partial phase, the behaviour is qualitatively similar to the confined phase in fig.~\ref{fig:Dirac_lowT}.
}\label{fig:Dirac_N8N12N16-T029}
\end{figure}

\begin{figure}[htbp]
\begin{center}
\begin{subfigure}{.5\textwidth}
\scalebox{0.45}{
\includegraphics{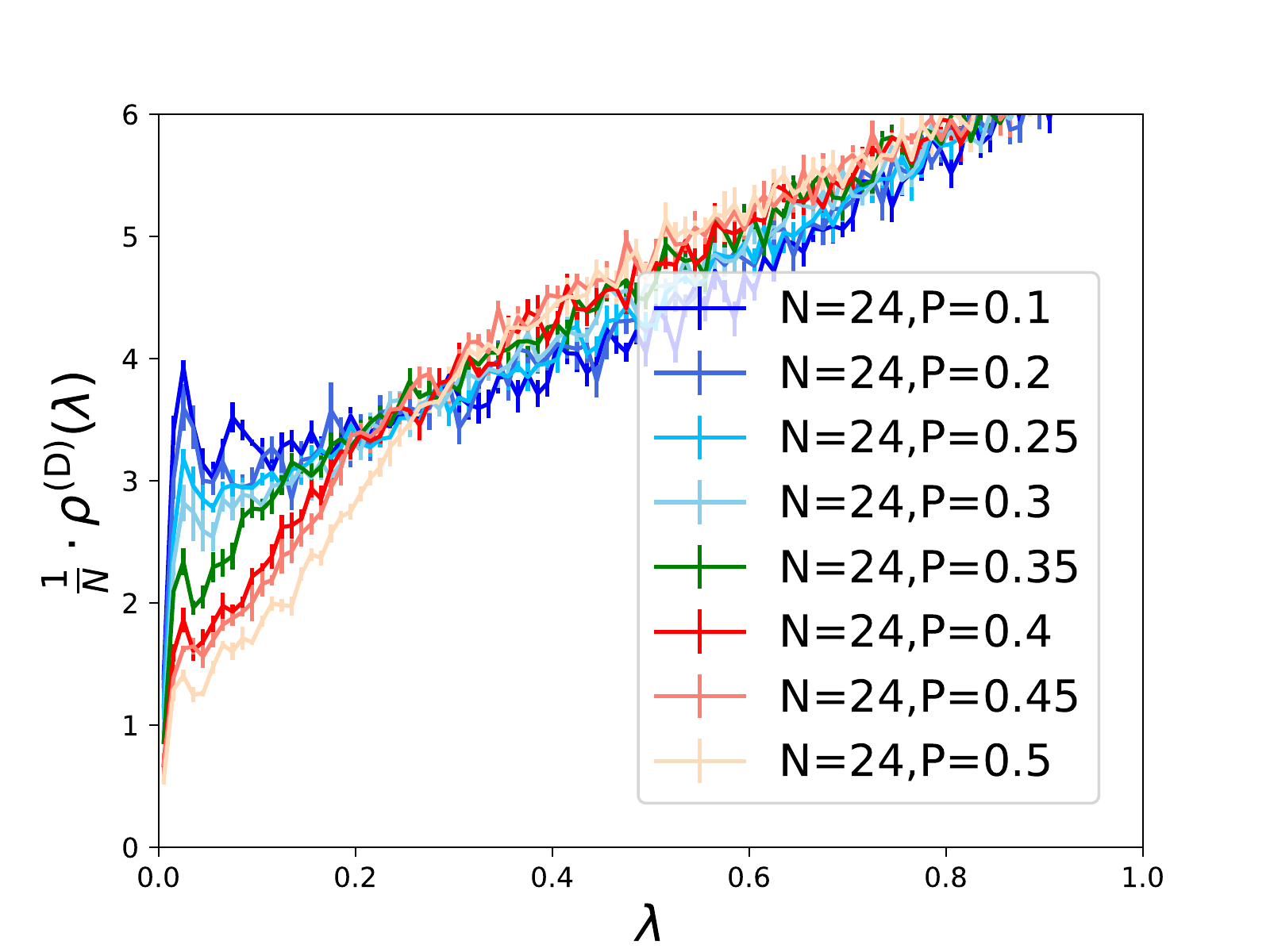}}
   \caption{}
   \end{subfigure}%
   \begin{subfigure}{.5\textwidth}
\scalebox{0.45}{
\includegraphics{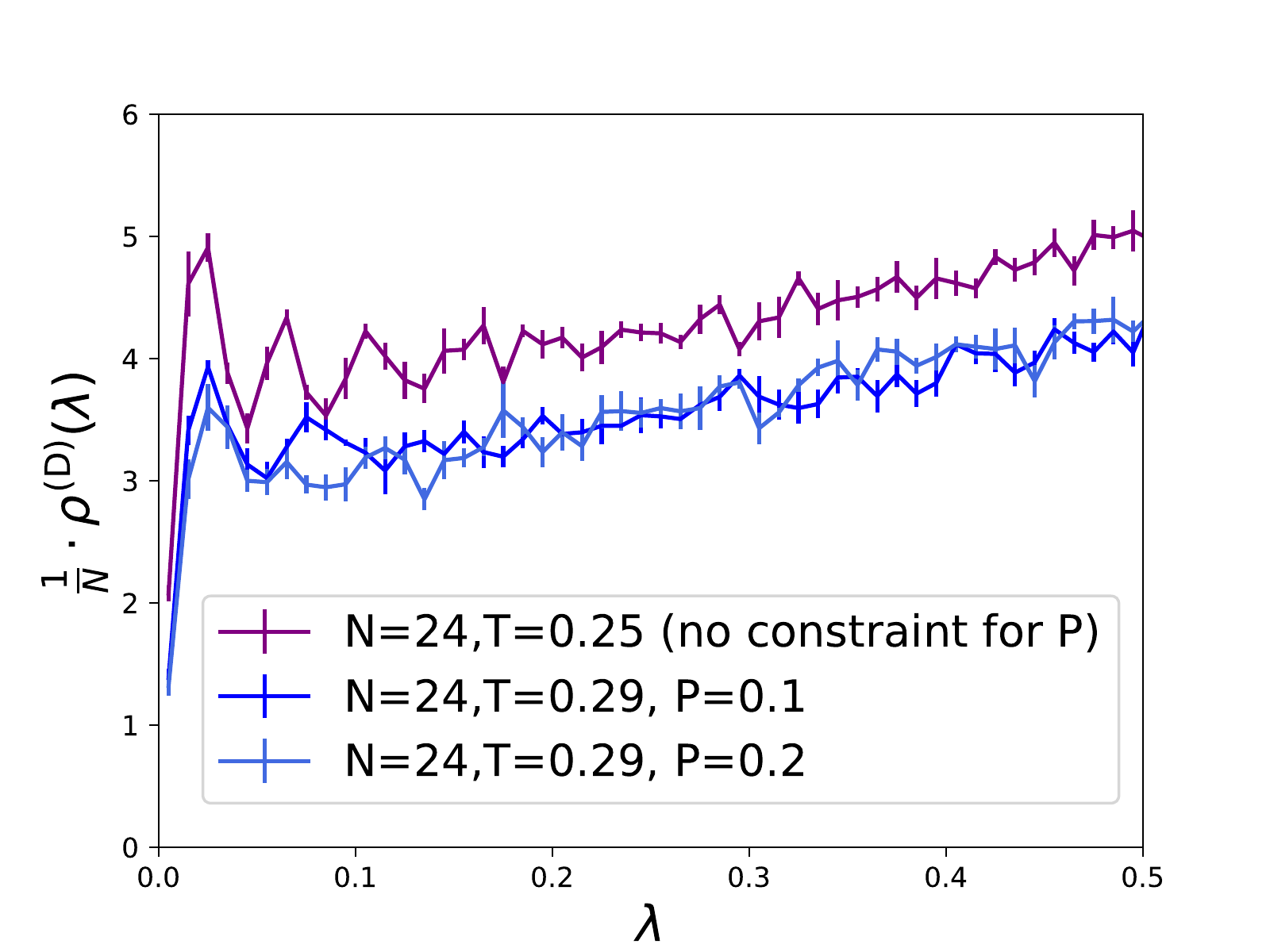}}
   \caption{}
   \end{subfigure}
\end{center}
\caption{(a) The density of Dirac eigenvalues, $\rho^{\rm (D)}(\lambda)$, divided by $N$, for $N=24$, $n_t=24$, $T=0.29$, and various $P$ above and below the GWW point, $P\simeq0.35$. In the partial phase, $P\lesssim 0.35$, the behaviour is similar to the confined phase in fig.~\ref{fig:Dirac_lowT}, whereas when $P\gtrsim0.35$, the behaviour is similar to that in the deconfined phase in fig.~\ref{fig:Dirac_highT}. (b) The same, but for $P=0.1$ and $0.2$ only, and now including $T=0.25$ with no $P$-constraint, i.e. the confined phase, to highlight the similarity.
}\label{fig:completely_confined_vs_partially_confined}
\end{figure}

Fig.~\ref{fig:Dirac_highT} shows our numerical results for $\frac{1}{N}\rho^{\rm (D)}(\lambda)$ for various $T$ in the deconfined phase for $N=8$, $12$ and $16$. For all of these $N$, we clearly observe that $\rho^{\rm (D)}(0)=0$, and hence $\langle \bar{\psi}\psi\rangle=0$ and chiral symmetry is preserved, as expected. Furthermore, by comparing $\frac{1}{N}\rho^{\rm (D)}(\lambda)$ for $N=8$, $12$ and $16$, we find no significant $N$-dependence, so in the $N\to\infty$ limit we again expect $\frac{1}{N}\rho^{\rm (D)}(\lambda)=0$, and hence $\langle \bar{\psi}\psi\rangle=0$ and chiral symmetry restoration. Notice the slope of $\frac{1}{N}\rho^{\rm (D)}(\lambda)$ as $\lambda \to 0$ increases as $T$ approaches the GWW-point. 

Fig.~\ref{fig:Dirac_lowT} shows our numerical results for $\frac{1}{N}\rho^{\rm (D)}(\lambda)$ at $T=0.25$ in the confined phase. In contrast to the deconfined phase, here we find that $\frac{1}{N}\rho^{\rm (D)}(0)\neq0$, and hence $\langle \bar{\psi}\psi\rangle\neq0$ and chiral symmetry is spontaneously broken, as expected. In fact, $\frac{1}{N}\rho^{\rm (D)}(\lambda)$ exhibits a small maximum and then a steep slope down as $\lambda \to 0$, both of which increase as $N$ increases.

Fig.~\ref{fig:Dirac_N8N12N16-T029} shows our numerical results for $\frac{1}{N}\rho^{\rm (D)}(\lambda)$ at $T=0.29$, near the GWW transition, for $N=12$, $16$, and $24$, and various $P$ below and above the GWW point at $P \simeq 0.35$. Figs.~\ref{fig:completely_confined_vs_partially_confined} shows the same, for a larger range of $P$, and also shows $T=0.25$, which is the confined phase.

In fig.~\ref{fig:Dirac_N8N12N16-T029} we see a clear difference as $P$ increases through $P\simeq0.35$. When $P\lesssim 0.35$, which is the partial phase, our results are similar to those of the confined phase in fig.~\ref{fig:Dirac_lowT}. In particular, $\frac{1}{N}\rho^{\rm (D)}(\lambda)$ exhibits a small maximum and a steep slope near $\lambda=0$, and most importantly, $\frac{1}{N}\rho^{\rm (D)}(0)\neq0$, hence $\langle \bar{\psi}\psi\rangle\neq0$ and chiral symmetry is spontaneously broken. This is our main result. In contrast, when $P\gtrsim0.35$, which is the deconfined phase, our results are qualitatively more similar to those in fig.~\ref{fig:Dirac_highT}, including a less pronounced maximum and slope near $\lambda=0$. Moreover, as $P$ increases we see that $\frac{1}{N}\rho^{\rm (D)}(0)$ approaches zero, suggesting chiral symmetry restoration.

Our results are insufficient to determine whether, as we increase $P$, the GWW transition and chiral symmetry breaking occur simultaneously, although that is a natural expectation. In any case, we have shown that complete confinement is not necessary for chiral symmetry breaking. Most importantly, we have demonstrated our main point in this section: for at least some range of $P$ in the partial phase, chiral symmetry is spontaneously broken, and hence $\langle \bar{\psi}\psi\rangle$ is an order parameter that can distinguish the partial and deconfined phases.

\section{Summary and Outlook}
\label{sec:conclusion}
\hspace{0.51cm}
In this paper, we studied the partial phase in two large-$N$ YM theories, namely weakly-coupled, softly-broken $\N=1$ SYM with $\theta=\pi$ and strongly-coupled lattice YM. In each case, we showed that in the partial phase both centre symmetry and a second global symmetry, either CP symmetry or chiral symmetry, respectively, are spontaneously broken. As a result, in each case an order parameter exists that can distinguish the partial phase from both the confined and deconfined phases. For $\N=1$ SYM we also presented finite-$N$ numerical evidence for the same phenomenon.

Our results raise some immediate questions for the two YM theories we studied. For example, in $\N=1$ SYM the adiabatic continuity conjecture~\cite{Poppitz:2012sw} proposes that the phase transitions that occur in the softly-broken $\N=1$ SYM theory are smoothly connected to those in the pure YM theory, i.e. that the phase structure is preserved with respect to increasing the gaugino mass. Do our results for the partial phase extend to larger gaugino masses? Could we find the same global symmetry breaking pattern and a GWW point at large gaugino mass values, or infinite gaugino mass, i.e. in pure YM with $\theta=\pi$? If so, then such agreement would provide a more stringent test of the adiabatic continuity conjecture than the existing tests, many of which could just be coincidences.

Looking farther afield, our numerical results, combined with anomaly-based arguments such as those in refs.~\cite{Gaiotto:2017yup,Chen:2020syd}, motivate the conjecture that the partial phase can be defined by spontaneous breaking of global symmetries in general, and in particular may apply for any $N$, not just for large $N$. Indeed, our results suggest that the novel phase found in ref.~\cite{Chen:2020syd}, for the weakly-coupled, softly-broken $\N=1$ SYM with $\theta=\pi$ with gauge group $SU(2)$, is a partial phase. Spontaneous breaking of global symmetries may provide a definition of the GWW transition beyond the large-$N$ limit, and a definition that, in practical terms, may be easier to calculate in many theories. 

Spontaneous breaking of global symmetries also means the partial phase's spectrum will be distinct from both the confined or deconfined phases, implying the existence of a either a Goldstone boson, if the global symmetry is continuous (like chiral symmetry in our lattice YM example) or domain walls, if the global symmetry is discrete (like CP symmetry in our $\N=1$ SYM example). An obvious and important task is thus to formulate effective field theory descriptions of partial phases, and to explore their physical consequences.

An especially important question is whether global symmetries can be used to identify partial phases in theories closer to QCD, including $N=3$, $\theta=0$, and physical quark masses. If a partial phase exists, then a crucial question is what observable effects it may have for the quark-gluon plasma created in heavy ion collider and the early universe, as well as for neutron stars and many other systems.

The concept of partial phases was originally motivated by holography, to explain the field theory dual to certain phase transitions in the bulk gravity theory~\cite{Hanada:2016pwv,Hanada:2021ipb}. For instance, the small black hole phase in global $\mathrm{AdS}_5$ is conjectured to be dual to a partial phase in ${\cal N}=4$ SYM on ${\rm S}^1\times{\rm S}^3$. If this is the case, then the point at which the specific heat of the AdS black hole phase becomes negative would be a GWW point. Following the findings of this paper, an obvious question is whether probe fermions added to the dual ${\cal N}=4$ SYM exhibit any global symmetry breaking when the specific heat becomes negative. This could provide compelling evidence for the conjecture, and a better understanding of the holographic mapping. A closely related question is whether a partial phase appears in the deconfinement transition observed in the index of ${\cal N}=4$ SYM on ${\rm S}^1\times{\rm S}^3$~\cite{Choi:2018vbz}. 

Clearly partial phases deserve further study, and need to be understood in greater detail. Although here we focused on global symmetries, another possible characterisation of the partial phase at finite $N$ and infinite volume is based on the spontaneous breaking of \textit{gauge} symmetry~\cite{Hanada:2019czd}. When $SU(M)$ inside $SU(N)$ deconfines, we must specify the embedding of $SU(M)$ into $SU(N)$ as a boundary condition, where different boundary conditions correspond to different superselection sectors. Equivalently, the partial phase can be characterised as partial breaking of the genuine gauge symmetry~\cite{Hanada:2020uvt,Hanada:2021ipb}. A key question is whether the partial phase defined in this way agrees with a definition based on breaking of global symmetries.

We intend to study many of these, and other related, questions, in the future.

\section*{Acknowledgements}
\hspace{0.51cm}
The authors thank Yuya Tanizaki and Hiromasa Watanabe for stimulating discussions and collaborations at an early stage, and Hidenori Fukaya and Kostas Skenderis for very useful discussions. MH would like to thank Ramy Brustein, Jordan Cotler and Ira Wolfson for discussions regarding the Eguchi-Kawai model in a somewhat related context, which was useful for this work. MH also thanks the Yukawa Institute for Theoretical Physics at Kyoto University, where this work was discussed during the YITP-W-21-09 ``QCD phase diagram and lattice QCD''. We acknowledge support from STFC through Consolidated Grant ST/P000711/1. MH was supported by the STFC Ernest Rutherford Grant ST/R003599/1. JH thanks STFC for the support of a studentship. A.O'B. is a Royal Society University Research Fellow. All numerical calculations are done on laptops and home PC's, while we worked from home following the UK government's lockdown measure. We thank our families for sharing the electricity bills with us.

\appendix 
\section{Second order Taylor method for softly-broken $\N=1$ SYM large-$N$ numerical calculations}
\hspace{0.51cm} \label{Appendix:2ndOrderTaylorMethod}
The second order Taylor method of solving an initial value problem is the natural extension of the Euler method to a second order Taylor expansion. That is, using analytic expressions for the first and second derivatives at the $i$-th iteration, $X'_i\equiv X'(X_i)$ and $X''_i\equiv X''(X_i)$, from eqs.~\eqref{eq:continuumDiffEqn} and~\eqref{eq:dXdt} respectively, we find the value of the $(i+1)$-th iteration by second order Taylor expansion,
\begin{align}
\label{eq:2ndOrderTaylor}
    X_{i+1} = X_{i} + X'_{i} \delta t + X''_{i} \frac{\delta t ^ 2}{2},
\end{align}
for small $\delta t$. The local truncation error is $\mathcal{O}(\delta t ^ 3)$, rather than the $\mathcal{O}(\delta t^2)$ associated with Euler's method, and we avoid incurring large errors near the minimum, $X_i=X(-1/2)$, where $X'$ is zero but $X''$ is very large. Moreover, in our case, near the GWW transition point, typical solutions exhibit a sharp gradient. To handle these sharp gradients, we limit the maximum change $\delta X(t)$ at each step of the Taylor method to $1\%$ of the current magnitude of $X(t)$ by decreasing the step size $\delta t$ as necessary.

Numerical uncertainties in $V$ and $M_0$ (as reported for large $N$ in appendix~\ref{Appendix:theta-0-table}), arise from three places: from errors in the numerical integration of the constraint eq.~\eqref{eq:continuumConstraint}, from the numerical integration of the continuum limit of eq.~\eqref{effective-action} when calculating $V$, and from the approximations made by the second order Taylor method. Uncertainties in the numerical integrals are calculated using concave and convex properties of the integrands. Below, we discuss how we determine uncertainties arising from the second order Taylor approximation and compose these with the errors from numerical integration. We assume throughout that floating point precision errors are negligible.

We can write the exact relation, 
\begin{align}
X_{i+1}^{\rm exact} = X_i + X'_t\delta t_i + \int_{t_i}^{t_{i+1}}ds\int_{t_i}^s du \, X''(u)
\end{align}
By differentiating eq.~\eqref{eq:continuumDiffEqn} we find that the third derivative is always negative between the start and end points (defined by $X'=0$). We therefore have $X''(t_i)\geq X''(t) \geq X''(t_{i+1})$ for $t_i\leq t \leq t_{i+1}$, and from this we can derive upper and lower bounds, $X^{\rm UB}_{i+1}$ and $X^{\rm LB}_{i+1}$, on $X_{i+1}^{\rm exact}$, namely
\begin{subequations}
\begin{align}
    &X^{\rm UB}_{i+1} \equiv X_i + X'_i \delta t_i +  X''_i \frac{\delta t_i^2}{2} \geq X_{i+1}^{\rm exact} \\
    &X^{\rm LB}_{i+1} \equiv X_i + X'_i \delta t_i +  X''_{i+1} \frac{\delta t_i^2}{2} \leq X_{i+1}^{\rm exact} \label{eq:X_LB}
\end{align}
\end{subequations}
These hold even if $\delta t$ is not small. The upper bound $X^{UB}$ coincides with our definition eq.~\eqref{eq:2ndOrderTaylor}. The approximate value of $X_{i+1}$ found at each iteration of our original computation is thus an overestimate. Since the terminating value of $X$ will be reached sooner, the parameter $\tau=\sum_i \delta t_i$ evaluated in this way is underestimated. More precisely, for the choices of $X_0=X(-1/2)$ and $X''_0=X''(-1/2)$ that give the correct solution as $\delta t\to 0$, the maximum of $X$ is obtained slightly below $\tau=1/2$ due to the discretisation effect at finite $\delta t$. 

To obtain lower bounds on $X_i$, and subsequently an upper bound on the value of $\tau$, we use eq.~\eqref{eq:X_LB}. We use the values of $X_i$ obtained in our first computation, and solve the quadratic equation to obtain values of $\delta t_i$ at each iteration $i$ such that eq.~\eqref{eq:X_LB} is satisfied. For any given $X_0$ and $X''_0$, each $\delta t_i$, and hence their sum $\tau$, will be overestimated by this procedure. We thus have a lower bound on $X(t)$ and upper bound on $\tau$.

We can find bounds on our evaluation of the second constraint, eq.~\eqref{eq:continuumConstraint}, by feeding both the upper and lower bounds on $X_i$ into the numerical integration, and compounding each result with the uncertainty introduced by the integration.

Next, we find the uncertainty in $X_0$ and $X_0''$. For any given value of $X_0$ and $X_0''$, we have a range of uncertainty around the computed values of $\tau$ and the constraint. If we can obtain values of $X_0$ and $X_0''$ such that the most pessimistic part of these uncertainty ranges lies just around the target values ($1/2$ for $\tau$ and $0$ for the constraint), then we obtain bounds on $X_0$ and $X_0''$. We must do this for each of the four combinations of signs, i.e. above and below $1/2$ and $0$ for $\tau$ and the constraint, respectively.

We achieve this using the 2D Newton-Raphson method, as follows. We define
\beq
c \equiv \int_{-1/2}^{1/2} dt \log X(t),
\eeq
so that the constraint in eq.~\eqref{eq:continuumConstraint} is satisfied when $c = i\theta$, which for us means $c=0$. We then compute the values $\frac{\partial\tau}{\partial X_0}$, $\frac{\partial\tau}{\partial X_0''}$, $\frac{\partial c}{\partial X_0}$, and $\frac{\partial c}{\partial X_0''}$, and calculate $\Delta X_0$ and $\Delta X_0''$ as 
\begin{align*}
    \begin{pmatrix}
    \Delta X_0 \\
    \Delta X_0''
    \end{pmatrix}
    = \frac{1}{\rm det}
    \begin{pmatrix}
    \frac{\partial c}{\partial X_0''} & -\frac{\partial \tau}{\partial X_0''}\\
    -\frac{\partial c}{\partial X_0} & \frac{\partial \tau}{\partial X_0}
    \end{pmatrix}
    \begin{pmatrix}
    \tau-\frac{1}{2} \\
    \mathcal{N}
    \end{pmatrix},
\end{align*}
where ``$\rm det$'' represents the determinant of the $2\times 2$ matrix shown on the right-hand side. We then update the values of $X_0$ and $X_0''$ as $X_0\rightarrow X_0-\Delta X_0$ and $X_0''\rightarrow X_0''-\Delta X_0''$, respectively. We repeat this such that $X_0$ and $X''_0$ approach the solution of $\tau=1/2$ and $c=0$, until sufficiently good precision is achieved. We repeat this for all four combinations of signs for $\tau-1/2$ and $c$. We thus obtain upper and lower values of $X_0$ and $X_0''$.

Finally, we find errors on $V$ by combining all the previous uncertainties with the uncertainties from the numerical integration, which are calculated using concave and convex properties of the integrand.

\section{Numerical data for the partial phase in softly-broken $\N=1$ SYM}
\hspace{0.51cm}
\label{Appendix:numericaldata}
In this appendix we summarise the numerical data that we used for the partial-phase saddle in mass-deformed $\N=1$ SYM. For finite $N$, we performed the minimisation procedure described in sec.~\ref{sec:nummethods} until the values cease to change within the machine precision. One should not take last few digits too seriously, because there can be round-off errors. 
\subsection{$\theta=0$} \label{Appendix:theta-0-table}
\hspace{0.51cm}
\begin{center}
{\small
\begin{supertabular}{|c||c|c|} 
\hline
$\tilde{\gamma}$ & $N$ & $V$\\
\hline
\hline
35& 30 & $ -0.8478043903335438$\\
& 50 & $ -0.5102166657650717$\\
& 70 & $ -0.3595906166171574$\\
& 100 & $ -0.25107574410395184$\\
\hline
40& 30 & $ -1.0649718464701392$\\
& 50 & $ -0.6332206873888775$\\
& 70 & $ -0.45061468574478286$\\
& 100 & $ -0.3145884066304803$\\
\hline
45& 30 & $ -1.2785511307705075$\\
& 50 & $ -0.761118429321458$\\
& 70 & $ -0.5420426727252557$\\
& 100 & $ -0.3787049502744858$\\
\hline
50& 30 & $ -1.4907670060318092$\\
& 50 & $ -0.8894724152036553$\\
& 70 & $ -0.6342449477861769$\\
& 100 & $ -0.44359705462577126$\\
\hline
55& 30 & $ -1.702131639184869$\\
& 50 & $ -1.0181099952602852$\\
& 70 & $ -0.7266752877179159$\\
& 100 & $ -0.5084837953775918$\\
\hline
     \end{supertabular}
}
\\
 \vspace{3mm}
$N=30, 50, 70$ and $100$, partial phase, $\theta=0$
\end{center}
\newpage
\begin{center}
{\small
\begin{supertabular}{|c||c|c|}
\hline
$\tilde{\gamma}$ & $V$ & $M_0$\\
\hline
\hline
$ 34.105 $&$ -0.8077784649468457 $&$ 1.6912265451456194 \times 10^{ -3 } $\\ 
\hline 
$ 34.11 $&$ -0.808006900961617 $&$ 1.825013252705753 \times 10^{ -3 } $\\ 
\hline 
$ 34.12 $&$ -0.8084628031642389 $&$ 1.9478656004707329 \times 10^{ -3 } $\\ 
\hline 
$ 34.13 $&$ -0.8089179187701646 $&$ 2.036195524651749 \times 10^{ -3 } $\\ 
\hline 
$ 34.15 $&$ -0.8098264888913062 $&$ 2.1755532782433296 \times 10^{ -3 } $\\ 
\hline 
$ 34.2 $&$ -0.8120911012726231 $&$ 2.4397435964616405 \times 10^{ -3 } $\\ 
\hline 
$ 34.3 $&$ -0.8166006952471729 $&$ 2.8419172665191623 \times 10^{ -3 } $\\ 
\hline 
$ 34.4 $&$ -0.8210917533707988 $&$ 3.1768860264097065 \times 10^{ -3 } $\\ 
\hline 
$ 34.6 $&$ -0.8300338142941242 $&$ 3.759119403517365 \times 10^{ -3 } $\\ 
\hline 
$ 34.8 $&$ -0.8389351432399467 $&$ 4.281329673944455 \times 10^{ -3 } $\\ 
\hline 
$ 35.0 $&$ -0.8478043903335438 $&$ 4.7702137949790195 \times 10^{ -3 } $\\ 
\hline 
$ 36.0 $&$ -0.8918117613679025 $&$ 6.995520728491199 \times 10^{ -3 } $\\ 
\hline 
$ 37.0 $&$ -0.9354339738397822 $&$ 9.08813053438817 \times 10^{ -3 } $\\ 
\hline 
$ 38.0 $&$ -0.9787946802571408 $&$ 1.1156329329176953 \times 10^{ -2 } $\\ 
\hline 
$ 39.0 $&$ -1.0219599197166551 $&$ 1.3242392404282448 \times 10^{ -2 } $\\ 
\hline 
$ 40.0 $&$ -1.0649718464701392 $&$ 1.5370757634584244 \times 10^{ -2 } $\\ 
\hline 
$ 45.0 $&$ -1.2785511307705075 $&$ 2.722990757508838 \times 10^{ -2 } $\\ 
\hline 
$ 50.0 $&$ -1.4907670060318092 $&$ 4.288292912186151 \times 10^{ -2 } $\\ 
\hline 
$ 55.0 $&$ -1.702131639184869 $&$ 6.710582512934164 \times 10^{ -2 } $\\ 
\hline 
$ 60.0 $&$ -1.9116692104565214 $&$0.10823277553764502$\\ 
\hline 
$ 65.0 $&$ -2.1161089139314724 $&$0.18747909204305171$\\ 
\hline 
$ 70.0 $&$ -2.311793371129293 $&$0.3079225504046863$\\ 
\hline

     \end{supertabular}
}
\\
\vspace{3mm}
$N=30$, partial phase, $\theta=0$
\end{center}

\begin{center}
{\small
\begin{supertabular}{|c||c|c|}
\hline
$\tilde{\gamma}$ & $NV$ & $M_0$\\
\hline
\hline
30 &$ 	-18.74626162(8)$&$ 	6.3142(0)\times 10^{-49}$\\
\hline
31&$ 	-19.98133943(4)$&$ 	5.55392(3)\times 10^{-16}$\\	
\hline
32&$ 	-21.2204182(5)$&$ 	5.20898(7)\times 10^{-10}$\\
\hline
33&$ 	-22.4637859(1)$&$ 	1.62375(1)\times 10^{-7}$\\
\hline
34&$ 	-23.711824(6)$&$	3.72298(3)\times 10^{-6}$\\	
\hline
35&$ 	-24.9650239(7)$&$	2.64232(4)\times 10^{-5}$\\  
\hline
36&$ 	-26.2238647(7)$&$	1.0095981(8)\times 10^{-4}$\\  
\hline
37&$ 	-27.4886645(9)$&$ 	2.6907652(0)\times 10^{-4}$\\  
\hline
38&$ 	-28.7595114(5)$&$	5.726169(6)\times 10^{-4}$\\	
\hline
39&$ 	-30.0362787(7)$&$	1.04973585(9)\times 10^{-3}$\\  
\hline
40&$ 	-31.3186700(6)$&$	1.7345736(7)\times 10^{-3}$\\  
\hline
41&$ 	-32.6062631(6)$&$	2.6581096(1)\times 10^{-3}$\\   
\hline
42&$ 	-33.8985461(4)$&$	3.8491581(7)\times 10^{-3}$\\   
\hline
43&$ 	-35.1949437(6)$&$	5.3352114(3)\times 10^{-3}$\\
\hline
44&$ 	-36.4948366(6)$&$	7.1430956(1)\times 10^{-3}$\\   
\hline
45&$ 	-37.7975750(8)$&$ 	9.2994790(6)\times 10^{-3}$\\ 
\hline
46&$ 	-39.1024887(4)$&$	1.183127(7)\times 10^{-2}$\\   
\hline
47&$ 	-40.4088940(9)$&$	1.47659901(3)\times 10^{-2}$\\  
\hline
48&$ 	-41.7160994(9)$&$ 	1.81319999(2)\times 10^{-2}$\\  
\hline
49&$ 	-43.0234090(4)$&$	2.19588494(7)\times 10^{-2}$\\   
\hline
50&$ 	-44.3301255(2)$&$	2.62775163(4) \times 10^{-2}$\\
\hline
51&$ 	-45.6355526(2)$&$	3.11206952(7) \times 10^{-2}$\\
\hline     
52&$ 	-46.9389962(8)$&$	3.65231002(2) \times 10^{-2}$\\
\hline     
53&$ 	-48.2397661(0)$&$	4.25217969(9) \times 10^{-2}$\\
\hline     
54&$ 	-49.5371762(1)$&$	4.91565774(0) \times 10^{-2}$\\
\hline     
55&$ 	-50.8305458(6)$&$	5.64703882(0)\times 10^{-2}$\\
\hline     
56&$ 	-52.1191999(9)$&$	6.4509830(3)\times 10^{-2}$\\
\hline     
57&$ 	-53.4024695(7)$&$	7.3325746(4)\times 10^{-2}$\\
\hline     
58&$ 	-54.6796918(6)$&$	8.2973921(7) \times 10^{-2}$\\
\hline     
59&$ 	-55.9502105(9)$&$	9.3515928(8) \times 10^{-2}$\\
\hline     
60&$ 	-57.2133759(5)$&$	0.105020155(6)$\\
\hline     
     \end{supertabular}
}
\\
\vspace{3mm}
Large $N$, partial phase, $\theta=0$
\end{center}

\subsection{$\theta\sim \pi$}
\hspace{0.51cm}

\begin{center}
{\small
\begin{supertabular}{|c|c||c|c|c|}
\hline
$\tilde{\gamma}$ & $\theta$ & $V$ & ${\rm Re}M_0$ & ${\rm Im}M_0$\\
\hline
\hline
$32.2$&$3.14159$&$-0.7170743511036082$&$ -1.3600120580223471 \times 10^{ -3 } $&$ 2.5812519554220423 \times 10^{ -4 } $ \\ 
&$3.14155$&$-0.7170743628838336$&$ -1.3609155565518565 \times 10^{ -3 } $&$ 2.4784827954029973 \times 10^{ -4 } $ \\ 
&$3.14149$&$-0.7170743795687697$&$ -1.3624561102521084 \times 10^{ -3 } $&$ 2.2927790368880996 \times 10^{ -4 } $ \\ 
\hline 
$32.21$&$3.14159$&$-0.717533767794569$&$ -1.3321141681612805 \times 10^{ -3 } $&$ 4.350584609566822 \times 10^{ -4 } $ \\ 
&$3.14155$&$-0.7175337879330195$&$ -1.3326243195853749 \times 10^{ -3 } $&$ 4.3165909238074365 \times 10^{ -4 } $ \\ 
&$3.14149$&$-0.717533817841014$&$ -1.3334057063425766 \times 10^{ -3 } $&$ 4.264017413306071 \times 10^{ -4 } $ \\ 
\hline 
$32.22$&$3.14159$&$-0.7179929889543566$&$ -1.3043571146458457 \times 10^{ -3 } $&$ 5.57310975551346 \times 10^{ -4 } $ \\ 
&$3.14155$&$-0.7179930147641919$&$ -1.3047479039944685 \times 10^{ -3 } $&$ 5.552750211788038 \times 10^{ -4 } $ \\ 
&$3.14149$&$-0.717993053301827$&$ -1.305339812640364 \times 10^{ -3 } $&$ 5.5217786311380666 \times 10^{ -4 } $ \\ 
\hline 
$32.23$&$3.14159$&$-0.7184520170186798$&$ -1.276726078298016 \times 10^{ -3 } $&$ 6.562673111173997 \times 10^{ -4 } $ \\ 
&$3.14155$&$-0.7184520473852677$&$ -1.2770527104979121 \times 10^{ -3 } $&$ 6.5481765509790115 \times 10^{ -4 } $ \\ 
&$3.14149$&$-0.7184520928099734$&$ -1.2775455721013584 \times 10^{ -3 } $&$ 6.526246943145053 \times 10^{ -4 } $ \\ 
\hline 
$32.24$&$3.14159$&$-0.7189108543080747$&$ -1.2492167707726551 \times 10^{ -3 } $&$ 7.413419732325752 \times 10^{ -4 } $ \\ 
&$3.14155$&$-0.7189108885694004$&$ -1.2495016117582143 \times 10^{ -3 } $&$ 7.402186698172351 \times 10^{ -4 } $ \\ 
&$3.14149$&$-0.7189109398649564$&$ -1.2499306258614793 \times 10^{ -3 } $&$ 7.385239606680516 \times 10^{ -4 } $ \\ 
\hline 
$32.25$&$3.14159$&$-0.7193695030812562$&$ -1.2218263069640964 \times 10^{ -3 } $&$ 8.169019109646184 \times 10^{ -4 } $ \\ 
&$3.14155$&$-0.719369540783399$&$ -1.2220810489925724 \times 10^{ -3 } $&$ 8.159864591338513 \times 10^{ -4 } $ \\ 
&$3.14149$&$-0.719369597258429$&$ -1.2224643426274118 \times 10^{ -3 } $&$ 8.146073893530602 \times 10^{ -4 } $ \\ 
\hline 
$32.3$&$3.14159$&$-0.721659994897846$&$ -1.0865777302750799 \times 10^{ -3 } $&$ 1.1133356734658564 \times 10^{ -3 } $ \\ 
&$3.14155$&$-0.7216600459096133$&$ -1.0867518649578847 \times 10^{ -3 } $&$ 1.112865030557661 \times 10^{ -3 } $ \\ 
&$3.14149$&$-0.7216601223879526$&$ -1.08701339135787 \times 10^{ -3 } $&$ 1.1121579353577529 \times 10^{ -3 } $ \\ 
\hline 
$32.35$&$3.14159$&$-0.7239460866385014$&$ -9.539872225466342 \times 10^{ -4 } $&$ 1.336510099405864 \times 10^{ -3 } $ \\ 
&$3.14155$&$-0.7239461474331051$&$ -9.541226194560151 \times 10^{ -4 } $&$ 1.3361958057500578 \times 10^{ -3 } $ \\ 
&$3.14149$&$-0.7239462385992684$&$ -9.543258607049511 \times 10^{ -4 } $&$ 1.3357239597633632 \times 10^{ -3 } $ \\ 
\hline 
$32.4$&$3.14159$&$-0.7262280139802966$&$ -8.238336317452523 \times 10^{ -4 } $&$ 1.5198967085948667 \times 10^{ -3 } $ \\ 
&$3.14155$&$-0.7262280826292161$&$ -8.239449058808901 \times 10^{ -4 } $&$ 1.519661261553193 \times 10^{ -3 } $ \\ 
&$3.14149$&$-0.7262281855836519$&$ -8.241119012794183 \times 10^{ -4 } $&$ 1.5193078955209953 \times 10^{ -3 } $ \\ 
\hline 
$32.45$&$3.14159$&$-0.7285059915660974$&$ -6.95920795442486 \times 10^{ -4 } $&$ 1.6772622461292543 \times 10^{ -3 } $ \\ 
&$3.14155$&$-0.7285060668033074$&$ -6.960151327319915 \times 10^{ -4 } $&$ 1.6770737855282707 \times 10^{ -3 } $ \\ 
&$3.14149$&$-0.7285061796442005$&$ -6.96156691043507 \times 10^{ -4 } $&$ 1.67679098903428 \times 10^{ -3 } $ \\ 
\hline 
$32.5$&$3.14159$&$-0.7307802155542057$&$ -5.700733563845978 \times 10^{ -4 } $&$ 1.815922434919294 \times 10^{ -3 } $ \\ 
&$3.14155$&$-0.7307802964681885$&$ -5.701549451005102 \times 10^{ -4 } $&$ 1.8157648292755948 \times 10^{ -3 } $ \\ 
&$3.14149$&$-0.7307804178268624$&$ -5.702773652676167 \times 10^{ -4 } $&$ 1.8155283552895594 \times 10^{ -3 } $ \\ 
\hline 
$32.55$&$3.14159$&$-0.7330508657711814$&$ -4.46133941861139 \times 10^{ -4 } $&$ 1.9403360418338673 \times 10^{ -3 } $ \\ 
&$3.14155$&$-0.7330509516677158$&$ -4.462054880739285 \times 10^{ -4 } $&$ 1.940200026377375 \times 10^{ -3 } $ \\ 
&$3.14149$&$-0.7330510805020409$&$ -4.463128344139536 \times 10^{ -4 } $&$ 1.9399959606733754 \times 10^{ -3 } $ \\ 
\hline 
$32.6$&$3.14159$&$-0.7353181075532847$&$ -3.2396082753576216 \times 10^{ -4 } $&$ 2.0534664742781557 \times 10^{ -3 } $ \\ 
&$3.14155$&$-0.7353181978832048$&$ -3.240242052503035 \times 10^{ -4 } $&$ 2.053346256161984 \times 10^{ -3 } $ \\ 
&$3.14149$&$-0.7353183333689319$&$ -3.2411929223816873 \times 10^{ -4 } $&$ 2.05316590012776 \times 10^{ -3 } $ \\ 
\hline 
$32.7$&$3.14159$&$-0.7398429640031353$&$ -8.441353218502272 \times 10^{ -5 } $&$ 2.2536927081194666 \times 10^{ -3 } $ \\ 
&$3.14155$&$-0.7398430619362288$&$ -8.446433117245173 \times 10^{ -5 } $&$ 2.2535937029615045 \times 10^{ -3 } $ \\ 
&$3.14149$&$-0.7398432088284732$&$ -8.454054168750659 \times 10^{ -5 } $&$ 2.2534451813331 \times 10^{ -3 } $ \\ 
\hline 
$32.8$&$3.14159$&$-0.7443558729386548$&$ 1.4945625418492055 \times 10^{ -4 } $&$ 2.427763658793177 \times 10^{ -3 } $ \\ 
&$3.14155$&$-0.7443559772114889$&$ 1.4941476202572692 \times 10^{ -4 } $&$ 2.427677878174378 \times 10^{ -3 } $ \\ 
&$3.14149$&$-0.7443561336144284$&$ 1.4935251527640314 \times 10^{ -4 } $&$ 2.42754919912249 \times 10^{ -3 } $ \\ 
\hline 
$33.0$&$3.14159$&$-0.7533494785206036$&$ 6.029864611902578 \times 10^{ -4 } $&$ 2.722097577747968 \times 10^{ -3 } $ \\ 
&$3.14155$&$-0.7533495928971548$&$ 6.029579386735867 \times 10^{ -4 } $&$ 2.722026565347336 \times 10^{ -3 } $ \\ 
&$3.14149$&$-0.7533497644568607$&$ 6.029151504548835 \times 10^{ -4 } $&$ 2.721920043916067 \times 10^{ -3 } $ \\ 
\hline 
$33.5$&$3.14159$&$-0.7756774553170565$&$ 1.6768822556210576 \times 10^{ -3 } $&$ 3.2794358050263086 \times 10^{ -3 } $ \\ 
&$3.14155$&$-0.7756775865578464$&$ 1.6768711746605895 \times 10^{ -3 } $&$ 3.2793770628401324 \times 10^{ -3 } $ \\ 
&$3.14149$&$-0.7756777834149097$&$ 1.6768545514918272 \times 10^{ -3 } $&$ 3.2792889492241932 \times 10^{ -3 } $ \\ 
\hline 
$34.0$&$3.14159$&$-0.7978302202388292$&$ 2.6962105102698835 \times 10^{ -3 } $&$ 3.707440095081149 \times 10^{ -3 } $ \\ 
&$3.14155$&$-0.7978303624083353$&$ 2.6962083806895687 \times 10^{ -3 } $&$ 3.707383288903868 \times 10^{ -3 } $ \\ 
&$3.14149$&$-0.797830575658689$&$ 2.696205185554002 \times 10^{ -3 } $&$ 3.707298079813313 \times 10^{ -3 } $ \\ 
\hline 
$35.0$&$3.14159$&$-0.8417550290603878$&$ 4.652679404706676 \times 10^{ -3 } $&$ 4.397709927810709 \times 10^{ -3 } $ \\ 
&$3.14155$&$-0.8417551856349794$&$ 4.652686740494166 \times 10^{ -3 } $&$ 4.397650104350867 \times 10^{ -3 } $ \\ 
&$3.14149$&$-0.8417554204929414$&$ 4.652697743717927 \times 10^{ -3 } $&$ 4.397560369302386 \times 10^{ -3 } $ \\ 
\hline 
$40.0$&$3.14159$&$-1.0573792156622257$&$ 1.4412981881233673 \times 10^{ -2 } $&$ 7.189654944745767 \times 10^{ -3 } $ \\ 
&$3.14155$&$-1.0573794090113$&$ 1.441300645625501 \times 10^{ -2 } $&$ 7.189564703253496 \times 10^{ -3 } $ \\ 
&$3.14149$&$-1.0573796990302957$&$ 1.4413043318185594 \times 10^{ -2 } $&$ 7.189429340916259 \times 10^{ -3 } $ \\ 
\hline 
$45.0$&$3.14159$&$-1.2699610845858604$&$ 2.580531090072105 \times 10^{ -2 } $&$ 1.0251017087791112 \times 10^{ -2 } $ \\ 
&$3.14155$&$-1.2699613031848664$&$ 2.580534712424157 \times 10^{ -2 } $&$ 1.025088921290766 \times 10^{ -2 } $ \\ 
&$3.14149$&$-1.2699616310781705$&$ 2.5805401458661206 \times 10^{ -2 } $&$ 1.0250697400390536 \times 10^{ -2 } $ \\ 
\hline 
$50.0$&$3.14159$&$-1.4812961563094476$&$ 4.094947544666873 \times 10^{ -2 } $&$ 1.4243699194655928 \times 10^{ -2 } $ \\ 
&$3.14155$&$-1.4812963972886606$&$ 4.0949524558640515 \times 10^{ -2 } $&$ 1.424352143939323 \times 10^{ -2 } $ \\ 
&$3.14149$&$-1.4812967587517454$&$ 4.094959822543871 \times 10^{ -2 } $&$ 1.42432548062422 \times 10^{ -2 } $ \\ 
\hline 
$55.0$&$3.14159$&$-1.6918402420861214$&$ 6.4604328760748135 \times 10^{ -2 } $&$ 2.0052826308490346 \times 10^{ -2 } $ \\ 
&$3.14155$&$-1.691840503923804$&$ 6.460439236308307 \times 10^{ -2 } $&$ 2.0052575094367473 \times 10^{ -2 } $ \\ 
&$3.14149$&$-1.6918408966740976$&$ 6.460448770552378 \times 10^{ -2 } $&$ 2.00521982780708 \times 10^{ -2 } $ \\ 
\hline 
$60.0$&$3.14159$&$-1.9005649352418637$&$ 0.10809898207091294 $&$ 2.8902218845528007 \times 10^{ -2 } $ \\ 
&$3.14155$&$-1.9005652175955352$&$ 0.10809905918554755 $&$ 2.890185488245071 \times 10^{ -2 } $ \\ 
&$3.14149$&$-1.9005656411193246$&$ 0.10809917465036076 $&$ 2.890130889764398 \times 10^{ -2 } $ \\ 
\hline 
$65.0$&$3.14159$&$-2.104169423293366$&$ 0.18387620479734793 $&$ 4.060558125607354 \times 10^{ -2 } $ \\ 
&$3.14155$&$-2.1041697270523865$&$ 0.18387629641317327 $&$ 4.060506857755586 \times 10^{ -2 } $ \\ 
&$3.14149$&$-2.104170182683694$&$ 0.18387643368075676 $&$ 4.060429953799894 \times 10^{ -2 } $ \\ 
\hline 
$70.0$&$3.14159$&$-2.298983857096773$&$ 0.30372551405890685 $&$ 5.480071091863914 \times 10^{ -2 } $ \\ 
&$3.14155$&$-2.29898418298924$&$ 0.30372562079881626 $&$ 5.480001782656032 \times 10^{ -2 } $ \\ 
&$3.14149$&$-2.298984671820188$&$ 0.30372578090795543 $&$ 5.4798978188303415 \times 10^{ -2 } $ \\ 
\hline 

\end{supertabular}
}
\\
\vspace{3mm}
$N=30$, partial phase at finite $\theta$
\end{center}

\newpage

\begin{center}
{\small 
\begin{supertabular}{|c|c||c|c|c|}
\hline
$\tilde{\gamma}$ & $\theta$ & $V$ & ${\rm Re}M_0$ & ${\rm Im}M_0$\\
\hline
\hline
$31.6$&$3.14159$&$-0.4155123088833649$&$ -1.8782005991409765 \times 10^{ -4 } $&$ 6.330578970860719 \times 10^{ -4 } $ \\ 
&$3.14155$&$-0.4155123251919556$&$ -1.8784922965670887 \times 10^{ -4 } $&$ 6.330043923677622 \times 10^{ -4 } $ \\ 
&$3.14149$&$-0.4155123496523715$&$ -1.878929966005143 \times 10^{ -4 } $&$ 6.32924113843514 \times 10^{ -4 } $ \\ 
\hline 
$31.7$&$3.14159$&$-0.4181552328004205$&$ -4.600790872308327 \times 10^{ -5 } $&$ 7.559655923117715 \times 10^{ -4 } $ \\ 
&$3.14155$&$-0.41815525200604925$&$ -4.602804756640492 \times 10^{ -5 } $&$ 7.559296156993012 \times 10^{ -4 } $ \\ 
&$3.14149$&$-0.4181552808128721$&$ -4.605826079233592 \times 10^{ -5 } $&$ 7.558756446524675 \times 10^{ -4 } $ \\ 
\hline 
$31.78$&$3.14159$&$-0.4202649022150887$&$ 6.326245956469611 \times 10^{ -5 } $&$ 8.332766036474448 \times 10^{ -4 } $ \\ 
&$3.14155$&$-0.42026492316425357$&$ 6.324671926096537 \times 10^{ -5 } $&$ 8.332471502032584 \times 10^{ -4 } $ \\ 
&$3.14149$&$-0.4202649545866925$&$ 6.322310593859195 \times 10^{ -5 } $&$ 8.33202967268706 \times 10^{ -4 } $ \\ 
\hline 
$31.802$&$3.14159$&$-0.42084439582102184$&$ 9.275084219401403 \times 10^{ -5 } $&$ 8.522634299284915 \times 10^{ -4 } $ \\ 
&$3.14155$&$-0.420844417187829$&$ 9.273606185621027 \times 10^{ -5 } $&$ 8.522352193851644 \times 10^{ -4 } $ \\ 
&$3.14149$&$-0.4208444492367917$&$ 9.271388887576544 \times 10^{ -5 } $&$ 8.521929013447016 \times 10^{ -4 } $ \\ 
\hline 
$31.82$&$3.14159$&$-0.4213183239133828$&$ 1.1671237306609559 \times 10^{ -4 } $&$ 8.671909558256232 \times 10^{ -4 } $ \\ 
&$3.14155$&$-0.4213183456054398$&$ 1.1669831738462748 \times 10^{ -4 } $&$ 8.671636395208687 \times 10^{ -4 } $ \\ 
&$3.14149$&$-0.4213183781423188$&$ 1.1667723159806762 \times 10^{ -4 } $&$ 8.671226631305268 \times 10^{ -4 } $ \\ 
\hline 
$31.85$&$3.14159$&$-0.42210780928338515$&$ 1.5633301651780533 \times 10^{ -4 } $&$ 8.909669581169569 \times 10^{ -4 } $ \\ 
&$3.14155$&$-0.4221078314876758$&$ 1.5632006291092861 \times 10^{ -4 } $&$ 8.909409280614963 \times 10^{ -4 } $ \\ 
&$3.14149$&$-0.42210786479296847$&$ 1.5630063057297492 \times 10^{ -4 } $&$ 8.909018814793864 \times 10^{ -4 } $ \\ 
\hline 
$31.9$&$3.14159$&$-0.42342256294940883$&$ 2.215471299404181 \times 10^{ -4 } $&$ 9.278918024374619 \times 10^{ -4 } $ \\ 
&$3.14155$&$-0.4234225859345137$&$ 2.2153576841864933 \times 10^{ -4 } $&$ 9.278674695536766 \times 10^{ -4 } $ \\ 
&$3.14149$&$-0.4234226204111069$&$ 2.2151872465800584 \times 10^{ -4 } $&$ 9.278309692573936 \times 10^{ -4 } $ \\ 
\hline 
$32.5$&$3.14159$&$-0.4391174957274477$&$ 9.496371234248153 \times 10^{ -4 } $&$ 1.2294221709574773 \times 10^{ -3 } $ \\ 
&$3.14155$&$-0.43911752431471013$&$ 9.496345451950141 \times 10^{ -4 } $&$ 1.229403158143993 \times 10^{ -3 } $ \\ 
&$3.14149$&$-0.4391175671948034$&$ 9.496306776463555 \times 10^{ -4 } $&$ 1.2293746389938343 \times 10^{ -3 } $ \\ 
\hline 
$34.0$&$3.14159$&$-0.4779612628170003$&$ 2.601455269143662 \times 10^{ -3 } $&$ 1.653288813560361 \times 10^{ -3 } $ \\ 
&$3.14155$&$-0.4779612968106036$&$ 2.601457687671169 \times 10^{ -3 } $&$ 1.6532672341409855 \times 10^{ -3 } $ \\ 
&$3.14149$&$-0.4779613478001801$&$ 2.6014613153755284 \times 10^{ -3 } $&$ 1.6532348650414865 \times 10^{ -3 } $ \\ 
\hline 
$45.0$&$3.14159$&$-0.7592450724009824$&$ 1.734899931471533 \times 10^{ -2 } $&$ 4.515391915490412 \times 10^{ -3 } $ \\ 
&$3.14155$&$-0.7592451200935519$&$ 1.7349010423645999 \times 10^{ -2 } $&$ 4.5153349993345975 \times 10^{ -3 } $ \\ 
&$3.14149$&$-0.7592451916312689$&$ 1.734902708674579 \times 10^{ -2 } $&$ 4.515249625049745 \times 10^{ -3 } $ \\ 
\hline 
$50.0$&$3.14159$&$-0.8874198012140362$&$ 3.0056225594155026 \times 10^{ -2 } $&$ 6.954488692854987 \times 10^{ -3 } $ \\ 
&$3.14155$&$-0.887419853467416$&$ 3.005624166450025 \times 10^{ -2 } $&$ 6.954400874243307 \times 10^{ -3 } $ \\ 
&$3.14149$&$-0.887419931846239$&$ 3.0056265769649566 \times 10^{ -2 } $&$ 6.954269146282227 \times 10^{ -3 } $ \\ 
\hline 

\end{supertabular}
}
\\
\vspace{3mm}
$N=50$, partial phase at finite $\theta$
\end{center}

\newpage
\begin{center}
{\small
\begin{supertabular}{|c|c||c|c|c|}
\hline
$\tilde{\gamma}$ & $\theta$ & $V$ & ${\rm Re}M_0$ & ${\rm Im}M_0$\\
\hline
\hline
$31.325$&$3.14159$&$-0.2918407654541345$&$ 7.98026871679131 \times 10^{ -5 } $&$ 4.424333155597573 \times 10^{ -4 } $ \\ 
&$3.14155$&$-0.291840773245185$&$ 7.979585621482465 \times 10^{ -5 } $&$ 4.42420336394707 \times 10^{ -4 } $ \\ 
&$3.14149$&$-0.29184078493135135$&$ 7.978560880788008 \times 10^{ -5 } $&$ 4.42400866869941 \times 10^{ -4 } $ \\ 
\hline 
$31.35$&$3.14159$&$-0.2923040923705275$&$ 1.0278346984026125 \times 10^{ -4 } $&$ 4.550621221981385 \times 10^{ -4 } $ \\ 
&$3.14155$&$-0.29230410035776927$&$ 1.0277720816889515 \times 10^{ -4 } $&$ 4.550497714063126 \times 10^{ -4 } $ \\ 
&$3.14149$&$-0.29230411233824344$&$ 1.0276781494419844 \times 10^{ -4 } $&$ 4.5503124467492855 \times 10^{ -4 } $ \\ 
\hline 
$31.4$&$3.14159$&$-0.29323011303276025$&$ 1.4803076391381091 \times 10^{ -4 } $&$ 4.78144033280332 \times 10^{ -4 } $ \\ 
&$3.14155$&$-0.29323012137172827$&$ 1.4802545804784515 \times 10^{ -4 } $&$ 4.781326262086492 \times 10^{ -4 } $ \\ 
&$3.14149$&$-0.29323013387982394$&$ 1.480174987299652 \times 10^{ -4 } $&$ 4.781155153336815 \times 10^{ -4 } $ \\ 
\hline 
$31.5$&$3.14159$&$-0.29507983238035607$&$ 2.360576804307633 \times 10^{ -4 } $&$ 5.175995910049874 \times 10^{ -4 } $ \\ 
&$3.14155$&$-0.2950798412980503$&$ 2.3605377995782924 \times 10^{ -4 } $&$ 5.175893000261233 \times 10^{ -4 } $ \\ 
&$3.14149$&$-0.2950798546742728$&$ 2.360479289004306 \times 10^{ -4 } $&$ 5.175738634434299 \times 10^{ -4 } $ \\ 
\hline 
$31.75$&$3.14159$&$-0.29969279644826335$&$ 4.455054068316847 \times 10^{ -4 } $&$ 5.921697420074003 \times 10^{ -4 } $ \\ 
&$3.14155$&$-0.29969280636979745$&$ 4.455035033503333 \times 10^{ -4 } $&$ 5.921603720842227 \times 10^{ -4 } $ \\ 
&$3.14149$&$-0.29969282125181285$&$ 4.45500648040117 \times 10^{ -4 } $&$ 5.921463172458622 \times 10^{ -4 } $ \\ 
\hline 
$32.0$&$3.14159$&$-0.30429298401236626$&$ 6.449889616569478 \times 10^{ -4 } $&$ 6.4801563646402665 \times 10^{ -4 } $ \\ 
&$3.14155$&$-0.30429299459764153$&$ 6.449881034089653 \times 10^{ -4 } $&$ 6.480063166785355 \times 10^{ -4 } $ \\ 
&$3.14149$&$-0.3042930104752741$&$ 6.449868159820404 \times 10^{ -4 } $&$ 6.479923370314297 \times 10^{ -4 } $ \\ 
\hline 
$32.5$&$3.14159$&$-0.3134658092525986$&$ 1.0279762018595526 \times 10^{ -3 } $&$ 7.34053348582117 \times 10^{ -4 } $ \\ 
&$3.14155$&$-0.313465820704502$&$ 1.0279764159264388 \times 10^{ -3 } $&$ 7.340435347025221 \times 10^{ -4 } $ \\ 
&$3.14149$&$-0.31346583788207316$&$ 1.0279767369974029 \times 10^{ -3 } $&$ 7.340288139099266 \times 10^{ -4 } $ \\ 
\hline 
$33.0$&$3.14159$&$-0.3226126313086142$&$ 1.4010179461914722 \times 10^{ -3 } $&$ 8.042442954538322 \times 10^{ -4 } $ \\ 
&$3.14155$&$-0.322612643339618$&$ 1.4010187169395665 \times 10^{ -3 } $&$ 8.042338161760316 \times 10^{ -4 } $ \\ 
&$3.14149$&$-0.3226126613858309$&$ 1.4010198732807868 \times 10^{ -3 } $&$ 8.04218097538276 \times 10^{ -4 } $ \\ 
\hline 
$34.0$&$3.14159$&$-0.3408572002751646$&$ 2.142071796713208 \times 10^{ -3 } $&$ 9.269612602029794 \times 10^{ -4 } $ \\ 
&$3.14155$&$-0.3408572131130061$&$ 2.1420731847348127 \times 10^{ -3 } $&$ 9.269493944101995 \times 10^{ -4 } $ \\ 
&$3.14149$&$-0.34085723236945964$&$ 2.142075266736319 \times 10^{ -3 } $&$ 9.269315957258254 \times 10^{ -4 } $ \\ 
\hline 
$35.0$&$3.14159$&$-0.3590631133644247$&$ 2.895574207226948 \times 10^{ -3 } $&$ 1.042688775208161 \times 10^{ -3 } $ \\ 
&$3.14155$&$-0.3590631268061734$&$ 2.8955759853564587 \times 10^{ -3 } $&$ 1.0426755056744041 \times 10^{ -3 } $ \\ 
&$3.14149$&$-0.3590631469684751$&$ 2.8955786523457916 \times 10^{ -3 } $&$ 1.0426556014402968 \times 10^{ -3 } $ \\ 
\hline 
$40.0$&$3.14159$&$-0.45000014388659393$&$ 7.225957517480029 \times 10^{ -3 } $&$ 1.6925857932954413 \times 10^{ -3 } $ \\ 
&$3.14155$&$-0.45000015953462313$&$ 7.225960775567354 \times 10^{ -3 } $&$ 1.6925643698951656 \times 10^{ -3 } $ \\ 
&$3.14149$&$-0.45000018300629346$&$ 7.225965662618408 \times 10^{ -3 } $&$ 1.6925322347811091 \times 10^{ -3 } $ \\ 
\hline 
$45.0$&$3.14159$&$-0.5413574476911213$&$ 1.3626015882620472 \times 10^{ -2 } $&$ 2.697613400021317 \times 10^{ -3 } $ \\ 
&$3.14155$&$-0.5413574651378426$&$ 1.3626021051768908 \times 10^{ -2 } $&$ 2.6975792579866846 \times 10^{ -3 } $ \\ 
&$3.14149$&$-0.5413574913075072$&$ 1.3626028805361086 \times 10^{ -2 } $&$ 2.6975280449202055 \times 10^{ -3 } $ \\ 
\hline 
$50.0$&$3.14159$&$-0.6334972947003091$&$ 2.6550506491932127 \times 10^{ -2 } $&$ 4.66305021094182 \times 10^{ -3 } $ \\ 
&$3.14155$&$-0.6334973137359372$&$ 2.6550514317777956 \times 10^{ -2 } $&$ 4.662991057611511 \times 10^{ -3 } $ \\ 
&$3.14149$&$-0.6334973422889246$&$ 2.6550526056381325 \times 10^{ -2 } $&$ 4.662902327597171 \times 10^{ -3 } $ \\ 
\hline 

\end{supertabular}
}
\\
\vspace{3mm}
$N=70$, partial phase at finite $\theta$
\end{center}

\bibliographystyle{utphys}
\bibliography{R3S1}

\providecommand{\href}[2]{#2}\begingroup\raggedright\begin{thebibliography}{10}

\bibitem{Witten:1998zw}
E.~Witten, ``{Anti-de Sitter space, thermal phase transition, and confinement
  in gauge theories},''
  \href{http://dx.doi.org/10.4310/ATMP.1998.v2.n3.a3}{{\em Adv. Theor. Math.
  Phys.} {\bfseries 2} (1998) 505--532},
  \href{http://arxiv.org/abs/hep-th/9803131}{{\ttfamily arXiv:hep-th/9803131}}.

\bibitem{Sundborg:1999ue}
B.~Sundborg, ``{The Hagedorn transition, deconfinement and N=4 SYM theory},''
  \href{http://dx.doi.org/10.1016/S0550-3213(00)00044-4}{{\em Nucl. Phys. B}
  {\bfseries 573} (2000) 349--363},
  \href{http://arxiv.org/abs/hep-th/9908001}{{\ttfamily arXiv:hep-th/9908001}}.

\bibitem{Aharony:2003sx}
O.~Aharony, J.~Marsano, S.~Minwalla, K.~Papadodimas, and M.~Van~Raamsdonk,
  ``{The Hagedorn - deconfinement phase transition in weakly coupled large N
  gauge theories},'' \href{http://dx.doi.org/10.4310/ATMP.2004.v8.n4.a1}{{\em
  Adv. Theor. Math. Phys.} {\bfseries 8} (2004) 603--696},
  \href{http://arxiv.org/abs/hep-th/0310285}{{\ttfamily arXiv:hep-th/0310285}}.

\bibitem{Hanada:2016pwv}
M.~Hanada and J.~Maltz, ``{A proposal of the gauge theory description of the
  small Schwarzschild black hole in AdS$_5\times$S$^5$},''
  \href{http://dx.doi.org/10.1007/JHEP02(2017)012}{{\em JHEP} {\bfseries 02}
  (2017) 012}, \href{http://arxiv.org/abs/1608.03276}{{\ttfamily
  arXiv:1608.03276 [hep-th]}}.

\bibitem{Berenstein:2018lrm}
D.~Berenstein, ``{Submatrix deconfinement and small black holes in AdS},''
  \href{http://dx.doi.org/10.1007/JHEP09(2018)054}{{\em JHEP} {\bfseries 09}
  (2018) 054}, \href{http://arxiv.org/abs/1806.05729}{{\ttfamily
  arXiv:1806.05729 [hep-th]}}.

\bibitem{Hanada:2018zxn}
M.~Hanada, G.~Ishiki, and H.~Watanabe, ``{Partial Deconfinement},''
  \href{http://dx.doi.org/10.1007/JHEP03(2019)145}{{\em JHEP} {\bfseries 03}
  (2019) 145}, \href{http://arxiv.org/abs/1812.05494}{{\ttfamily
  arXiv:1812.05494 [hep-th]}}. [Erratum: JHEP 10, 029 (2019)].

\bibitem{Hanada:2019czd}
M.~Hanada, A.~Jevicki, C.~Peng, and N.~Wintergerst, ``{Anatomy of
  Deconfinement},'' \href{http://dx.doi.org/10.1007/JHEP12(2019)167}{{\em JHEP}
  {\bfseries 12} (2019) 167}, \href{http://arxiv.org/abs/1909.09118}{{\ttfamily
  arXiv:1909.09118 [hep-th]}}.

\bibitem{Hanada:2019kue}
M.~Hanada and B.~Robinson, ``{Partial-Symmetry-Breaking Phase Transitions},''
  \href{http://dx.doi.org/10.1103/PhysRevD.102.096013}{{\em Phys. Rev. D}
  {\bfseries 102} no.~9, (2020) 096013},
  \href{http://arxiv.org/abs/1911.06223}{{\ttfamily arXiv:1911.06223
  [hep-th]}}.

\bibitem{Hanada:2019rzv}
M.~Hanada, G.~Ishiki, and H.~Watanabe, ``{Partial deconfinement in gauge
  theories},'' \href{http://dx.doi.org/10.22323/1.363.0055}{{\em PoS}
  {\bfseries LATTICE2019} (2019) 055},
  \href{http://arxiv.org/abs/1911.11465}{{\ttfamily arXiv:1911.11465
  [hep-lat]}}.

\bibitem{Hanada:2020uvt}
M.~Hanada, H.~Shimada, and N.~Wintergerst, ``{Color confinement and
  Bose-Einstein condensation},''
  \href{http://dx.doi.org/10.1007/JHEP08(2021)039}{{\em JHEP} {\bfseries 08}
  (2021) 039}, \href{http://arxiv.org/abs/2001.10459}{{\ttfamily
  arXiv:2001.10459 [hep-th]}}.

\bibitem{Watanabe:2020ufk}
H.~Watanabe, G.~Bergner, N.~Bodendorfer, S.~Shiba~Funai, M.~Hanada, E.~Rinaldi,
  A.~Schaefer, and P.~Vranas, ``{Partial Deconfinement at Strong Coupling on
  the Lattice},'' \href{http://arxiv.org/abs/2005.04103}{{\ttfamily
  arXiv:2005.04103 [hep-th]}}.

\bibitem{Gross:1980he}
D.~J. Gross and E.~Witten, ``{Possible Third Order Phase Transition in the
  Large N Lattice Gauge Theory},''
  \href{http://dx.doi.org/10.1103/PhysRevD.21.446}{{\em Phys. Rev. D}
  {\bfseries 21} (1980) 446--453}.

\bibitem{Wadia:2012fr}
S.~R. Wadia, ``{A Study of U(N) Lattice Gauge Theory in 2-dimensions},''
  \href{http://arxiv.org/abs/1212.2906}{{\ttfamily arXiv:1212.2906 [hep-th]}}.

\bibitem{Hanada:2021ipb}
M.~Hanada, ``{Bulk geometry in gauge/gravity duality and color degrees of
  freedom},'' \href{http://dx.doi.org/10.1103/PhysRevD.103.106007}{{\em Phys.
  Rev. D} {\bfseries 103} no.~10, (2021) 106007},
  \href{http://arxiv.org/abs/2102.08982}{{\ttfamily arXiv:2102.08982
  [hep-th]}}.

\bibitem{Asplund:2008xd}
C.~T. Asplund and D.~Berenstein, ``{Small AdS black holes from SYM},''
  \href{http://dx.doi.org/10.1016/j.physletb.2009.02.043}{{\em Phys. Lett. B}
  {\bfseries 673} (2009) 264--267},
  \href{http://arxiv.org/abs/0809.0712}{{\ttfamily arXiv:0809.0712 [hep-th]}}.

\bibitem{Gaiotto:2017yup}
D.~Gaiotto, A.~Kapustin, Z.~Komargodski, and N.~Seiberg, ``{Theta, Time
  Reversal, and Temperature},''
  \href{http://dx.doi.org/10.1007/JHEP05(2017)091}{{\em JHEP} {\bfseries 05}
  (2017) 091}, \href{http://arxiv.org/abs/1703.00501}{{\ttfamily
  arXiv:1703.00501 [hep-th]}}.

\bibitem{Dashen:1970et}
R.~F. Dashen, ``{Some features of chiral symmetry breaking},''
  \href{http://dx.doi.org/10.1103/PhysRevD.3.1879}{{\em Phys. Rev. D}
  {\bfseries 3} (1971) 1879--1889}.

\bibitem{Chen:2020syd}
S.~Chen, K.~Fukushima, H.~Nishimura, and Y.~Tanizaki, ``{Deconfinement and
  $\mathcal {CP}$ breaking at $\theta=\pi$ in Yang-Mills theories and a novel
  phase for SU(2)},'' \href{http://dx.doi.org/10.1103/PhysRevD.102.034020}{{\em
  Phys. Rev. D} {\bfseries 102} no.~3, (2020) 034020},
  \href{http://arxiv.org/abs/2006.01487}{{\ttfamily arXiv:2006.01487
  [hep-th]}}.

\bibitem{Witten:1982df}
E.~Witten, ``{Constraints on Supersymmetry Breaking},''
  \href{http://dx.doi.org/10.1016/0550-3213(82)90071-2}{{\em Nucl. Phys. B}
  {\bfseries 202} (1982) 253}.

\bibitem{Witten:1982im}
E.~Witten, ``{Supersymmetry and Morse theory},'' {\em J. Diff. Geom.}
  {\bfseries 17} no.~4, (1982) 661--692.

\bibitem{Poppitz:2012sw}
E.~Poppitz, T.~Sch\"afer, and M.~\"Unsal, ``{Continuity, Deconfinement, and
  (Super) Yang-Mills Theory},''
  \href{http://dx.doi.org/10.1007/JHEP10(2012)115}{{\em JHEP} {\bfseries 10}
  (2012) 115}, \href{http://arxiv.org/abs/1205.0290}{{\ttfamily arXiv:1205.0290
  [hep-th]}}.

\bibitem{Davies:1999uw}
N.~M. Davies, T.~J. Hollowood, V.~V. Khoze, and M.~P. Mattis, ``{Gluino
  condensate and magnetic monopoles in supersymmetric gluodynamics},''
  \href{http://dx.doi.org/10.1016/S0550-3213(99)00434-4}{{\em Nucl. Phys. B}
  {\bfseries 559} (1999) 123--142},
  \href{http://arxiv.org/abs/hep-th/9905015}{{\ttfamily arXiv:hep-th/9905015}}.

\bibitem{Davies_2003}
N.~M. Davies, T.~J. Hollowood, and V.~V. Khoze, ``{Monopoles, affine algebras
  and the gluino condensate},'' \href{http://dx.doi.org/10.1063/1.1586477}{{\em
  J. Math. Phys.} {\bfseries 44} (2003) 3640--3656},
  \href{http://arxiv.org/abs/hep-th/0006011}{{\ttfamily arXiv:hep-th/0006011}}.

\bibitem{Poppitz:2012nz}
E.~Poppitz, T.~Sch\"afer, and M.~\"Unsal, ``{Universal mechanism of
  (semi-classical) deconfinement and theta-dependence for all simple groups},''
  \href{http://dx.doi.org/10.1007/JHEP03(2013)087}{{\em JHEP} {\bfseries 03}
  (2013) 087}, \href{http://arxiv.org/abs/1212.1238}{{\ttfamily arXiv:1212.1238
  [hep-th]}}.

\bibitem{Anber_2014}
M.~M. Anber, E.~Poppitz, and B.~Teeple, ``{Deconfinement and continuity between
  thermal and (super) Yang-Mills theory for all gauge groups},''
  \href{http://dx.doi.org/10.1007/JHEP09(2014)040}{{\em JHEP} {\bfseries 09}
  (2014) 040}, \href{http://arxiv.org/abs/1406.1199}{{\ttfamily arXiv:1406.1199
  [hep-th]}}.

\bibitem{Poppitz:2021cxe}
E.~Poppitz, ``{Notes on Confinement on $\mathbf{R^3 \times S^1}$: From
  Yang-Mills, super-Yang-Mills, and QCD(adj) to QCD(F)},''
  \href{http://arxiv.org/abs/2111.10423}{{\ttfamily arXiv:2111.10423
  [hep-th]}}.

\bibitem{Poppitz:2008hr}
E.~Poppitz and M.~Unsal, ``{Index theorem for topological excitations on R**3 x
  S**1 and Chern-Simons theory},''
  \href{http://dx.doi.org/10.1088/1126-6708/2009/03/027}{{\em JHEP} {\bfseries
  03} (2009) 027}, \href{http://arxiv.org/abs/0812.2085}{{\ttfamily
  arXiv:0812.2085 [hep-th]}}.

\bibitem{Unsal_2009}
M.~Unsal, ``{Magnetic bion condensation: A New mechanism of confinement and
  mass gap in four dimensions},''
  \href{http://dx.doi.org/10.1103/PhysRevD.80.065001}{{\em Phys. Rev. D}
  {\bfseries 80} (2009) 065001},
  \href{http://arxiv.org/abs/0709.3269}{{\ttfamily arXiv:0709.3269 [hep-th]}}.

\bibitem{Gross:1980br}
D.~J. Gross, R.~D. Pisarski, and L.~G. Yaffe, ``{QCD and Instantons at Finite
  Temperature},'' \href{http://dx.doi.org/10.1103/RevModPhys.53.43}{{\em Rev.
  Mod. Phys.} {\bfseries 53} (1981) 43}.

\bibitem{Eguchi:1982nm}
T.~Eguchi and H.~Kawai, ``{Reduction of Dynamical Degrees of Freedom in the
  Large N Gauge Theory},''
  \href{http://dx.doi.org/10.1103/PhysRevLett.48.1063}{{\em Phys. Rev. Lett.}
  {\bfseries 48} (1982) 1063}.

\bibitem{Banks:1979yr}
T.~Banks and A.~Casher, ``{Chiral Symmetry Breaking in Confining Theories},''
  \href{http://dx.doi.org/10.1016/0550-3213(80)90255-2}{{\em Nucl. Phys. B}
  {\bfseries 169} (1980) 103--125}.

\bibitem{Duane:1987de}
S.~Duane, A.~D. Kennedy, B.~J. Pendleton, and D.~Roweth, ``{Hybrid Monte
  Carlo},'' \href{http://dx.doi.org/10.1016/0370-2693(87)91197-X}{{\em Phys.
  Lett. B} {\bfseries 195} (1987) 216--222}.

\bibitem{Hanada:2014noa}
M.~Hanada, J.~Maltz, and L.~Susskind, ``{Deconfinement transition as black hole
  formation by the condensation of QCD strings},''
  \href{http://dx.doi.org/10.1103/PhysRevD.90.105019}{{\em Phys. Rev. D}
  {\bfseries 90} no.~10, (2014) 105019},
  \href{http://arxiv.org/abs/1405.1732}{{\ttfamily arXiv:1405.1732 [hep-th]}}.

\bibitem{Choi:2018vbz}
S.~Choi, J.~Kim, S.~Kim, and J.~Nahmgoong, ``{Comments on deconfinement in
  AdS/CFT},'' \href{http://arxiv.org/abs/1811.08646}{{\ttfamily
  arXiv:1811.08646 [hep-th]}}.

\end{thebibliography}\endgroup

\end{document}